\documentstyle[preprint,aps,epsf]{revtex} 

\newcommand{\et}{\mbox{$E_{T}$}}
\newcommand{\met}{\mbox{$\protect \raisebox{.3ex}{$\not$}\et$}}

\newcommand{\pbarp}{\mbox{$\overline{p}p$}}

\newcommand{\zmumu}{\mbox{$Z \rightarrow \mu^+\mu^-$}}

\newcommand{\ee}    {\mbox{$e^+ e^-$}}
\newcommand{\zee}    {\mbox{$Z \rightarrow\ e^+ e^-$}}

\newcommand{\wenu}   {\mbox{$W \rightarrow\ e \nu$}}
\newcommand{\wmunu}  {\mbox{$W \rightarrow\ \mu \nu$}}
\newcommand{\wtaunu} {\mbox{$W \rightarrow\ \tau \nu$}}

\newcommand{\mumu}  {\mbox{$\mu^+\mu^-$}}
\newcommand{\ptw}  {\mbox{$p_T^W$}}
\newcommand{\ptm}    {\mbox{$p_{T}$}}

\newcommand{\upara}  {\mbox{$\rm{u_{\parallel}}$}}
\newcommand{\uperp}  {\mbox{$\rm{u_{\perp}}$}}

\newcommand{\sumet} {\mbox{$\Sigma E_T$}}
\newcommand{\dzero}  {\hbox{DO\kern-0.62em\raise+0.2ex\hbox{/}}}

\begin{document}

\begin{center}
{\bf Measurement of the $W$ Boson Mass
	with the Collider Detector at Fermilab}
\end{center}

\font\eightit=cmti8
\def\r#1{\ignorespaces $^{#1}$}
\hfilneg
\begin{sloppypar}
\noindent
T.~Affolder,\r {21} H.~Akimoto,\r {43}
A.~Akopian,\r {36} M.~G.~Albrow,\r {10} P.~Amaral,\r 7 S.~R.~Amendolia,\r {32} 
D.~Amidei,\r {24} K.~Anikeev,\r {22} J.~Antos,\r 1 
G.~Apollinari,\r {10}
% ...
\end{sloppypar}
\vskip .026in
\begin{center}
(CDF Collaboration)
\end{center}

\vskip .026in
\begin{center}
\r 1  {\eightit Institute of Physics, Academia Sinica, Taipei, Taiwan 11529, 
Republic of China} \\
\r 2  {\eightit Argonne National Laboratory, Argonne, Illinois 60439} \\
\r 3  {\eightit Istituto Nazionale di Fisica Nucleare, University of Bologna,
I-40127 Bologna, Italy} \\
\r 4  {\eightit Brandeis University, Waltham, Massachusetts 02254} \\
\r 5  {\eightit University of California at Los Angeles, Los 
Angeles, California  90024} \\  
\r 6  {\eightit Instituto de Fisica de Cantabria, CSIC-University of Cantabria, 
39005 Santander, Spain} \\
\r 7  {\eightit Enrico Fermi Institute, University of Chicago, Chicago, 
Illinois 60637} \\
\r 8  {\eightit Joint Institute for Nuclear Research, RU-141980 Dubna, Russia}
\\
\r 9  {\eightit Duke University, Durham, North Carolina  27708} \\
\r {10}  {\eightit Fermi National Accelerator Laboratory, Batavia, Illinois 
60510} \\
\r {11} {\eightit University of Florida, Gainesville, Florida  32611} \\
\r {12} {\eightit Laboratori Nazionali di Frascati, Istituto Nazionale di Fisica
               Nucleare, I-00044 Frascati, Italy} \\
\r {13} {\eightit University of Geneva, CH-1211 Geneva 4, Switzerland} \\
\r {14} {\eightit Harvard University, Cambridge, Massachusetts 02138} \\
\r {15} {\eightit Hiroshima University, Higashi-Hiroshima 724, Japan} \\
\r {16} {\eightit University of Illinois, Urbana, Illinois 61801} \\
\r {17} {\eightit The Johns Hopkins University, Baltimore, Maryland 21218} \\
\r {18} {\eightit Institut f\"{u}r Experimentelle Kernphysik, 
Universit\"{a}t Karlsruhe, 76128 Karlsruhe, Germany} \\
\r {19} {\eightit Korean Hadron Collider Laboratory: Kyungpook National
University, Taegu 702-701; Seoul National University, Seoul 151-742; and
SungKyunKwan University, Suwon 440-746; Korea} \\
\r {20} {\eightit High Energy Accelerator Research Organization (KEK), Tsukuba, 
Ibaraki 305, Japan} \\
\r {21} {\eightit Ernest Orlando Lawrence Berkeley National Laboratory, 
Berkeley, California 94720} \\
\r {22} {\eightit Massachusetts Institute of Technology, Cambridge,
Massachusetts  02139} \\   
\r {23} {\eightit Institute of Particle Physics: McGill University, Montreal 
H3A 2T8; and University of Toronto, Toronto M5S 1A7; Canada} \\
\r {24} {\eightit University of Michigan, Ann Arbor, Michigan 48109} \\
\r {25} {\eightit Michigan State University, East Lansing, Michigan  48824} \\
\r {26} {\eightit University of New Mexico, Albuquerque, New Mexico 87131} \\
\r {27} {\eightit The Ohio State University, Columbus, Ohio  43210} \\
\r {28} {\eightit Osaka City University, Osaka 588, Japan} \\
\r {29} {\eightit University of Oxford, Oxford OX1 3RH, United Kingdom} \\
\r {30} {\eightit Universita di Padova, Istituto Nazionale di Fisica 
          Nucleare, Sezione di Padova, I-35131 Padova, Italy} \\
\r {31} {\eightit University of Pennsylvania, Philadelphia, 
        Pennsylvania 19104} \\   
\r {32} {\eightit Istituto Nazionale di Fisica Nucleare, University and Scuola
               Normale Superiore of Pisa, I-56100 Pisa, Italy} \\
\r {33} {\eightit University of Pittsburgh, Pittsburgh, Pennsylvania 15260} \\
\r {34} {\eightit Purdue University, West Lafayette, Indiana 47907} \\
\r {35} {\eightit University of Rochester, Rochester, New York 14627} \\
\r {36} {\eightit Rockefeller University, New York, New York 10021} \\
\r {37} {\eightit Rutgers University, Piscataway, New Jersey 08855} \\
\r {38} {\eightit Texas A\&M University, College Station, Texas 77843} \\
\r {39} {\eightit Texas Tech University, Lubbock, Texas 79409} \\
\r {40} {\eightit Istituto Nazionale di Fisica Nucleare, University of Trieste/
Udine, Italy} \\
\r {41} {\eightit University of Tsukuba, Tsukuba, Ibaraki 305, Japan} \\
\r {42} {\eightit Tufts University, Medford, Massachusetts 02155} \\
\r {43} {\eightit Waseda University, Tokyo 169, Japan} \\
\r {44} {\eightit University of Wisconsin, Madison, Wisconsin 53706} \\
\r {45} {\eightit Yale University, New Haven, Connecticut 06520} \\
\end{center}

\begin{abstract}
We present a measurement of the $W$ boson mass using data collected
with the CDF detector during the 1994-95 collider run at the Fermilab
Tevatron.  A fit to the transverse mass spectrum of a sample of 30,115
$\wenu$ events recorded in an integrated luminosity of 84~pb$^{-1}$
gives a mass 
$M_W = {\rm 80.473\pm 0.065(stat.) \pm 0.092(syst.)}$~GeV/c$^2$.
A fit to the transverse mass spectrum of a sample of 14,740 $\wmunu$
events from 80~pb$^{-1}$ gives a mass
$M_W = {\rm 80.465\pm 0.100(stat.) \pm 0.103(syst.)}$~GeV/c$^2$.
The dominant contributions to the systematic uncertainties are the
uncertainties in the electron energy scale and the muon momentum scale, 
0.075~GeV/c$^2$ and 0.085~GeV/c$^2$, respectively.
The combined value for the electron and muon channel
is $M_W = 80.470 \pm 0.089$~GeV/c$^2$.
When combined with previously published CDF measurements,
we obtain $M_W = 80.433 \pm 0.079$~GeV/c$^2$.
\end{abstract}

\section {Introduction}
\label{intro}

This paper describes a measurement of the $W$ mass using $W$ boson
decays observed in antiproton-proton ($\pbarp$) collisions produced 
at the Fermilab Tevatron with a center-of-mass energy of  1800~GeV.  
The results are from an analysis of the decays of the $W$ into a muon 
and neutrino in a data sample of integrated
luminosity of 80~pb$^{-1}$, and the decays of the $W$ into an electron and
neutrino in a data sample of 84~pb$^{-1}$, 
collected by the Collider Detector at
Fermilab (CDF) from 1994 to 1995.  This time period is referred to as
Run IB whereas the period from 1992 and 1993 with about 20~pb$^{-1}$ 
of integrated luminosity is referred to as Run IA.

The relations among the masses and couplings of gauge bosons  allow incisive
tests of the Standard Model of the electroweak  interactions~\cite{sm}. 
These relations include higher-order radiative corrections 
which are sensitive to the top quark mass, $M_{\rm top}$, and the
Higgs boson mass, $M_{\rm Higgs}$~\cite{review}.
The $W$ boson mass provides a significant test of the Standard Model 
in the context of measurements of the properties of the $Z$ boson, 
measurements of atomic transitions, muon decay, neutrino interactions, 
and searches for the Higgs boson.

Direct measurement of the $W$ mass originated at
the  antiproton-proton collider at CERN~\cite{spps}.
Measurements at the Fermilab Tevatron collider by CDF~\cite{wmass_1a} and 
\dzero~\cite{d0} have greatly improved precision.  
At LEP II, the $W$ boson mass has been measured from the $W$ pair
production cross section near threshold~\cite{lepth} and 
by direct reconstruction of the two $W$s~\cite{lepdr}.
The average of direct measurements including the analysis in this paper 
is of $80.39 \pm0.06$ GeV/c$^2$~\cite{mwave}.

Indirect $W$ mass determinations involve $Z$ boson measurements at
LEP and SLC~\cite{lepewwg}, charged- and neutral-current neutrino
interactions at Fermilab~\cite{nutev}, and the top quark 
mass measurement at Fermilab~\cite{top}.
A recent survey~\cite{lepewwg} gives a $W$ mass
of $80.381 \pm0.026$ GeV/c$^2$ inferred from indirect measurements.

The paper is structured as follows. A description of the detector and an 
overview of the analysis are given in Section~\ref{overview}. The calibration
and alignment of the central tracking chamber, 
which provides the momentum scale, is described in Section~\ref{mumeas}. 
Section~\ref{mumeas} also describes muon identification and  the
measurement of the momentum resolution.  Section~\ref{emeas}  describes
electron identification, the calorimeter
energy scale, and the measurement of the energy resolution. 
The effects of backgrounds are described in Section~\ref{backgrounds}.
Section~\ref{wprod} describes a Monte Carlo simulation of $W$ production
and decay, and QED radiative corrections.
Section~\ref{recoil} describes the measurement of the detector response
to the hadrons recoiling against the $W$ in the event, necessary to infer the
neutrino momentum scale and resolution. 
The knowledge of the lepton and recoil responses is
incorporated in the Monte Carlo simulation of $W$ production and decay.
Section~\ref{results} gives a description of the fitting method used to 
extract the $W$ mass from a comparison of the data and the simulation.  
It also presents a global summary of the measured values and the 
experimental uncertainties. Finally, the measured $W$ mass is compared 
to previous measurements and current predictions.

\section {Overview}
\label{overview}

This section begins with a discussion of how the nature of $W$ boson
production and decay motivates the strategy used to measure the $W$
mass.  The aspects of the detector and triggers critical to the 
measurement are then described.  A brief description of the data samples
used for the calibrations and for the mass measurement follows.
A summary of the analysis strategy and comparison of this analysis
with our last analysis concludes the section.

\subsection{Nature of $W$ Events}

\begin{figure}[p]
\vspace{-5.0cm}
        \centerline{\epsfysize 24cm
                    \epsffile{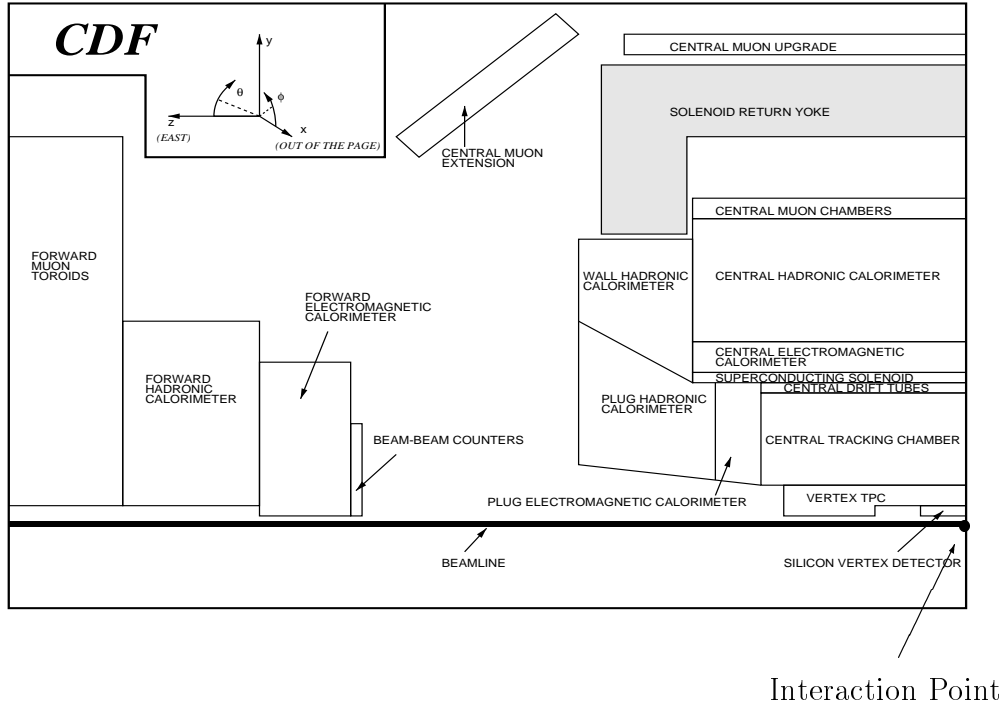}}
\vspace{-5.0cm}
\caption{One quarter of the CDF detector.  The detector is symmetric 
	about the interaction point. 
	CDF uses a cylindrical coordinate system with the $z$ 
	(longitudinal) axis along the proton  beam axis; $r$ is 
	the transverse coordinate, and $\phi$ is the azimuthal angle.  
	Pseudorapidity  ($\eta$) is defined as 
	$\eta\equiv-{\rm ln(tan}(\theta/2))$, where $\theta$ is the polar 
	angle relative to the proton-beam direction.}
\label{gencdfpic}
\end{figure}

\begin{figure}[p]
\vspace*{12cm}
\epsfysize=3.0in
\epsffile[125 146 500 345]{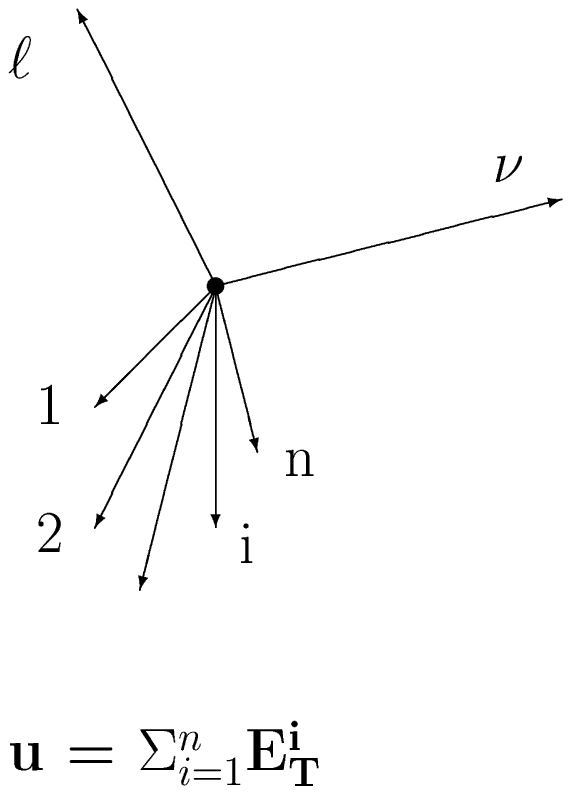}
\vspace*{-7.5cm}
\caption{Kinematics of $W$ boson production and decay for the events 
	used in this analysis, as viewed in the plane transverse to the 
	antiproton-proton beams. The recoil energy vector ${\bf u}$ 
	is the sum of the transverse energy vectors $\bf E_T^i$ 
	of the particles recoiling against the $W$. 
	Although energy is a scalar quantity, 
	``transverse energy'' commonly denotes the transverse component 
	of the vector whose $magnitude$ is the energy of the particle  
	and $direction$ is parallel to the momentum of the particle.}
\label{collide}
\end{figure}

The dominant mechanism for production of $W$ bosons in antiproton-proton 
collisions is antiquark-quark annihilation.  The $W$
is produced with momentum relative to the center-of-mass of the
antiproton-proton collision in the transverse ($x,y$) and longitudinal 
($z$) directions (see Figure~\ref{gencdfpic}). 
The transverse component of the momentum is
balanced by the transverse momentum of hadrons produced in association 
with the $W$, referred to as the ``recoil'', as  illustrated in 
Figure~\ref{collide}.


The $W$ boson decays  used in this analysis are the two-body 
leptonic decays producing 
an electron or muon and a neutrino.  Since the apparatus neither detects the
neutrino nor measures the $z$-component of the recoil momentum, much of
which is carried in fragments of the initial proton and antiproton 
at small angles relative to the beams, 
there is insufficient information to reconstruct the
invariant mass of the $W$ on an event-by-event basis.  This analysis
uses the  transverse mass of each $W$ event, which is analogous to 
the invariant mass except that only the components transverse to the 
beamline are used.  Specifically,
\begin{equation}
(M_{T}^{W})^{2}=(E_{T}^{\ell} + E_{T}^{\nu})^{2} - 
({\bf E}_{T}^{\ell}+{\bf E}_{T}^{\nu})^{2},
\label{mtrans}
\end{equation}
where $M_{T}^{W}$ is the transverse mass of the $W$,  $E_{T}^{\ell}$ is the
transverse energy (see Figure~\ref{collide})
of the electron or the transverse momentum of the muon,
and $E_{T}^{\nu}$ is
the transverse energy of the neutrino.
The boldface denotes two-component vector quantities. 
The transverse energy of the
neutrino is inferred from apparent energy imbalance in
the calorimeters,
\begin{equation}
{\protect \raisebox{.3ex}{$\not$}{\bf E}_T}
= {\bf E}_{T}^{\nu} = -({\bf E}_{T}^{\ell} + {\bf u}),
\label{nudef}
\end{equation} 
where ${\bf u}$ denotes the transverse energy vector of 
the recoil (see Figure~\ref{collide}) measured by the calorimeters.

\subsection{Detector and Triggers}

This section briefly describes those aspects of the CDF detector 
and triggers pertinent to the $W$ mass measurement. 
A more detailed detector description can be found in 
Reference~\cite{gennim}; recent detector upgrades are described in 
Reference~\cite{top_prd} and references therein.

The CDF detector is an  azimuthally and forward-backward symmetric
magnetic detector designed to study $\pbarp$ collisions  at the Tevatron. The
magnetic spectrometer consists of tracking devices inside a 3-m diameter, 5-m
long superconducting solenoidal magnet which operates  at 1.4~T. 
The calorimeter is
divided into a central region 
({\mbox{$30^{\circ} < \theta < 150^{\circ}$}) outside the solenoidal
magnet, end-plugs
(\mbox{$10^{\circ} < \theta < 30^{\circ}$},  
\mbox{$150^{\circ} < \theta < 170^{\circ}$}),  
which form the pole pieces for the solenoidal magnet, and  forward
and backward regions
(\mbox{$2^{\circ} < \theta < 10^{\circ}$},
\mbox{$170^{\circ} < \theta < 178^{\circ}$}).
Muon chambers are placed outside (at larger radius) 
of the hadronic calorimeters
in the central region and behind added shielding.
An elevation view of one quarter of
the CDF detector is shown in  Figure~\ref{gencdfpic}. 

\subsubsection{Tracking Detectors}
A four-layer silicon microstrip vertex detector 
(SVX$^\prime$)~\cite{svxprimenim} is used in
this analysis to provide a precision measurement of the 
location of the beam axis (luminous region). The SVX$^\prime$ is located
directly outside the 1.9-cm radius beryllium beampipe. The four layers of the
SVX$^\prime$ are at radii of 2.9, 4.3, 5.7, and 7.9~cm from the beamline. 
Outside the SVX$^\prime$
is a set of vertex time projection chambers (VTX)~\cite{vtxnim}, which provides
$r$-$z$ tracking information out to a radius of 22 cm for $|\eta| < 3.25$.  The
VTX is used in this analysis for finding the $z$ position of the
antiproton-proton interaction (the event vertex). The event vertex is 
necessary for event selection, lepton track reconstruction, and the calculation
of $E_T$. 

Both the SVX$^\prime$ and VTX are mounted inside the central tracking chamber
(CTC)~\cite{ctcnim}, a 3.2-m long drift chamber that extends in radius from 
31.0 cm to 132.5 cm.  The CTC has 84 sampling wire
layers, organized in 5 axial and 4
stereo ``super-layers''.   Axial super-layers have 12 radially
separated layers of sense wires, parallel to the $z$ axis, that  measure the
$r$-$\phi$\ position of a track.   Stereo super-layers have 6 sense wire
layers, with a $\sim\!2.5^\circ$ stereo angle, 
that  measure a combination of $r$-$\phi$\
and $z$ information.  The stereo angle direction alternates at each stereo
super-layer. Axial and stereo data are combined to form a 3-dimensional track.
Details of the calibration and alignment
of the CTC are given in  Section~\ref{mumeas}.

Track reconstruction uses $r$-$\phi$ information from the beam axis and the
CTC axial layers, and $z$ information from the VTX $z$ vertex and 
the CTC stereo layers.
In this analysis, the electron or muon momentum is measured from the curvature,
azimuthal angle, and polar angle of the track as the particle traverses the
magnetic field.

\subsubsection{Calorimeters}
The electromagnetic and hadronic calorimeters subtend 2$\pi$ in azimuth and from
$-$4.2 to 4.2 in pseudorapidity ($\eta$). The calorimeters are constructed with
a projective tower geometry, with towers subtending approximately 0.1 in 
pseudorapidity by 15$^{\circ}$ in $\phi$ (central) or 5$^{\circ}$ 
in $\phi$ (plug and forward). Each tower consists of an electromagnetic
calorimeter followed by a hadronic calorimeter at larger radius. The energies
of central  electrons used in the mass  measurement are measured from the
electromagnetic shower produced in the central electromagnetic calorimeter
(CEM)~\cite{cemnim}. The central calorimeter is constructed as 24 ``wedges'' in
$\phi$  for each half of the detector ($-1.1 < \eta < 0$ and $0 < \eta < 1.1$).
Each wedge has 10 electromagnetic towers, which use lead as the absorber and
scintillator as the active medium, for a total of 480 CEM 
towers.\footnote{There are actually only 478 physical CEM towers; 
the locations of two towers are used for the cryogenic penetration 
for the magnet.}   
A proportional chamber (CES) measures the electron shower position 
in the $\phi$ and $z$ directions at a
depth of $\sim 6$ radiation lengths in the CEM~\cite{cemnim}.  
A fiducial region of uniform
electromagnetic response is defined by avoiding the edges of the wedges.
For the purposes of triggering and data sample selection, 
the CEM calibrations are derived from 
testbeam data taken during 1984-85;  the
tower gains were set in March 1994 using Cesium-137 gamma-ray sources. 
Details of the further calibration of the CEM are given in Section~\ref{emeas}.

The calorimeters measure the energy flow of particles produced in
association with the $W$. Outside the CEM is a similarly segmented hadronic
calorimeter (CHA)~\cite{chanim}. Electromagnetic and hadronic
calorimeters 
which use multi-wire proportional chambers as the active sampling medium extend
this coverage to $|\eta| = 4.2$~\cite{gasnim}. 
In this analysis, however, the recoil energy is
calculated only in the region of full azimuthal symmetry, $|\eta|<3.6$.
Understanding the response of these devices to the recoil from bosons is 
difficult from first principles 
as it depends on details of the flow and energy 
distributions of the recoil hadrons.  The energy response to recoil 
energy is parameterized primarily using $\zee$ and $\zmumu$ events. 
Details of the calibration of the calorimeters
to recoil energy  are given in Section~\ref{recoil}.

\subsubsection{Muon Detectors}
\label{muondet}
Four-layer drift chambers, embedded in the wedge directly outside  (in
radius) of the CHA, form the central muon detection system (CMU)~\cite{cmunim}.
The CMU covers the region $|\eta|<0.6$. Outside of these systems there is an 
additional absorber of 0.6 m of steel followed by a  system of
four-layer drift chambers (CMP). 
Approximately 84\% of the solid angle for $|\eta| < 0.6$
is covered by CMU, 63\% by CMP, and 53\% by both.  
Additional four-layer muon chambers (CMX) with partial (70 \%) 
azimuthal coverage subtend $0.6 < |\eta| < 1$.
Muons from $W$ decays are
required in this analysis to produce a track (stub) in the CMU or CMX 
that matches a track in
the CTC.   The CMP is used in this measurement only in the Level 1 and 
Level 2 triggers.
Details of the muon selection and reconstruction are given in 
Section~\ref{mumeas}.

\subsubsection{Trigger and Data Acquisition}
\label{triggers}
The CDF trigger is a three-level system that selects events for recording to
magnetic tape. The crossing rate of proton and antiproton bunches in the
Tevatron is 286~kHz, with a mean interaction rate of 1.7 interactions per
crossing at a luminosity of $\sim 1 \times 10^{31}$ cm$^{-2}$
sec$^{-1}$, 
which is 
typical of the data presented here. The first two levels of the trigger 
\cite{Level1and2} consist of dedicated electronics 
with data paths separate from
the data acquisition system. The third level~\cite{Level3},  which 
is initiated after
the event information is digitized and stored, uses a farm of commercial
computers to reconstruct events.  
The triggers selecting $\wenu$ and $\wmunu$ events are described below.

At Level~1,
electrons were selected by the presence of an electromagnetic 
trigger-tower with $E_T$ above 8 GeV 
(one trigger tower is two physical towers, which are longitudinally
adjacent, adjacent in pseudorapidity).
Muons were selected by the presence of a track stub in the CMU or CMX, 
and, where there is coverage, also in the CMP.  

At Level~2, electrons from $W$ decay could satisfy one of several 
triggers. Some required a track to be found in the
$r$-$\phi$ plane  by a fast hardware processor~\cite{CFT} and matched to a
calorimeter cluster; the most relevant required an electromagnetic 
cluster~\cite{Level1and2} with $E_T$ above 16 GeV and 
a track with $p_T$ above 12 GeV/c.
This was complemented by a trigger which required an electromagnetic
cluster with $E_T$ above 16 GeV matched with energy in the 
CES~\cite{karennim} and net missing transverse energy 
in the overall calorimeter of at least 20 GeV, with
no track requirements.  The muon Level 2 trigger required a 
track of at least 12 GeV/c that matches to a CMX stub (CMX triggers),
both CMU and CMP stubs (CMUP triggers), or a CMU stub but no 
CMP stub (CMNP triggers).
Due to bandwidth limitations, 
only about 43\% of the CMX triggers and 
about 39\% of the CMNP triggers were recorded.

At Level~3, reconstruction programs included  three-dimensional track 
reconstruction. 
The muon triggers required a track with $p_T$ above 18 GeV/c matched 
with a muon stub. There were three relevant electron triggers.  
The first required an electromagnetic cluster 
with $E_T$ above 18 GeV matched to a track with $p_T$ above 13 GeV/c 
with requirements on track and shower maximum matching, little 
hadronic energy behind the cluster, and transverse profile in
$z$ in both the towers and the CES. 
Because such requirements may create
subtle biases, the second trigger required only a cluster above 22 GeV 
with a track above 13 GeV/c 
as well as 22 GeV net missing transverse energy in the overall calorimeter.
The third trigger required an isolated 25 GeV cluster with no track 
requirement and with 25 GeV missing transverse energy.

Events that pass the Level 3 triggers were sorted and recorded. 
The integrated luminosity of the data sample is 
$\sim$80~pb$^{-1}$ in the muon sample and 
$\sim$84~pb$^{-1}$ in the electron sample.

\subsection{Data Samples}
\label{datasets}

Nine data samples are employed in this analysis.  These are  described
briefly below and in more detail in subsequent sections as they are used.
A list of the samples follows:
\begin{itemize}

\item{\bf The $\psi\rightarrow\mu^+\mu^-$ sample.}
A sample of $\sim500,000$ $\psi \rightarrow\mu^+\mu^-$ 
candidates with $2.7 < M_{\mu^+\mu^-} < 4.1$ GeV/c$^2$ is used  to
investigate the momentum scale determination and 
to understand systematic effects associated with track reconstruction.

\item{\bf The $\Upsilon\rightarrow\mu^+\mu^-$ sample.}
A sample of $\sim83,000$ $\Upsilon \rightarrow\mu^+\mu^-$ candidates
with $8.6 < M_{\mu^+\mu^-} < 11.3$ GeV/c$^2$ offers 
checks of the momentum scale determination that are 
statistically weaker but systematically better than those from
the $\psi \rightarrow\mu^+\mu^-$ sample.

\item{\bf The $Z\rightarrow\mu^+\mu^-$ sample.}
A sample of $\sim$1,900 dimuon candidates near the $Z$ mass determines the 
momentum scale and resolution, and is used to model 
the response of the calorimeters to the recoil particles against 
the $Z$ and $W$ boson, and to 
derive the $Z$ and $W$ $p_T$ distributions in the $\wmunu$ analysis.   

\item{\bf The $W\rightarrow \mu\nu$ sample.}
A sample of $\sim14,700$  $W\rightarrow \mu\nu$ candidates is used 
to measure the $W$ mass.  

\item{\bf The inclusive electron sample.}
A sample of $\sim$750,000 central electron candidates 
with $E_{T}>8$~GeV is used to
calibrate the relative response of the  central electromagnetic 
calorimeter (CEM) towers. 

\item{\bf The Run IA inclusive electron sample.}
A sample of $\sim$210,000 central electron candidates 
with $E_{T}>9$~GeV is used to
measure the magnitude and the distribution of the material, in radiation
lengths, between the interaction point and the CTC tracking volume.

\item{\bf The $W\rightarrow e\nu$ sample.}
A sample of $\sim$30,100 $W\rightarrow e\nu$ candidates is used 
to align the CTC, to compare the CEM energy scale to the momentum scale,
and to measure the $W$ mass.

\item{\bf The $\zee$ sample.}
A sample of $\sim$1,500 dielectron candidates near the $Z$ mass is 
used to determine the electron energy scale and resolution, to model 
the response of the calorimeters to the recoil particles against the 
$Z$ and $W$ boson, and to derive the $Z$ and $W$ $p_T$ distributions 
in the $\wenu$ analysis.   

\item{\bf The minimum bias sample.}
A total of $\sim2,000,000$ events triggered only on a coincidence of
two luminosity counters is used to help understand 
underlying event.

\end{itemize}

\subsection{Strategy of the Analysis}

The determination of the momentum and energy scales\footnote{Throughout 
this paper, momentum measurements using the CTC are denoted as $p$, and 
calorimeter energy measurements are denoted as $E$.} is crucial to the 
$W$ mass measurement. Momentum is the kinematic quantity measured for
muons; for electrons, the energy measured in the calorimeter is the 
quantity of choice as it has better resolution and
is much less sensitive than the momentum to the effects of
bremsstrahlung~\cite{Zmass}.
The spectrometer measures the momentum $(p)$ of muons and electrons,
and the calorimeter measures the energy $(E)$ of electrons. 
This configuration allows {\em in situ}  calibrations of both the 
momentum and energy scales directly from
the collider data.   The final alignment of the CTC wires is done with high
momentum electrons, exploiting  the charge independence of the  
electromagnetic calorimeter measurement since both positives and 
negatives should give the same momentum for a given energy.  
The momentum scale of the magnetic spectrometer is then studied using 
the reconstructed mass of the $\psi\rightarrow\mu^+\mu^-$
and $\Upsilon \rightarrow\mu^+\mu^-$ resonances, exploiting the 
uniformity, stability, and linearity of the magnetic spectrometer. 
Similar studies for the calorimeter are done using the average 
calorimeter response to electrons (both $e^+$ and $e^-$) of 
a given momentum.
The momenta of lepton tracks from $W$ decays reconstructed with
the final CTC calibration typically change from the initial values 
used for data sample selection by less than 10\%; their mean changes 
by less than 0.1\%. 
The final CEM calibration differs from the initial 
source/testbeam calibration in early runs 
on average by less than 2\%, with a gradual 
decline of $\sim$5\% during the data-taking period.
Fits to the reconstructed $\zmumu$ and $\zee$ masses, along with
linearity studies, provide the final momentum and energy scales.
The mass distributions are also used to determine the momentum
and energy resolutions.

The detector response to the recoil ${\bf u}$ is calibrated 
primarily using 
$\zmumu$ and $\zee$ decays in the muon and electron analyses, 
respectively. These are input to fast Monte Carlo programs which 
combine the production model and detector simulation.

The observed transverse mass lineshape also depends on the transverse and
longitudinal $W$ momentum spectra. 
The $p_T^W$ spectrum is derived from the $\zee$ and $\zmumu$ data
and the theoretical calculations.
The $p_T^Z$ spectrum is measured from the leptons in the $Z$ decays
by taking into account the lepton momentum and energy resolution. 
The theoretical calculations are used to correct the difference between
the $p_T^Z$ and $p_T^W$ distributions.
The observed $\bf u$ distributions provide consistency checks. 
The longitudinal spectrum is constrained
by restricting the  choice of parton distribution functions (PDFs) to those 
consistent with data.

To extract the $W$ mass, the measured $W$ transverse mass spectrum 
is fit to fast Monte Carlo spectra generated at a range of $W$ masses. 
Electromagnetic radiative processes and backgrounds
are included in the simulated lineshapes. 
The uncertainties associated with known systematic effects are 
estimated by varying the magnitude of these effects in the 
Monte Carlo simulation and refitting the data.

\subsection{Comparison with Run IA Analysis}

This analysis is similar to that of our last (Run IA) 
measurement~\cite{wmass_1a}, with datasets $\sim4.5$ times larger.
The direct use of the $Z$ events in modeling $W$ production 
and recoil hadrons against the $W$~\cite{wmass_1a,trisch} is
replaced with a more sophisticated parameterization~\cite{ag_thesis}. 
In this analysis our efforts to set a momentum
scale using the $\psi$ and $\Upsilon$ dimuon masses and 
then to transfer that
to an energy scale using $E/p$ for $W$ electrons did not produce a 
self-consistent picture, particularly the reconstructed mass of the $Z$
with electron pairs. 
Instead we choose to normalize the electron energy and 
muon momentum scales to the
$Z$ mass, in order to minimize the systematic effects, at the cost of a
modest increase in the overall scale uncertainty due to the limited $Z$
statistics.
A discussion of this problem is given in Appendix~A.
The instantaneous luminosity of this dataset is a factor of $\sim$2 
larger,  resulting in higher probability of having additional 
interactions within the same beam crossing.
Also, we have included
muon triggers from a wider range of polar angle.  

\section{Muon Measurement}
\label{mumeas}

In the muon channel, the $W$ transverse mass depends primarily on
the muon momentum measurement in the central tracking chamber (CTC).
This section begins with a description of the reconstruction of 
charged-particle trajectories and describes the CTC calibration
and alignment.  It then describes the selection criteria to identify
muons and the criteria to select the $\wmunu$ and $\zmumu$ candidates.
The momentum scale is set by adjusting the measured mass 
from $\zmumu$ decays to the world-average value of the $Z$ 
mass~\cite{pdg}. The muon momentum resolution is extracted 
from the width of the $\zmumu$ peak in the same dataset. The muon momentum
scale is checked by comparing the $\Upsilon$ and $\psi$ masses
with the world-average values.
Since the average muon momentum is higher in $Z$ decays than $W$ decays,
a correction would be necessary for the $W$ mass determination
if there were a momentum nonlinearity.
Studies of the $Z$, $\Upsilon$, and $\psi$ mass measurements 
indicate that the size of the nonlinearity is negligible.
 
\subsection{Track Reconstruction}
\label{trkrecon}

\subsubsection{Helical Fit}

The momentum of a charged particle is determined from its trajectory 
in the CTC.
The CTC is operated in a nearly (to within $\sim\!$1\%) uniform axial
magnetic field. 
In a uniform field, charged particles follow a helical trajectory. 
This helix is parametrized by:
curvature, $C$ (inverse diameter of the circle in $r$-$\phi$); impact
parameter, $D_0$ (distance of closest approach to $r=0$); $\phi_0$ (azimuthal
direction at the point of closest approach to $r=0$); $z_0$ (the $z$ position
at the point of closest approach to $r=0$); and $\cot\theta$, where $\theta$ is
the polar angle with respect to the proton direction.
The helix parameters are determined taking into account the 
nonuniformities of the magnetic field using the magnetic field map.
The magnetic field was measured by 
NMR probes at two reference points on the endplates of the CTC
during the data-taking period as shown in Figure~\ref{f_bfield}, 
and corrections are made on the magnetic field run-by-run to convert 
curvatures to momenta.
\begin{figure}
	\centerline{\epsfysize 20cm
                    \epsffile{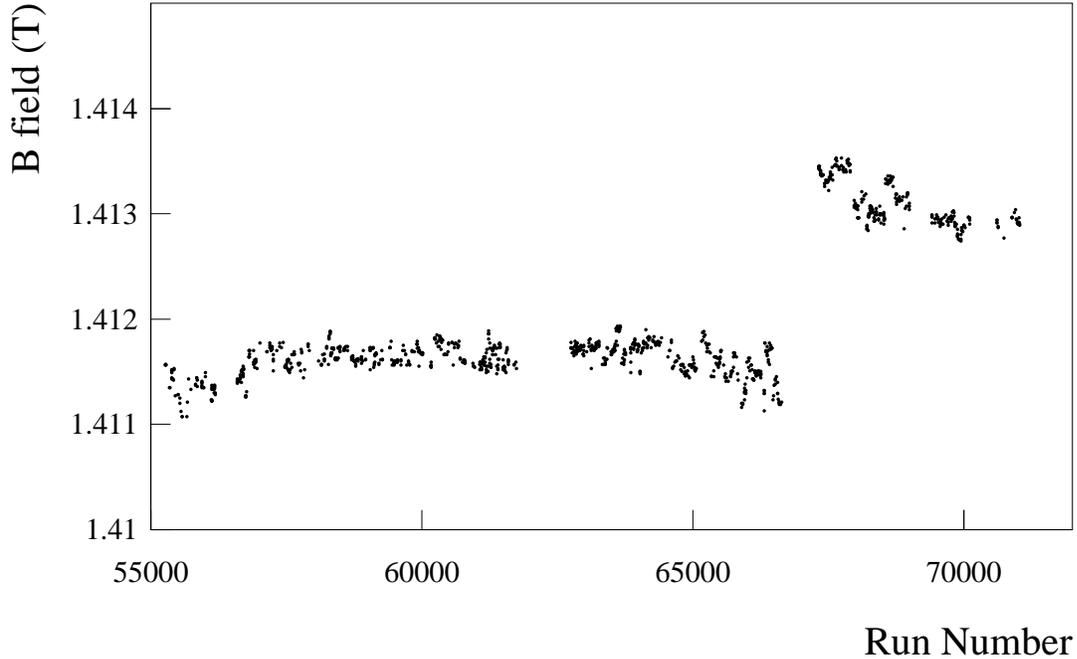}}
\vspace*{-9.0cm}
\caption{Variation of the average magnetic field 
	 as a function of run number.
	 The left side of the plot corresponds to January 1994 and 
	 the right side of the plot to July 1995.}
\label{f_bfield}
\end{figure}

The momentum resolution is improved by a factor of 
$\sim$2 by constraining  tracks to originate from the interaction point 
(``beam-constraint'').
The $z$ location of the interaction point is determined
using the VTX for each event with a precision of 1~mm.
The distribution of these interaction points
has an RMS spread of 
25$-$30~cm, depending on accelerator conditions.  
The $r$-$\phi$ location of the beam axis is measured with 
the SVX$^\prime$, 
as a function of $z$, to a precision of 10~$\mu$m.
The beam axis is tilted with respect to the CTC axis by a
slope that is typically about 400~microns per~meter.

\subsubsection{Material Effects on Helix Parameters}
\label{dedx}

The material between the interaction region and the CTC tracking
volume leads to the helix parameters measured in the CTC 
that are different than those at the interaction point.
For example, in traversing 7\% of a radiation length, 
muons lose about 5 MeV on average due to $dE/dx$ energy loss, 
which is significant for low $p_T$ tracks.
Because of its small mass, electrons passing through the material have a 
large amount of
(external) bremsstrahlung which changes both the curvature and
impact parameter of the electrons.
The beam constraint fit accounts for the $dE/dx$, and 
restores some of the energy loss due to the external bremsstrahlung.
In order to make accurate corrections for the $dE/dx$, 
and properly simulate biases from external bremsstrahlung, 
the magnitude and distribution of the material need to be understood.

The material distribution is measured using 
a Run IA sample of 210,000 photon conversions,  
where the conversion rate is proportional to the traversed
depth in radiation lengths.\footnote{The Run IA and Run IB detectors 
are identical except for the SVX. This difference, 
estimated to be less than 0.1\% of a radiation 
length, is negligible compared to the total radiation length.} 
Conversion candidates are selected from 
the 9~GeV inclusive electron sample.  
An electron associated with an oppositely-charged partner track
close in $\theta$ and distance at the point of conversion (the point
at which the two helices are parallel in azimuth) 
is identified as a $\gamma \rightarrow \ee$ candidate. 
To optimize the resolution on the measured conversion location, a
two-constraint fit is applied to the helix parameters of the two tracks:
the separation is constrained to vanish, and the angle $\phi$ from the
beam spot to the conversion point is constrained to match the $\phi$ of
the photon momentum vector.  These constraints give an average observed
resolution of 0.41~cm on the conversion radius, to be compared with an
expected resolution of 0.35~cm.
The radial distributions for conversions and backgrounds up to the
innermost superlayer in the CTC are shown in Figure~\ref{f_material_1}.
The prominent peak at 28~cm is due to the inner support structure of 
the CTC. Other structures such
as the silicon layers of the SVX and the VTX walls can be clearly resolved.
This resolution is important since we need to fix the proportionality
constant between conversions and radiation lengths by calibrating on a
feature of known composition. The CTC inner support is chosen for this
purpose since its construction is well-documented.  Its thickness at
normal incidence is $(1.26 \pm 0.06)$\% of a radiation length.
The result for
the integrated material thickness before the CTC volume, averaged over
the vertex distribution and angular distribution, is $(7.20 \pm 0.38)$\% 
of a radiation length~\footnote{This value is for electrons
from $W$ decay.  Due to difference in the detector 
acceptance between electrons and muons, the material thickness 
for muons is  $(7.10 \pm 0.38)$\%.}.
Variations in conversion-finding efficiency and electron trigger
efficiency as a function of the conversion point are taken into
account.  Other choices for the ``standard radiator'' such as
the wires of the innermost superlayer in the CTC, 
as shown in Figure~\ref{f_material_2},
give consistent results.

\begin{figure}
\vspace{2.0cm}
	\centerline{\epsfysize 14cm
                    \epsffile{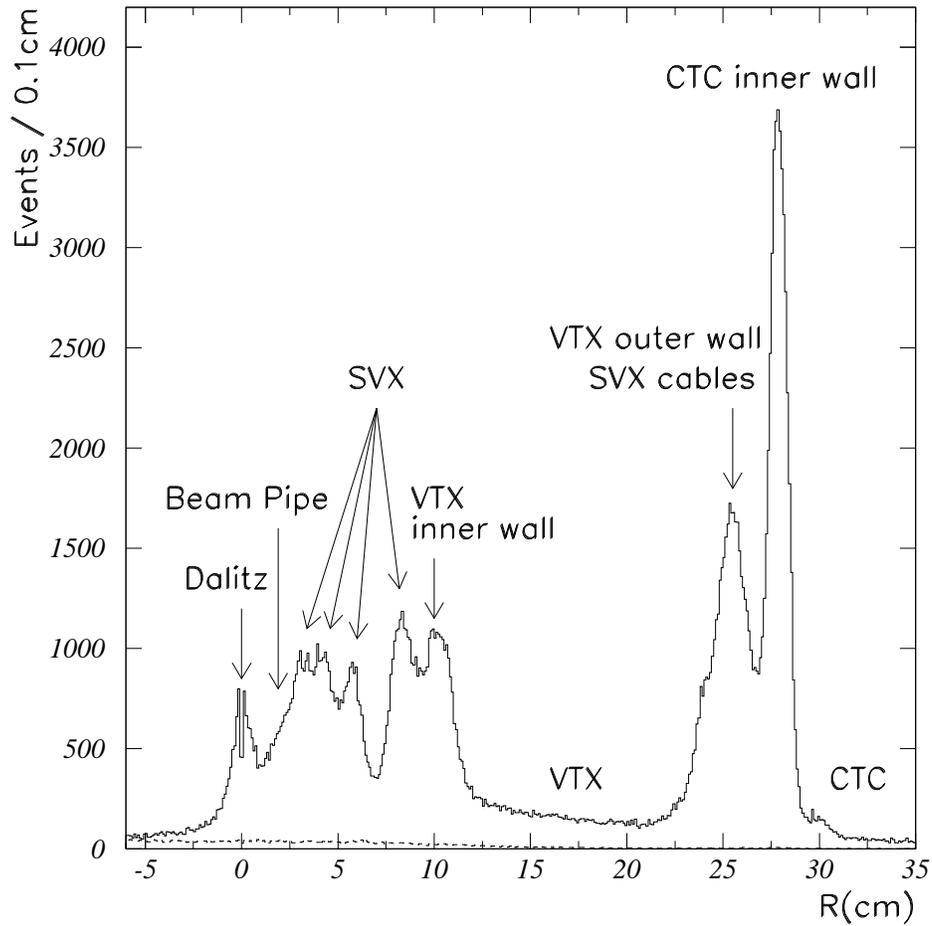}}
\caption{The radial (R) distributions for conversions (solid line) and
	background (dashed line) for the Run IA inclusive electron
	sample. 
	R is negative when the photon momentum direction is opposite
	to the vector from the beam spot to the conversion position 
	due to the detector resolution.}
\label{f_material_1}
\end{figure}

\begin{figure}
\vspace{2.0cm}
	\centerline{\epsfysize 14cm
		    \epsffile{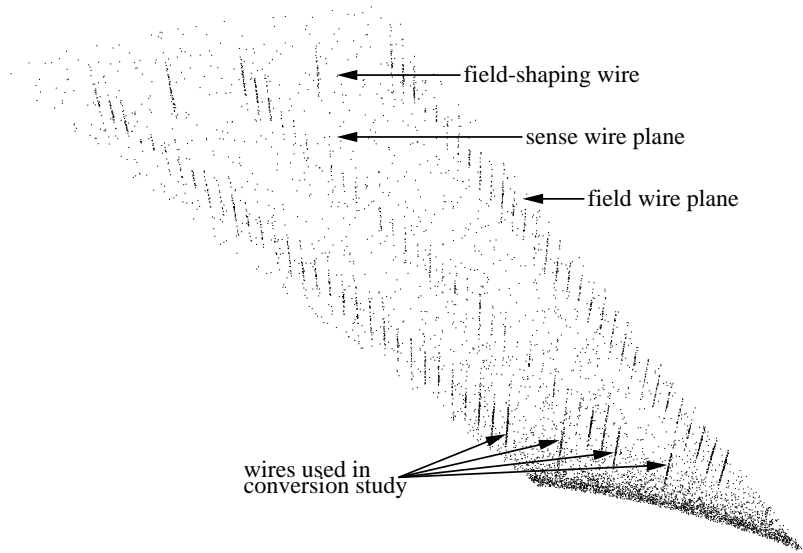}}
\caption{Reconstructed photon conversion vertex density in the $r-\phi$
	plane for the innermost superlayer in the CTC, folded into 1/30 
	of the circumference (this layer has 30-fold symmetry).  
	Each point represents one reconstructed vertex.}
\label{f_material_2}
\end{figure}

Another check is provided by the $E/p$ distribution
\footnote{ For convenience, the requisite factor of $c$ is 
dropped in the ratio $E/p$.}  
of electrons from $W$ decay (see Figure~\ref{f_eoverp}),
where $E$ is the electron energy measured by the CEM and
$p$ is the electron momentum measured by the CTC. External
bremsstrahlung photons~\cite{tsai} are collinear with the 
electron track at emission
and typically point at the calorimeter tower struck by the electron
track so that the calorimeter collects the full energy. Since the track
momentum is reduced by the radiated energy, the $E/p$ distribution
develops a high-side tail. 
Final state radiation from electron production (internal bremsstrahlung)
is about a 20 \% contribution to this tail. We define 
the fraction of events in the tail, $f_{tail}$,  
to be the fraction of events in the region $1.4 < E/p < 1.8$.
The lower bound is far enough away from the peak to be insensitive 
to resolution effects. After a small QCD background correction, 
we find :
$$f_{tail} = 0.0488 \pm 0.0014({\rm stat.}) \pm 0.0004({\rm syst.}).$$
The Monte Carlo simulation, including internal radiative effects, 
reproduces this value when the material equals $(7.55 \pm 0.37)$\% 
of a radiation length, in good agreement with the value from conversion 
photons above. 

An appropriate material distribution is applied to 
muon and electron tracks on a track-by-track basis.

\begin{figure}
	\centerline{\epsfysize 17cm
		    \epsffile{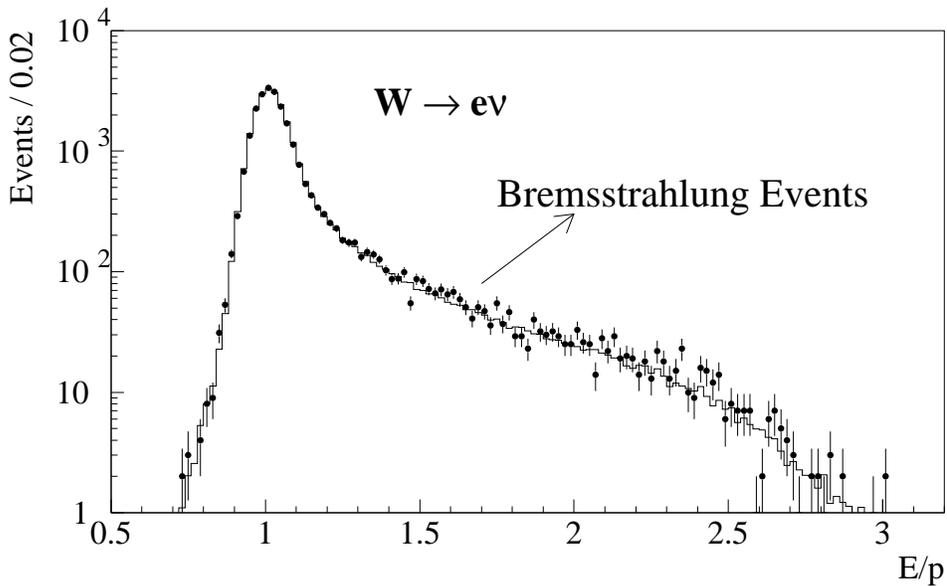}}
	\vspace{-8cm}
\caption{$E/p$ distribution of electrons in the $\wenu$ sample.
	The histogram indicates the simulation.}
\label{f_eoverp}
\end{figure}

\subsection{CTC Calibration and Alignment}
\label{calalg}

The CTC calibration and alignment
proceeds in two steps. First, the relationship
between the measured drift time and the distance to the sense wire 
is established. Second, the  relative alignment of wires and layers in
the CTC is performed.  
Small misalignments left after these procedures are removed with 
parametric corrections.

\subsubsection{Time-to-distance calibration}
Electronic pulsing, performed periodically during the data-taking
period, gives relative time pedestals for each sense wire.  
Variations in drift properties for each super-layer are removed  
run-by-run.
Additional corrections for nonuniformity in the drift trajectories 
are made based on data from many runs.  
After the calibration and alignment described in Section~\ref{ctc_align}, 
the CTC drift-distance resolution is determined 
to be 155~$\mu$m  (outer layers) to 215~$\mu$m (inner layers), 
to be compared with $\sim120$~$\mu$m expected from diffusion alone, 
and $\sim200$~$\mu$m expected from test-chamber results.

\subsubsection{Wire and layer alignment} 
\label{ctc_align}

The initial individual wire positions are taken to be the nominal positions
determined during the CTC construction~\cite{ctcnim}.  The distribution of 
differences between these nominal positions and the positions determined
with an optical survey has an RMS of 25 $\mu$m. The 84 layers of sense 
wires are azimuthally aligned relative to each other by requiring the 
ratio of energy to momentum $E/p$ for electrons to be independent of 
charge.
A physical model for these misalignments is a coherent twist of each
endplate as a function of radius.  
A sample of about $40,000$ 
electrons with $0.8 < E/p < 1.2$ from the $W\rightarrow e\nu$ sample 
(see Figure~\ref{f_eoverp}) is used for the alignment.   
The alignment consists of rotating each entire layer on each end of
the CTC by a different amount $r \times \Delta \phi$ with respect to
the outermost superlayer (superlayer 8) where the relative rotation of
two endplates is expected to be the smallest 
according to the chamber construction. 
The stereo alignment is adjusted to account for the 
calculated endplate deflection due to wire tension.
The measured deviation of each layer from its nominal position 
after this alignment is shown in Figure~\ref{wpo}. 

\begin{figure}[p]
\epsfysize=6.0in
\epsffile[54 162 531 675]{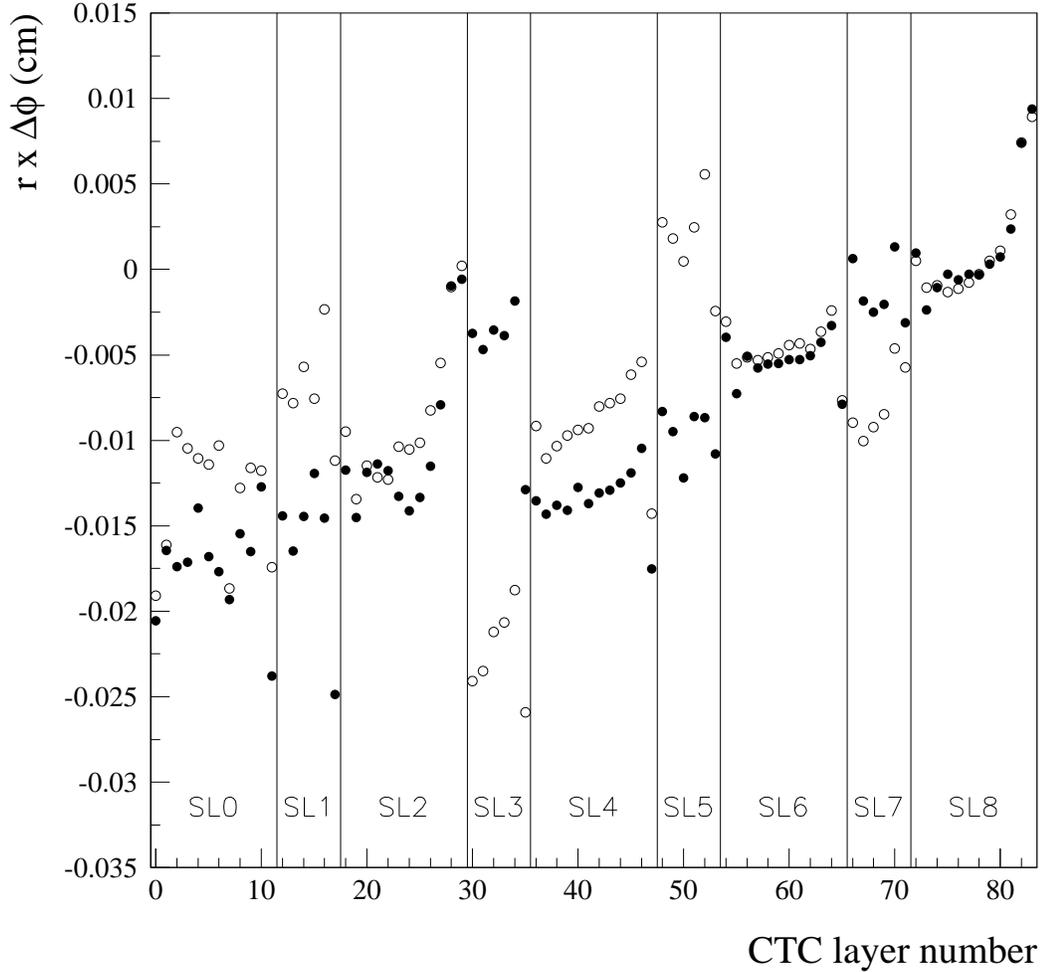}
\caption{The deviation ($r\times\Delta\phi$) 
         of each CTC layer from its nominal 
	 position at the end plates ($|z|$ = 150~cm) in cm, versus 
         the layer number.  The solid (open) circles represent
	 the west (east) CTC endplate.}
\label{wpo}
\end{figure}

Figure~\ref{f_jpsimass_dcot} demonstrates the elimination of 
misalignment after the alignment (open circles). 
A small residual dependence of the $J/\psi$ mass on cot$\theta$ remains,
which is removed with the correction, 
\begin{equation}
{\rm cot}\theta \rightarrow 1.0004 \times {\rm cot}\theta.
\label{cotmod}
\end{equation}

The only significant remaining misalignments are an 
azimuthally($\phi$)-modulated charge difference in $<E/p>$ 
and a misalignment between the magnetic field direction and the axial
direction of the CTC.
The $\phi$ modulation is removed with the correction 
\begin{equation}
C \rightarrow  
	C - 0.00031 \times \mbox{sin}~(\phi_{0} \ - \ 3.0) 
\label{phimod}
\end{equation}
where $C$ equals to $Q \times 1/p_T$ (GeV/c)$^{-1}$, $Q$ is the
charge of the lepton, the coefficient corresponds to a 
nominal beam position displacement of 37~$\mu$m, and $\phi$ is in
radians.  The magnetic field
misalignment is removed with the correction 
\begin{equation}
|C| \rightarrow |C| \cdot (1-0.0017 \cdot 
		{\rm cot}\theta \cdot {\rm sin}(\phi_0 -1.9)).
\end{equation}

\begin{figure}[p]
\epsfysize=5.in
\epsffile[54 162 531 675]{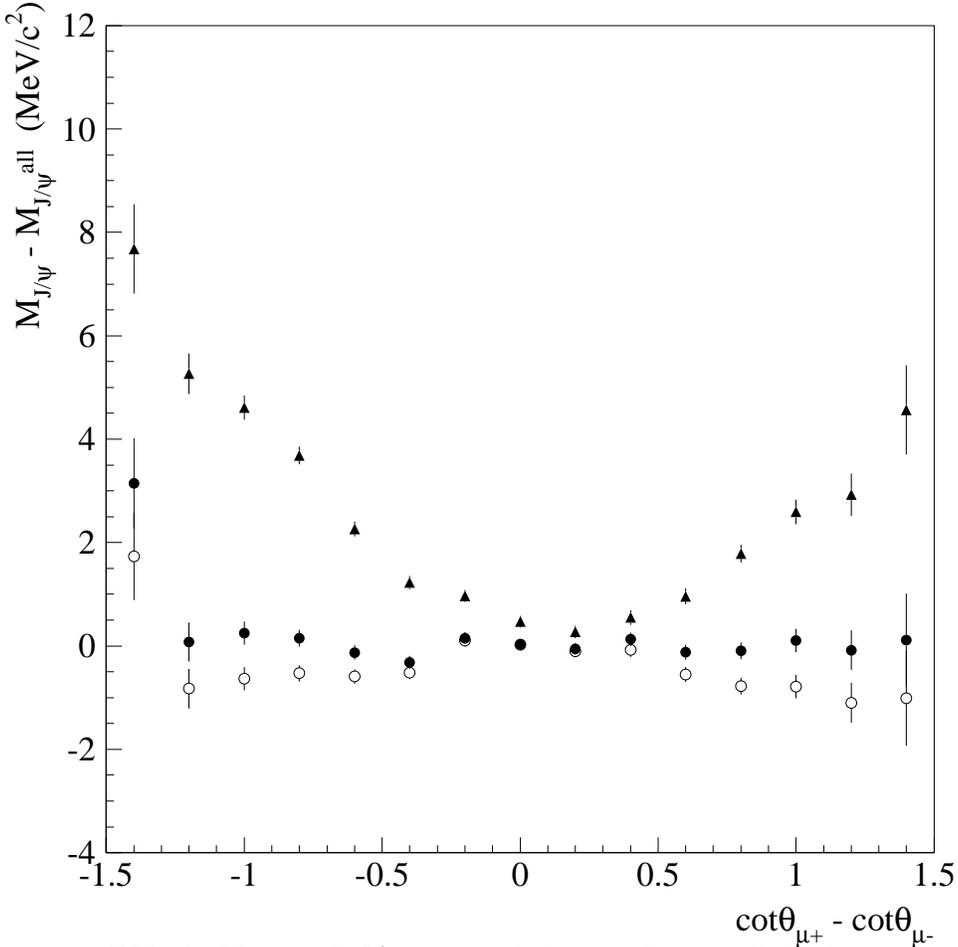}
\vspace{0.3cm}
\caption{Measured $J/\psi$ mass relative to the overall 
final mass measurement
as a function of $\Delta$cot$\theta$ = cot$\theta_{\mu^+}$ $-$ 
cot$\theta_{\mu^-}$.
The solid triangles and open circles are 
before and after the Run IB calibration and alignment, respectively.
Solid circles show the distribution with
the cot$\theta$ correction of 1.0004$\times$cot$\theta$.}
\label{f_jpsimass_dcot}
\end{figure}

\subsection{Muon Identification}
\label{muonid}

The $W$ mass analysis uses muons traversing the central muon system
(CMU) and the central muon extension system (CMX).

The CMU covers the region $|\eta|~<0.6$.  The CMX extends the coverage
to $|\eta|~<1$.
There are approximately five to eight hadronic absorption
lengths of material between the CTC and the muon chambers.
Muon tracks are reconstructed 
using the drift chamber time-to-distance relationship 
in the transverse ($\phi$) direction, and charge division in the longitudinal 
($z$) direction. Resolutions of 250 $\mu$m in the drift direction 
and 1.2 mm in $z$ are determined from cosmic-ray studies~\cite{cmunim}. 
Track segments consisting 
of hits in at least three layers are found separately in the
$r$-$\phi$ and $r$-$z$ planes.  These two sets of segments are merged
and a linear fit is performed to generate three-dimensional track 
segments (``stubs'').
Figure~\ref{f_muon_acc} shows the effects of the
bandwidth limitation of the CMX and CMNP triggers (see 
Section~\ref{triggers}) and partial azimuthal
coverage (see Section~\ref{muondet}).

\begin{figure}
	\centerline{\epsfysize 20cm
                    \epsffile{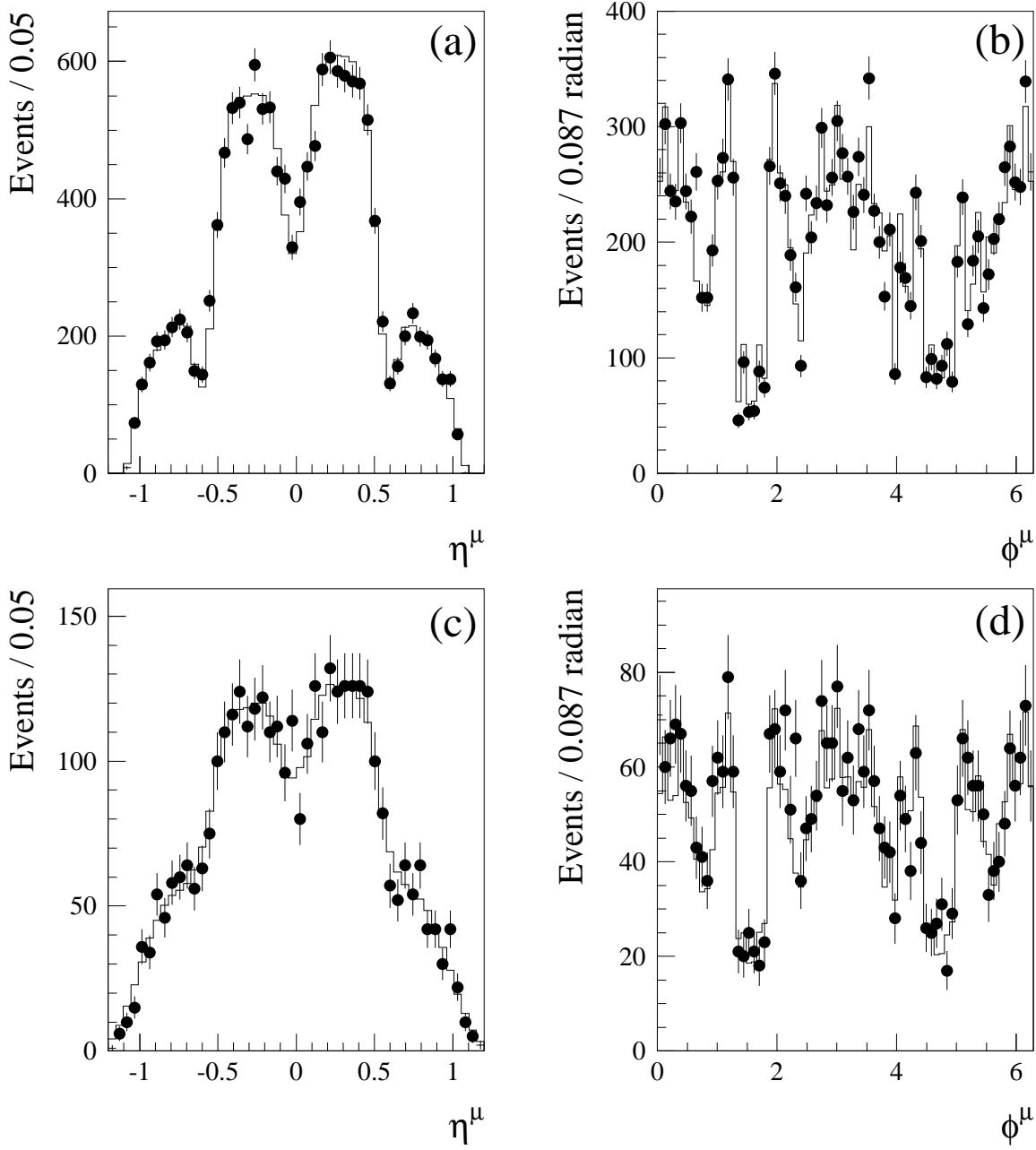}}
\vspace{-1.3cm}
\caption{The $\eta$ and $\phi$ distributions
	of muons are shown in (a) and (b) for $W$ decays, and
	(c) and (d) for $Z$ decays.
	Points (histograms) show the data (the simulation) with
	statistical uncertainties.}
\label{f_muon_acc}
\end{figure}

\vspace{0.6cm}
Muons from $W$, $Z$, $\Upsilon$, and $\psi$ decays are identified 
in the following manner.
The muon track is extrapolated to the muon chambers  
through the electromagnetic and hadronic calorimeters. 
The extrapolation must match to a track segment in the CMU or CMX.
For high $p_T$ muons from $W$ or $Z$ decays, 
the $r \times \Delta \phi$ matching is required to be within
2~cm; the RMS spread of the matching is 0.5~cm.
For low $p_T$ muons from $\Upsilon$ and $\psi$ decays, 
a $p_T$ dependent matching is required to allow for multiple scattering
effects.  Since the energy in the CEM tower(s) traversed by the muon is
0.3~GeV on average, the CEM energy is required to be less
than 2~GeV for $W$ and $Z$ muons. This cut is not applied
to muons from $\Upsilon$ or $\psi$ decays since $\Upsilon$'s and
$\psi$'s are often produced with particles associated with the
same initial partons.
Since the energy in the CHA tower(s) traversed by the muon is 2~GeV 
on average, the CHA energy is required to be less than 6~GeV.
In order to remove events with badly measured tracks,
muon tracks are required to pass through all nine superlayers of
the CTC, and to have the number of CTC stereo hits greater than
or equal to 12.
Muon tracks in the $\wmunu$ and $\zmumu$ data samples must satisfy 
$|D_0| <$ 0.2~cm, where $D_0$ is the impact parameter 
in the $r$-$\phi$ plane of 
the muon track with respect to the beam spot.
This reduces backgrounds from cosmic rays and QCD dijet events.
Additional cosmic ray background events are removed from the 
$\wmunu$ and $\zmumu$ samples when the hits of the muon track and 
the hits on the opposite side of the beam pipe, back-to-back in $\phi$, 
can be fit as one continuous trojectory.

\subsection{Event Selection: 
	$\wmunu$; $Z, \Upsilon, \psi \rightarrow \mu^+\mu^-$}

\subsubsection{$W\rightarrow \mu\nu$ and $Z \rightarrow \mu^+\mu^-$ 
		event selection}

The event selection criteria for the $W \rightarrow \mu\nu$ mass
measurement are intended to produce a sample with low background 
and with well-understood muon and neutrino kinematics. 
These criteria yield a sample that can be accurately modeled 
by simulation, and also preferentially choose those events with 
a good resolution for the transverse mass. 
The $Z$ sample is used to calibrate the muon momentum scale and 
resolution, to model the energy recoiling against the $Z$ and $W$, 
and to derive the $Z$ and $W$ transverse momentum spectra
($p_T^Z$ and $p_T^W$).
In order to minimize biases in these measurements,
the $\zmumu$ event selection is chosen to be as similar as possible
to the $\wmunu$ event selection.  

Both $\wmunu$ and $\zmumu$ sample extractions begin with events 
that pass a Level 3 high-$p_T$ muon trigger as discussed in Section~2.
From these, a final sample is
selected with the criteria listed in Table~\ref{wmunucuts}
and described in detail below.
	The event vertex chosen is the one reconstructed by the VTX 
	closest in $z$ to the origin of the muon track, and it is
	required to be within 60~cm in $z$ of the origin of the 
	detector coordinates.
	For the $Z$ sample, the two muons are required to be 
	associated either with the same vertex or with vertices within 
	5~cm of each other.
	For the $W$ sample, in order to reduce backgrounds from 
	$\zmumu$ and cosmic rays, events containing any oppositely 
	charged track with $p_T >$ 10~GeV/c and 
	$M_{\mu, track} > 50$~GeV/c$^2$ are rejected.
	Candidate $\wmunu$ events are required to have a muon CTC track
	with $p_T >$~25~GeV/c and a neutrino transverse energy
	$E_T^\nu >$~25~GeV.  
	A limit on recoil energy of  ${|\bf u|} < 20$~GeV reduces
	QCD background and improves transverse mass resolution.
	Candidate $\zmumu$ events are required to have
	two muons with $p_T > 25$~GeV/c.
	The two muon tracks must be oppositely charged.
	This requirement removes no events, indicating that
	the background in the $Z$ sample is negligible.
	The transverse mass in the region $65 < M_T < 100$~GeV/c$^2$ and 
	the mass in the region $80 < M < 100$~GeV/c$^2$ are used for
	extracting the $W$ mass and the $Z$ mass, respectively.
	These mass cuts apply only for mass fits and are absent 
	when we otherwise refer to the $W$ or $Z$ sample.
	The final $W$ sample contains 23,367 events, of which 14,740
	events are in the region 65 $< M_T <$ 100~GeV/c$^2$.
	The final $Z$ sample contains 1,840 events which are used
	for modeling the recoil energy against the $W$ and for deriving
	$p_T^W$, of which 
	1,697 events are in the region $80 < M <100$~GeV/c$^2$.
	
\begin{table}
\begin{center}
\begin{tabular}{|l|rr|}
Criterion	& W events after cut	& Z events after cut \\ 
\hline
Initial sample
with $Z$ vertex requirement		& 60,607 & 4,787 	\\
$E_T^{\rm CEM} < 2$~GeV			& 56,489 & 3,349	\\
Not a cosmic candidate			& 42,296 & 2,906   	\\
Impact parameter $|D_0| < 0.2$ cm 	& 37,310 & 2,952 	\\
Track - muon stub match			& 36,596 & 2,752	\\
Stereo hits $\ge$ 12    		& 34,062 & 2,442	\\
Tracks through all CTC superlayers	& 33,887 & 1,991 	\\
$p_T >$ 25 GeV/c			& 28,452 & 1,966	\\ 
$E_T^\nu >$ 25 GeV 			& 24,881 & N/A		\\
$|\bf u| <$ 20 GeV 			& 23,367 & N/A		\\
$p_T^{\mu\mu} < 45$~GeV/c, $70 < M^{\mu\mu} <110$~GeV/c$^2$	
					& N/A		& 1,840 \\ \hline
Mass fit region 			& 14,740  	& 1,697 \\
\end{tabular}
\caption{Criteria used to select the $\wmunu$ 
	 and $\zmumu$ samples.}
\label{wmunucuts}
\end{center}
\end{table}

\subsubsection{$\Upsilon, \psi \rightarrow \mumu$ event selection}

Samples of $\Upsilon$(1S, 2S, 3S) $\rightarrow \mumu$ events
and $\psi$(1S, 2S) $\rightarrow \mumu$ events are used
to check the momentum scale determined by $Z \rightarrow \mumu$
events.
The sample extraction begins with events 
that pass a Level 2 and 3 dimuon trigger with muon $p_T > 2$~GeV/c.
The requirement on the event vertex is identical to that for the $\zmumu$
selection.  Both muons are required to have opposite charges.

\begin{table}
\begin{center}
\begin{tabular}{|c|r|}
Sample & \# of events \\ \hline
$\Upsilon$(1S) 	&  12,800   	\\ 
$\Upsilon$(2S) 	&   3,500	\\ 
$\Upsilon$(3S) 	&   1,700	\\ 
$J/\psi$ 	& 228,900 	\\ 
$\psi$(2S) 	&   7,600 	\\
\end{tabular}
\caption{The number of events in the $\Upsilon$ and $\psi$ samples
	 after background subtraction.}
\label{t_upscuts}
\end{center}
\end{table}

Backgrounds are estimated from the dimuon invariant mass distributions
in the sidebands (regions outside the mass peaks).
The numbers of $\Upsilon$ and $\psi$ events after background
subtraction are listed in Table~\ref{t_upscuts}.
The average $p_T$ of muons in the $\Upsilon$ sample is 5.3~GeV/c,
and that in the $\psi$ sample is 3.5~GeV/c.  
The distributions of muon $p_T$ and the opening angle between the two
muons in $\phi$ are shown in Figure~\ref{f_lowpt}.  For comparison, 
the average $p_T$ of the muons and the average opening 
angle in the $Z$ sample are
43~GeV/c and 165$^\circ$, respectively.

\begin{figure}[p]
\epsfysize=6.0in
\epsffile[54 162 531 675]{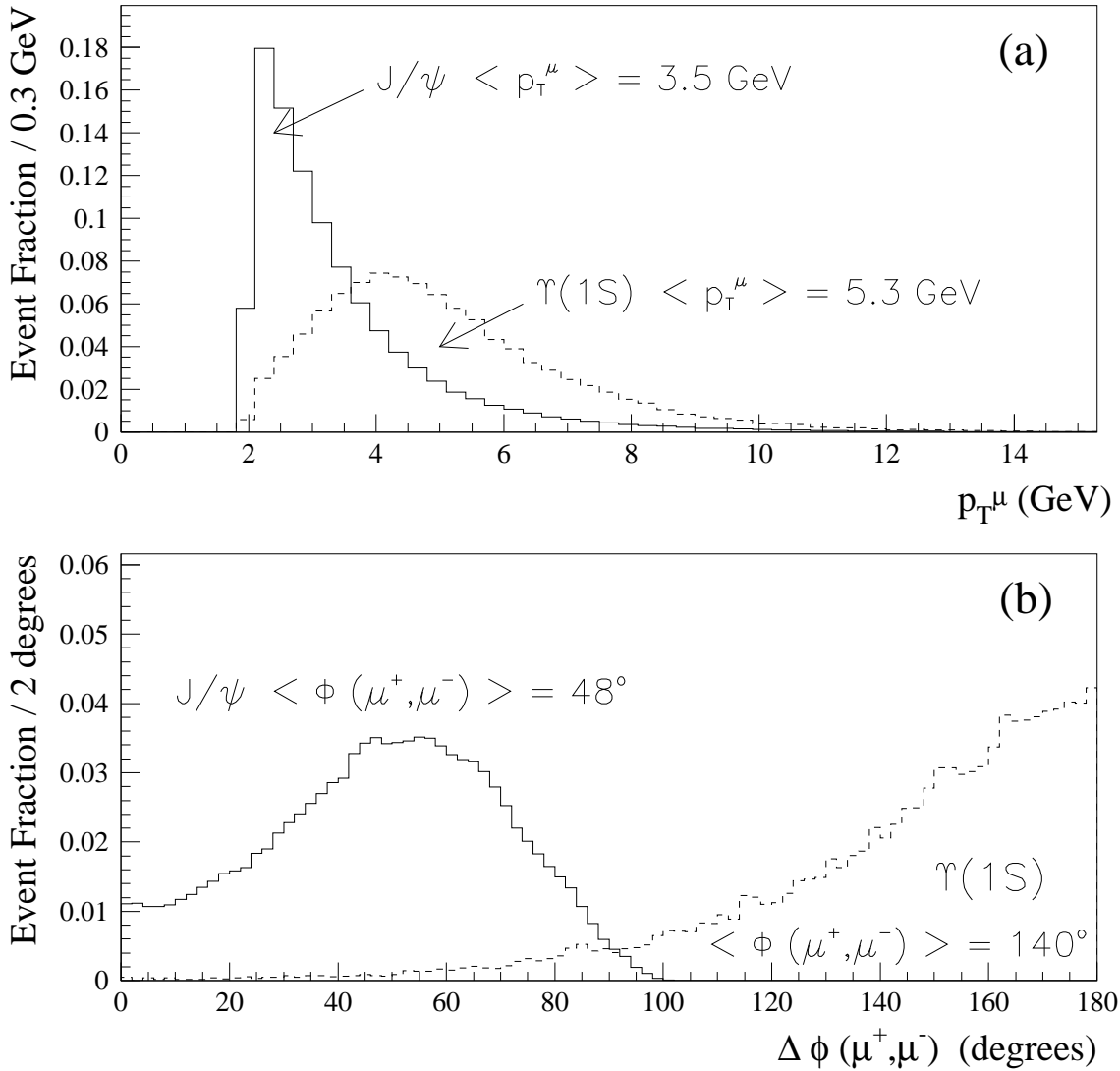}
\vspace{1.5cm}
\caption{(a) Transverse momentum distributions of muons and
	 (b) opening angle distributions between $\mu^+$ and 
	$\mu^-$ in the $\Upsilon$(1S) and $J/\psi$ samples.
	The histograms are normalized to unit area.}
\label{f_lowpt}
\end{figure}

\subsection{Event Selection Bias on $M_W$}
\label{trigger}

The $\wmunu$ selection requires muons at all three trigger levels.
Of these, only the level-2 trigger has a significant dependence on the
kinematics of the muon; its efficiency varies by $\sim$5\% with $\eta$
of the tracks.  This variation, however, leads to a negligible variation
($\sim$2 MeV/c$^2$) on the $W$ mass since the $M_T$ distribution 
is approximately invariant under $p_Z$ boosts.  The $W$ mass would be 
more sensitive to the $p_T$ dependence of the inefficiency 
since $M_T$ is directly related to $p_T$.  
No $p_T$ dependence is seen, but the statistical limitation on
measuring such a dependence leads to a 15~MeV/c$^2$ uncertainty on the
$\wmunu$ mass.

The muon identification requirements may also introduce a bias on
the $W$ mass.  For example, if the $W$ decays such that the muon travels
close to the recoil, there is greater opportunity for
the recoil particles to cause the muon identification to fail.  These
biases are investigated by tightening the muon identification
requirements and measuring the subsequent shifts in $M_W$.
The maximum shift observed of 10~MeV/c$^2$ is taken 
as a systematic uncertainty.

\subsection{Momentum Scale and Resolution}
\label{ctcscale}

A sample of $\zmumu$ events is used to determine the momentum scale 
by normalizing the reconstructed $Z \rightarrow \mumu$ mass to 
the world-average mass~\cite{pdg}, 
and to measure the momentum resolution in the high-$p_T$ region.
Since the muon tracks from $Z$ decays have curvatures comparable to those
for the $W$ mass determination, the systematic uncertainty 
from extrapolating the momentum scale from the $Z$ mass to the $W$ mass 
is small. The measurement is limited by the finite statistics in the 
$Z$ peak.

The $\zmumu$ Monte Carlo events are generated at various values of
$Z$ mass with the $Z$ width fixed to the world average~\cite{pdg}.
The generation program includes the $\gamma \rightarrow \mumu$ events 
and QED radiative effects, 
$Z \rightarrow \mu\mu\gamma$~\cite{wgamma,rgwgamma}, but uses 
a QCD leading order calculation so that the $Z$ is generated at $p_T^Z = 0$.
The $Z$ is then given a transverse momentum whose spectrum is 
extracted from the $\zmumu$ data (see Section~\ref{wprod}).
The generated muons are reconstructed by the detector simulation 
where CTC wire hit patterns, measured from
the real $\wenu$ data, are used to determine a covariance matrix of
the muon track, and the track parameters are smeared according to this
matrix. A beam constraint is then performed with 
the identical procedure as is used for the real data. 
The final covariance error matrix is scaled up by a free parameter  
to make the beam constraint momentum resolution agree with the data.
The detector acceptance is modeled according to the nominal
geometry.  The simulation includes the effects of the bandwidth
limitation of the CMX triggers.
Figure~\ref{f_muon_acc} illustrates how well 
the effects of the acceptance and the bandwidth limitation 
are simulated.
The mass distribution of the $\zmumu$ data, shown in
Figure~\ref{f_zmumu}, is then fit to simulated lineshapes, 
where the input $Z$ mass and the scale parameter to the covariance
matrix (or the momentum resolution) are allowed to vary.

\begin{figure}[p]
\epsfysize=6.0in
\epsffile[54 162 531 675]{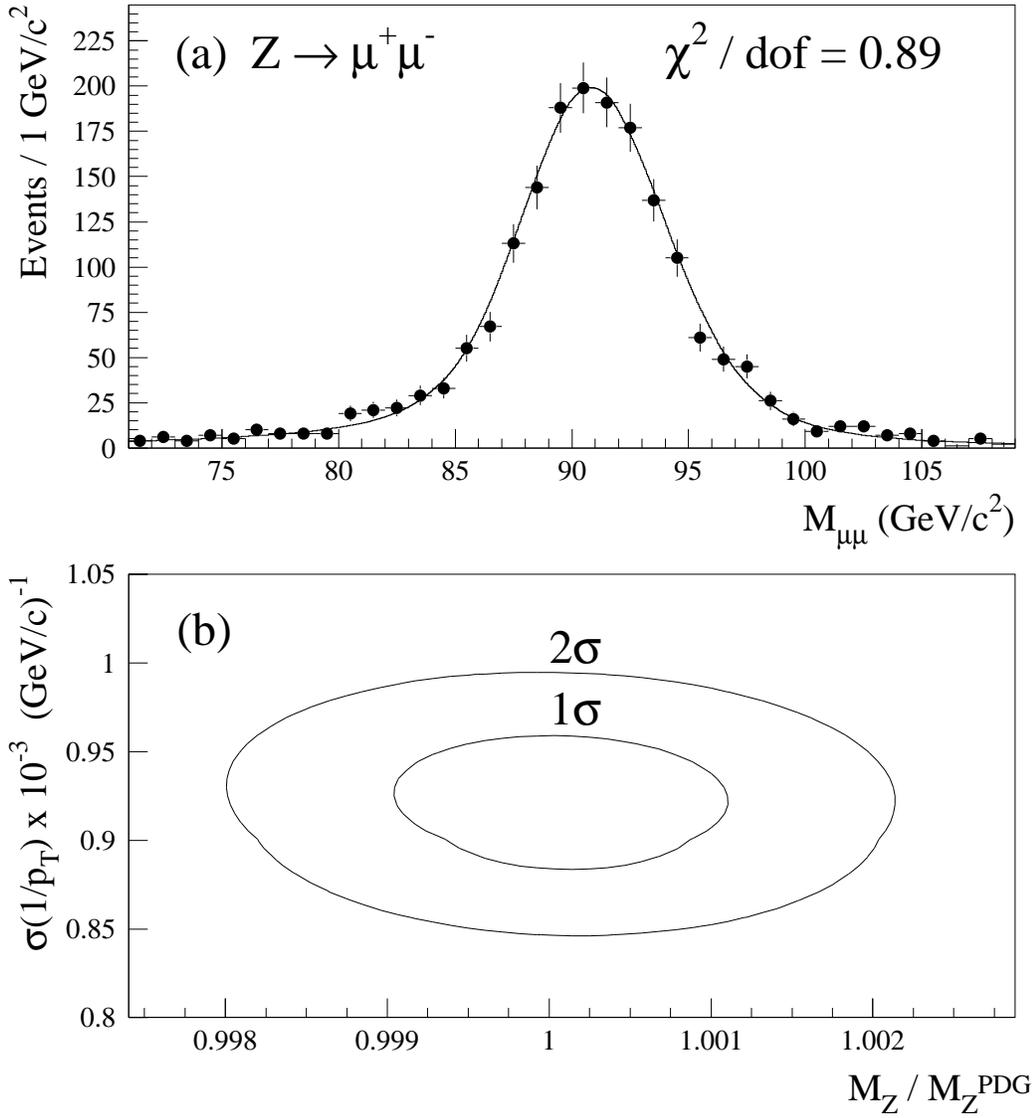}
\vspace{1.5cm}
\caption{Results of fit to $Z$ mass and momentum resolution.
	(a) Invariant mass distribution. The points are the data, 
	and the solid line is the Monte Carlo simulation 
	(normalized to the data) with best fit.
	(b) Correlation between the scale factor and the momentum 
	resolution.}
\label{f_zmumu}
\end{figure}

Fitting the invariant mass distribution in the region 
$80 < M_{\mu\mu} < 100$~GeV/c$^2$ with a fixed $\Gamma_Z$~\cite{pdg} 
yields 
\begin{equation}
M_Z = \rm 91.110 \pm 0.097(stat.) \pm 0.020(syst.)~GeV/c^2,
\label{zmumunumber}
\end{equation}
and momentum resolution 
\begin{equation}
\delta (1 / p_T) = (0.091 \pm 0.004(\rm stat.))
	\times 10^{-2}~\mbox{(GeV/c)$^{-1}$.}
\label{mumures}
\end{equation}
Equation~\ref{zmumunumber} results in the momentum scale factor
\begin{equation}
	{M_Z^{\rm PDG} \over M_Z^{\rm CDF}} 
	= 1.00085 \pm 0.00106
\label{muscale}
\end{equation}
which is applied to momenta of muons and electrons.
The fit is shown in Figure~\ref{f_zmumu}.  The two parameters,
$\delta(1/p_T)$ and $M_Z^{\rm PDG} /  M_Z^{\rm CDF}$, are largely
uncorrelated, as shown.

Table~\ref{zmumumass} contains a list of the 
systematic uncertainties on the $Z$ mass.
The largest uncertainty is from the radiative effects 
due to using the incomplete theoretical calculation~\cite{wgamma}; 
the calculation 
includes the final state radiation only and has a maximum of one 
radiated photon.
The effect arising from the missing diagrams is evaluated by
using the PHOTOS package~\cite{photos} which allows 
two photon emissions, and by using the calculation by U.~Baur 
{\it et al.}~\cite{baur} who have recently developed a complete
$O(\alpha)$ Monte Carlo program which incorporates the initial state QED
radiation from the quark-lines and the interference of the initial
and final state radiation, and includes a correct treatment of the 
final state soft and virtual photonic corrections.
When the PHOTOS package is used in the simulation instead, the change in 
the $Z$ mass is less than 10~MeV/c$^2$.
The effect of the initial state radiation and the initial and final
state interference is estimated to be 10~MeV/c$^2$~\cite{baur}.
To be conservative these changes are added linearly and 
20~MeV/c$^2$ is thus included in the systematic uncertainty.
The choice of parton distribution functions and that of the $p_T^Z$
spectrum contribute negligible uncertainties. 

A number of checks are performed to ensure that these results are
robust and unbiased.
The masses and resolutions at low and high $\eta$ are measured 
to be consistent.
The resolution is cross-checked using the $E/p$ distribution
in $W \rightarrow e\nu$ events, which is sensitive to the combined $E$
and $p$ resolution (see Section~\ref{epscale} and Figure~\ref{f_eoverpfit}).
Consistent results are found when much simpler techniques are used,
that is, comparing the mean $M_Z$, in the interval 86 -- 96 GeV/c$^2$, 
between the data and the Monte Carlo simulation or
fitting the invariant mass distribution with a Gaussian
distribution.
To address mis-measured tracks, a second Gaussian term is added to smear
track parameters for 8\% of the Monte Carlo events.
The change in $M_Z$ is negligible.

\begin{table}
\begin{center}
\begin{tabular}{|l|c|}
Effect                          & Uncertainty on $M_{Z}^{\mu}$ 
                         	  (MeV/c$^2$) 			\\ \hline
Statistics                      &  97 \\
Radiative corrections           &  20 \\
Fitting                         &  negligible \\
Parton distribution functions   &  negligible \\
$p_T^Z$ spectrum                &  negligible \\
Detector acceptance, triggers	&  negligible \\
\hline
Total                           & 100 \\
\end{tabular}
\caption{
\label{zmumumass}
Summary of uncertainties in measuring the $Z$ mass.}
\end{center}
\end{table}

\subsection{Checks of Momentum Scale}
\label{scalecheck}

The momentum scale is checked using $\psi$ and $\Upsilon$ masses, 
extracted by fitting the dimuon invariant mass distributions
to simulated lineshapes which include QED radiative processes and
backgrounds as shown in Figure~\ref{f_psimumu}.
The muon momenta are corrected by the momentum scale factor shown
in Eq.~\ref{muscale}.
The measured masses are summarized in Table~3.4.
Table 3.5 compares the measured masses with the world-average values.  
Within the momentum scale uncertainty, the agreement is very good.

\begin{table}
\begin{center}
\begin{tabular}{|c|c|}
 Resonance & Mass (MeV/c$^2$) \\ \hline\hline
$\Upsilon$(1S) 	& ${\rm 9464.3 \pm 0.7(stat.) \pm 1.6(syst.) 
		\pm 10.1(scale)}$  \\ \hline
$\Upsilon$(2S) 	& ${\rm10028.1 \pm 2.1(stat.) \pm 1.6(syst.) 
		\pm 10.7(scale)}$  \\ \hline
$\Upsilon$(3S) 	& ${\rm10358.9 \pm 3.6(stat.) \pm 1.6(syst.) 
		\pm 11.0(scale)}$  \\ \hline
$J/\psi$ 	& ${\rm 3098.4 \pm 0.1(stat.) \pm 1.1(syst.) 
		\pm  3.3(scale)}$  \\ \hline
$\psi$(2S) 	& ${\rm 3687.6 \pm 0.5(stat.) \pm 1.1(syst.) 
		\pm  3.9(scale)}$  \\
\end{tabular}
\end{center}
\label{t_masssummary}
\caption{Measured masses of the $\Upsilon$ 
	and $\psi$ resonances with the momentum scale
	correction.}
\end{table}

\begin{table}
\begin{center}
\label{oops}
\begin{tabular}{|c||c|c|}
 Resonance & World-Average Mass $M^{\rm PDG}$ (MeV/c$^{2}$)
	   & $M^{\rm CDF}/M^{\rm PDG}$ ~$-$ ~1 (\%) \\ \hline\hline
$\Upsilon$(1S) 	& $9460.4  \pm 0.2$ 
		& $0.041 \pm 0.018 \pm 0.106$ \\ \hline
$\Upsilon$(2S) 	& $10023.30 \pm 0.31$
		& $0.048 \pm 0.026 \pm 0.106$ \\ \hline
$\Upsilon$(3S) 	& $10355.3 \pm 0.5$
		& $0.035 \pm 0.038 \pm 0.106$ \\ \hline
$J/\psi$ 	& $3096.88 \pm 0.04$  
		& $0.050 \pm 0.035 \pm 0.106$ \\ \hline
$\psi$(2S) 	& $3686.00 \pm 0.09$  
		& $0.042 \pm 0.033 \pm 0.106$ \\
\end{tabular}
\caption{Measured masses of the $\Upsilon$ 
	and $\psi$ resonances with the momentum scale
	correction are compared to the world averages. 
	The second uncertainty in the last column is the momentum
	scale uncertainty, and the first uncertainty includes the 
	statistical and the other systematic uncertainties.}
\end{center}
\end{table}

\begin{figure}[p]
\epsfysize=6.0in
\epsffile[54 162 531 675]{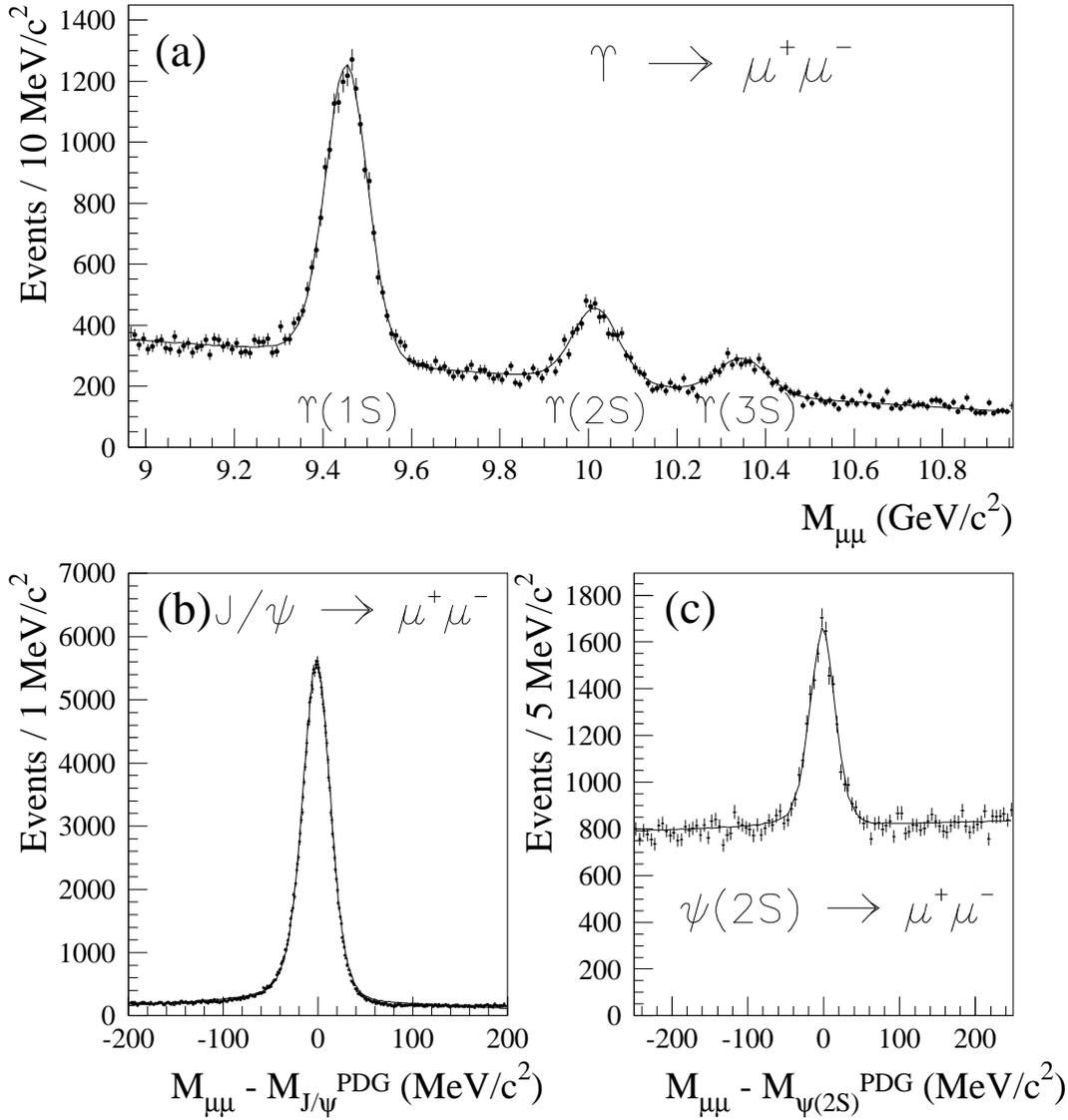}
\vspace{1.5cm}
\caption{The measured dimuon mass spectra near the 
	(a) $\Upsilon$ masses, 
	(b) $J/\psi$ mass, and (c) $\psi$(2S) mass.
	The curves are the best fits of lineshapes from the Monte 
	Carlo simulation.}
\label{f_psimumu}
\end{figure}

\begin{table}
\begin{center}
\begin{tabular}{|l||c|c|}
Source of Uncertainty 
	& Uncertainty on $M_\Upsilon$ (MeV/c$^{2}$)
	& Uncertainty on $M_\psi$     (MeV/c$^{2}$)\\ \hline
Muon energy loss	  & 1.5	& 1.0 	\\
Kinematics 		  & 0.4	& 0.1 	\\
Momentum Resolution	  & 0.3	& 0.1 	\\
Non-Prompt Production	  & - 	& 0.3 	\\
Misalignment		  & 0.2	& 0.1 	\\
Background		  & 0.1	& 0.1 	\\
Time variation		  & - 	& -  	\\
QED Radiative Effects	  & 0.4	& 0.2 	\\
Fitting Procedure, Window & -	& -  	\\ \hline
Total			  & 1.6 & 1.1 	\\
\end{tabular}
\caption{
\label{psisyst}
	Systematic uncertainties in $\Upsilon$ and $\psi$ mass
	measurements.}
\end{center}
\end{table}
A list of the systematic uncertainties on the $\psi$ and $\Upsilon$
masses is given in Table~\ref{psisyst}. 
The entries in the table are described below.

\vspace{2ex}
\noindent 
{\bf Muon Energy Loss:}
The momentum of each muon is corrected for energy loss in the material
traversed by the muon as described in Section~\ref{dedx}.  
Uncertainties in the energy loss come from
uncertainty in the total radiation length measurement
and in material type.
The measured $\Upsilon$ and $\psi$ masses vary 
by 0.8~MeV/c$^2$ and 0.3~MeV/c$^2$, respectively, when the average radiation 
length is changed by its uncertainty.
Uncertainty due to material type is estimated to be 0.6~MeV/c$^2$ per 
muon track.  This leads to 1.1~MeV/c$^2$ uncertainty in the $\Upsilon$ mass 
and 0.5~MeV/c$^2$ uncertainty in the $\psi$ mass.
There is a 0.8~MeV/c$^2$ variation in the observed $\psi$ mass, 
which is not understood, when the mass is plotted 
as a function of the radiation length traversed.  No statistically 
significant dependence ($<$ 0.7 MeV/c$^2$) on the total radiation
length is observed in the $\Upsilon$ mass.  
These variations of 0.7~MeV/c$^2$ in $M_\Upsilon$ and 0.8~MeV/c$^2$ 
in $M_\psi$ are taken as systematic uncertainties.
Adding the uncertainties described above in quadrature, 
the total uncertainty is 1.5~MeV/c$^2$ in $M_\Upsilon$ and 
1.0~MeV/c$^2$ in $M_\psi$.

\vspace{2ex}
\noindent 
{\bf Kinematics:}
Variation of the $p_T^\Upsilon$ and $p_T^\psi$ distributions allowed by
the data and $p_T^\mu$ cuts results in uncertainties of 
0.4~MeV/c$^2$ and 0.1~MeV/c$^2$ in $M_\Upsilon$ and $M_\psi$, respectively.

\vspace{2ex}
\noindent 
{\bf Momentum Resolution:}
Variation of the momentum resolution allowed by the data results
in uncertainties of 0.3~MeV/c$^2$ and 0.1~MeV/c$^2$ in 
$M_\Upsilon$ and $M_\psi$, respectively.

\vspace{2ex}
\noindent 
{\bf Non-Prompt Production:}
About 20\% of $\psi$'s come from decays
of $B$ mesons, which decay at some distance from the primary vertex.
The measured $\psi$ peak may be shifted by the application of 
the beam constraint.
The difference in the $\psi$ mass between a fit using the beam 
constraint and a fit using a constraint that the two muons 
originate from the same vertex point is 0.3 MeV/c$^{2}$.  
This difference is taken as an uncertainty.

\vspace{2ex}
\noindent 
{\bf Misalignment:}
\vspace{0.2cm}
\noindent
The CTC alignment eliminates most of the effects.  The
residual effects are measured by $\psi$ and $W$
samples and are removed by corrections as described in 
Section~\ref{calalg}.  The corrections and 
corresponding mass shifts on $M_\Upsilon$ are summarized 
in Table~\ref{t_align}.
The overall effects of 0.17~MeV/c$^2$ in $M_\Upsilon$
and less than 0.1~MeV/c$^2$ in $M_\psi$
are taken as a systematic uncertainty.

\begin{table}
\begin{center}
\begin{tabular}{|c|c|c|}
Source & Correction Formula & $\Delta M_\Upsilon$ (MeV/c$^2$)\\ \hline 
B-field direction       & 
$|C| \rightarrow |C| \cdot (1-0.0017 \cdot 
{\rm cot}\theta \cdot {\rm sin}(\phi_0 -1.9))$
                        & $+0.01$       \\ 
$\phi_0$ dependence & 
$C \rightarrow C - 0.00031 \cdot {\rm sin}(\phi_0 - 3.0)$ 
                        & $-0.24$       \\ 
cot$\theta$ dependence  & 
${\rm cot}\theta \rightarrow 1.0004 \cdot {\rm cot}\theta$ 
                        & $+0.40$       \\ \hline
Total correction        &               & $+0.17$ \\
\end{tabular}
\caption{
\label{t_align}
	Systematic uncertainties in $\Upsilon$ and $\psi$ mass
	 measurements.}
\end{center}
\end{table}

\vspace{2ex}
\noindent 
{\bf Background:}
The backgrounds in the $\Upsilon$ and $\psi$ mass peak regions are 
estimated by fitting the invariant mass distributions in the 
sideband regions (regions away from the peaks) 
with quadratic, linear and exponential distributions.  
The backgrounds are included in the templates used to fit the masses.
By varying the background shape, 
$M_\psi$ changes by less than 0.1~MeV/c$^2$ 
and $M_\Upsilon$ changes by 0.1~MeV/c$^2$.

\vspace{2ex}
\noindent 
{\bf Time Variation:}
As shown in Figure~\ref{f_runnum}, there is no indication of 
a time variation in the measured mass over the data-taking period, 
even though the resolution worsens due to high occupancy
in the CTC at high instantaneous luminosity during the latter 
portion of the data-taking period.

\begin{figure}[p]
\epsfysize=6.0in
\epsffile[54 162 531 675]{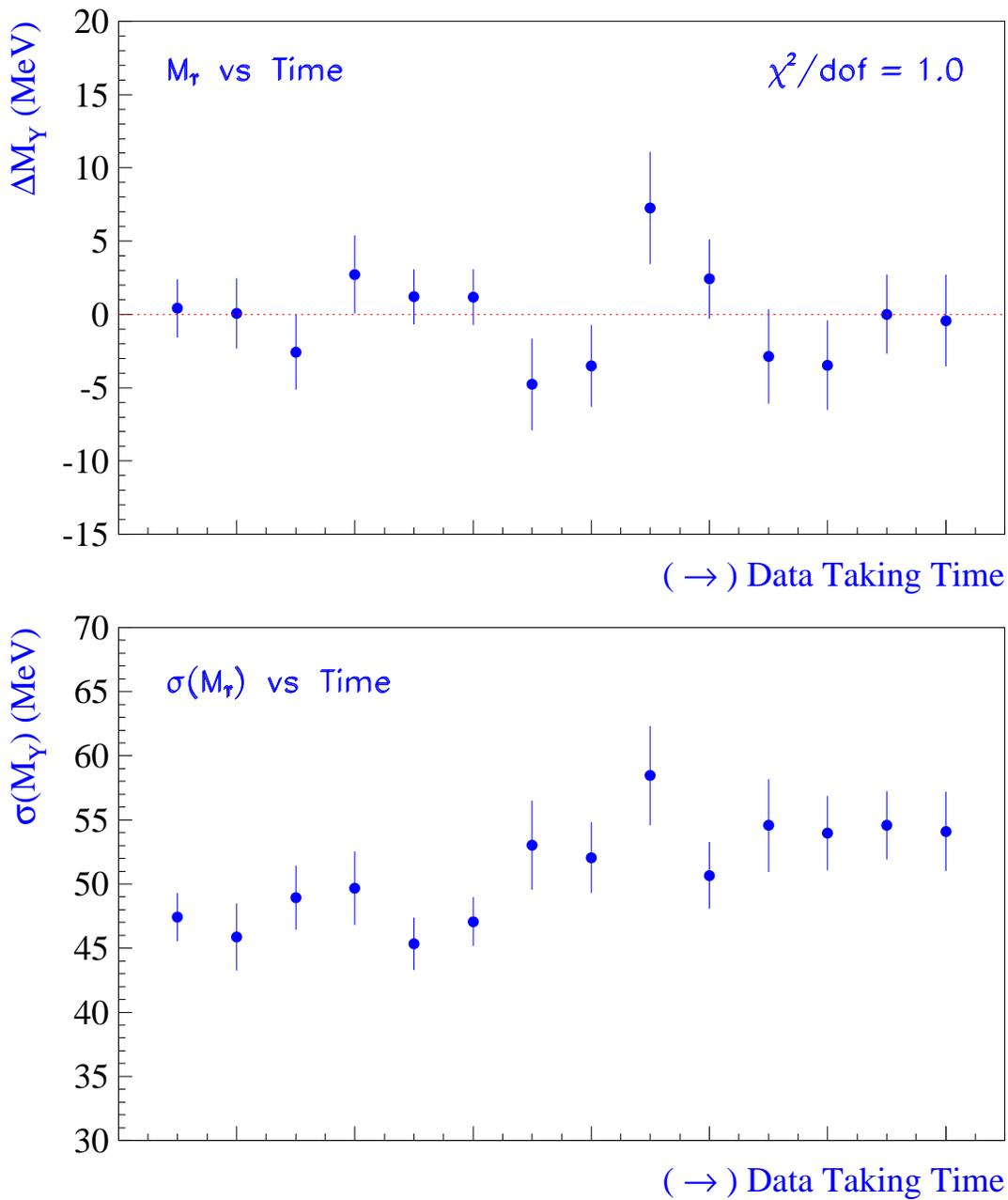}
\vspace{1.5cm}
\caption{Variation of the measured $\Upsilon$(1S) mass (Top)
	 and width (Bottom) as a function of time.
	 The left side of the plot corresponds to January 1994 and 
	 the right side of the plot to July 1995.
	 $\Delta M_\Upsilon$ is difference between 
	 the measured mass for a given time
	 period and the mass using all the data.}
\label{f_runnum}
\end{figure}

\vspace{2ex}
\noindent 
{\bf QED Radiative Effects:} The Monte Carlo program includes
final state QED radiation from muons.   The systematic uncertainties
of 0.4~MeV/c$^2$ in $M_\Upsilon$ and 0.2~MeV/c$^2$ in $M_\psi$  
represent missing diagrams such as two photon emission and the 
interference between the initial and final state radiation.

\vspace{2ex}
\noindent 
{\bf Fitting Procedure, Window:}
Consistent results are found when fitting windows are varied or 
much simpler fitting techniques are used, that is, 
comparing the mean $M_\Upsilon$ and $M_\psi$ and comparing 
the fit results with Gaussian plus linear distributions 
between the data and the Monte Carlo simulation.

\subsection{Momentum Nonlinearity}
\label{pnonlin}

The average $p_T$ for $Z$ decay muons is about 4.5~GeV/c higher than
that for $W$ decay muons.  Since the momentum is calibrated with the
$Z$ mass, any nonlinearity in the momentum measurement would translate
into an incorrect momentum scale for the $W$ mass measurement.
The momentum nonlinearity is studied using measured masses from a wide
range of curvatures --- the CTC does not directly measure momentum, 
but curvature, which is proportional to $1/p_T$.
The curvature ranges 
from 0.1 to 0.5~(GeV/c)$^{-1}$ in the $J/\psi$ data, 
from 0.1 to 0.3~(GeV/c)$^{-1}$ in the $\Upsilon$(1S) data, 
and 0.02 to 0.04~(GeV/c)$^{-1}$ in the $Z$ data.
Figure~\ref{f_p_non_linearity} shows the ratio of the measured mass
to the world-average value as a function of the average curvature of
two muons from these data.
The ratios are flat and all are well within statistical uncertainty 
of the ratio from the $Z$ data.
Since the curvature difference 0.003~(GeV/c)$^{-1}$ 
between the $W$ and $Z$ muons 
is much smaller than the range of curvature available in the 
$\psi$, $\Upsilon$, and $Z$ data, 
the nonlinearity effect in extrapolating from the $Z$ muon
momentum to the $W$ muon momentum is estimated to be negligible.

\begin{figure}[p]
\epsfysize=6.0in
\epsffile[54 162 531 675]{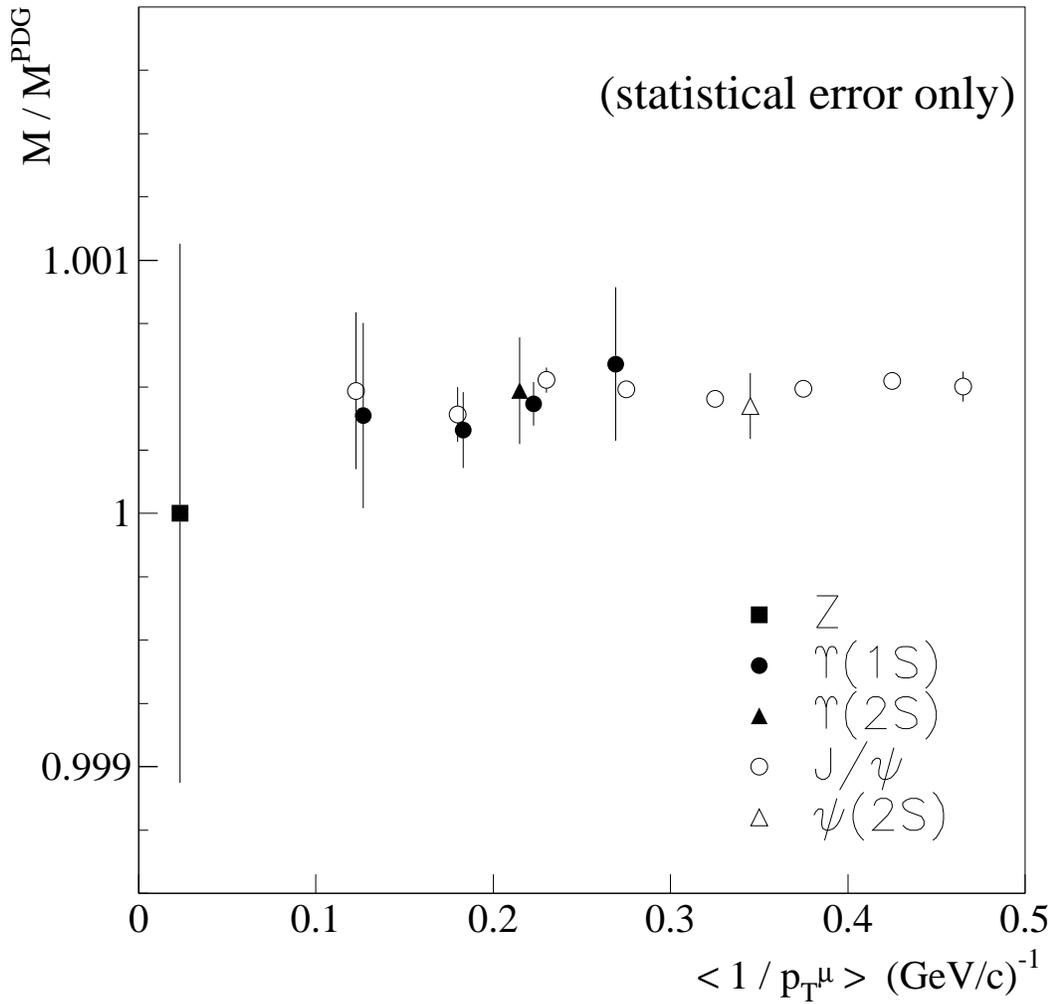}
\vspace{1cm}
\caption{The ratio of the measured mass to the world-average value 
	as a function of 
	the average curvature or inverse momentum
	for the $Z$, $\Upsilon$, and $\psi$ data.}
\label{f_p_non_linearity}
\end{figure}

\subsection{Summary}

The muon momentum scale is determined by normalizing the measured
$Z$ mass to the world-average mass. The scale in the data needs to be
corrected by a factor of $1.00084 \pm 0.00106$, the accuracy of which 
is limited by the finite statistics in the $Z$ peak. 
When the momentum scale is varied over its uncertainty in the simulation,
the measured $W$ mass changes by $\pm$85~MeV/c$^2$.
The scale is cross-checked by $M_{\psi}$ and $M_{\Upsilon}$.
The momentum resolution, 
$\delta (1 / p_T) = (0.091 \pm 0.004)
        \times 10^{-2}~{\rm (GeV/c)^{-1}}$, 
is measured from the width of the $\zmumu$ peak in the same dataset.
Lepton momenta in the Monte Carlo events are smeared according to 
this resolution.
When the momentum resolution is varied over its uncertainty
in the simulation,
the measured $W$ mass changes by 20~MeV/c$^2$.
Systematic uncertainties due to the triggers and the muon identification
requirements are estimated to be 15~MeV/c$^2$ and 10~MeV/c$^2$,
respectively.

\section{Electron Measurement}
\label{emeas}

This section begins with a 
description of the algorithm that associates calorimeter tower
responses with electron energy. It then describes the CEM relative
calibration procedure to correct for nonuniformity of the calorimeter
response and time dependence.
We discuss the selection criteria to identify
electrons and the criteria to select the $\wenu$ and $\zee$ candidates.
The electron energy scale is set by adjusting the reconstructed
mass in $\zee$ decays to the world-average value of the $Z$ mass.
The electron resolution is measured from the width of the $Z$ mass
distribution. The electron energy scale determined by using 
the $E/p$ distribution is discussed. 
A small calorimeter nonlinearity is observed, and
a correction is applied to the electron energy 
for the $W$ mass measurement.

\subsection{Electron Reconstruction}
The scintillation light for each tower in the CEM is
viewed by two phototubes, viewing light collected 
on each azimuthal side. The geometric mean 
of the two phototube charges, multiplied by an initial calibration, 
gives the tower energy. 
For electron candidates, 
the clustering algorithm finds a CEM ``seed'' tower with transverse energy 
above 5 GeV. The seed tower and the two adjacent towers 
in pseudorapidity form a cluster. One adjacent tower is not included 
if it lies on the opposite side of the $z = 0$
boundary from the seed tower. The total $E_T$ in the hadronic towers just 
behind the CEM cluster must be less than 12.5\% of the CEM cluster
$E_T$.  
The initial estimate of the electron energy is taken as the sum of the 
three (or two) CEM tower energies in the cluster.
There must be at least one CTC track that points to the CEM
cluster. The electron direction, used in the calculations of 
$E_T$ and the invariant mass, is defined by the highest $p_T$ track.  
The $W$ and $Z$ electron samples are further purified with 
additional cuts as discussed below in Section~\ref{evsel}.

\subsection{Uniformity Corrections}
\label{rel_corr}
To improve the CEM resolution, corrections are applied for known
variations in response of the towers, dependence on shower
position within the tower, and time variations over the course of the
data-taking period. 
For the present measurement, the nominal uniformity
corrections (testbeam)
are refined using two datasets -- the $W$ electrons and the
high-statistics inclusive electron dataset.
The reference for correcting the electron energy is
the track momentum as measured by the CTC. Uniformity is achieved by
adjusting the tower energy response (gain) until the mean $E/p$ is flat as
a function of time and $\phi$, and agrees with the Monte Carlo
simulation as a function of $\eta$.\footnote{
The material traversed by electrons increases with polar angle, so
$\langle E/p \rangle$ increases with $|\eta|$.}

The first step uses the inclusive electron data to set the
individual tower gains.  Tower gains are 
determined in four time periods. The time boundaries
correspond to natural breaks such as extended shutdowns or changes in
accelerator conditions, so the statistics for each time period are not the
same. The mean numbers of events per tower are 190, 190, 750, and 600, 
respectively, for the four time periods. These correspond to statistical
precisions on the tower gain determination of $\pm$0.64\%,  $\pm$0.64\%,
$\pm$0.33\%, and $\pm$0.38\%, respectively. 

Having determined the individual tower gains, long-term
drifts within each time period are measured 
by fitting to a line based on
run number (typically a run lasts about 12 hours).  These corrections
remove aging effects or seasonal temperature variations, but are
insensitive to short term variations such as thermal effects
caused by an access to the detector in the collision hall.

The next step uses the $W$ sample to update the mapping corrections which
describe the variation in response across the face of the towers.
The strip chamber determines 
the local $x$ (azimuthal) and $z$ (polar) coordinates within the wedge,
where $-24 < x <24$~cm is measured from the tower center and 
$-240 < z < 240$~cm from the
detector center. The $\langle E/p \rangle$ distribution as a function of $x$
is fitted to a quadratic function, which corrects primarily for 
non-exponential attenuation in 
the scintillator of the light seen by the two phototubes. 
Tower-$\eta$-dependent corrections are also made as a function of $z$.
The statistical uncertainty in the mapping corrections 
is 0.2\% in $x$ and 0.13\% in $z$.

Finally a very small correction takes into
account a systematic difference of the ``underlying event'' 
in the inclusive electron and $W$ datasets.
The underlying event consists of two components -- one due to additional
interactions within the same beam crossing (multiple interactions)
and the other due to the remnants of the protons and antiprotons that 
are involved in the inclusive or $W$ electron production. It
overlaps with the electron, contributing 
approximately 90~MeV on average to the electron $E_T$.
Because of the difference in $E_T$ between the inclusive electrons
($<E_T> \approx 10$~GeV) and the $W$ electrons ($<E_T> \approx 38$~GeV),
their underlying energy contribution is proportionately different.
This difference varies with the instantaneous luminosity,
which is strongly correlated with time.

All of the corrections applied to the $W$ electrons are shown 
in Figure~\ref{f_netcor}. 
The mean temporal correction is $+4.6$\% and the mean mapping correction
is $-2.5$\%. The corrections reduce the RMS width of the $E/p$
distribution from 0.0578 to 0.0497.

\begin{figure}[p]
\epsfysize=6.0in
\epsffile[54 162 531 675]{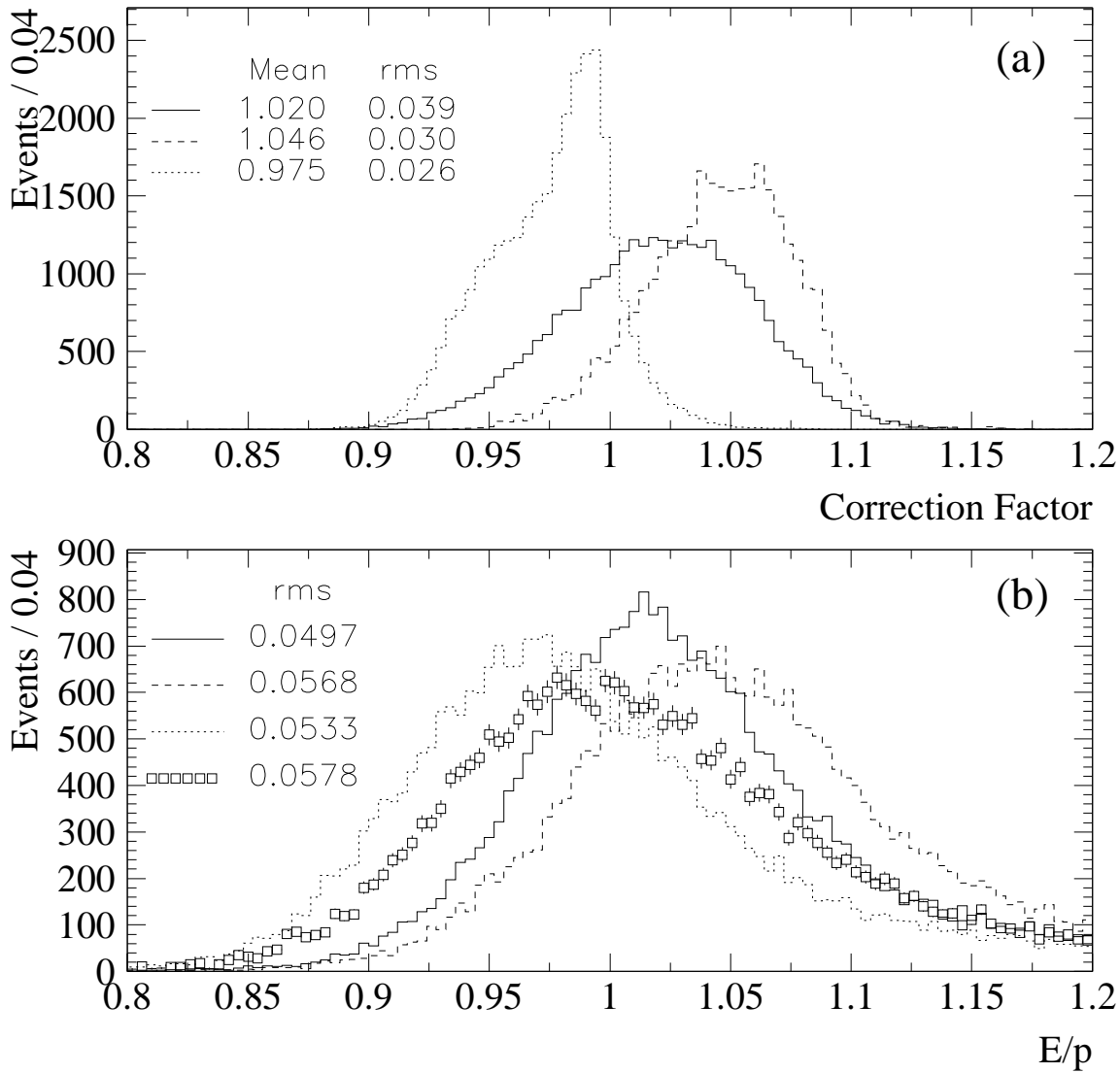}
\vspace*{0.5cm}
\caption{(a) Spatial and  temporal energy correction factors on the $W$ 
	electrons. The dotted curve shows the spatial corrections only, 
	the dashed curve the temporal corrections only, and the solid 
	curve the product of the two. 
	(b) The $E/p$ distributions of the $W$ electrons after the respective
	corrections. The squares show the data before any corrections 
	are applied. The improvement in the resolution after correction 
	is apparent.}
\label{f_netcor}
\end{figure}

\subsection{Event Selection: $\wenu$, $\zee$}
\label{evsel}
The $\wenu$ and $\zee$ selection criteria are chosen 
to produce datasets with low background and well-measured
electron energy and momentum.  They are identical to those for
the $\zmumu$ and $\wmunu$ datasets except for the charged lepton
identification and the criteria of removing $\zee$ events from
the $\wenu$ candidate sample.
The cuts and number of surviving events
are shown in Table~\ref{wenucuts} and the electron criteria
and the $Z$ removal criteria are described in detail below.
The samples begin with 108,455 $W$ candidate events and 19,527 $Z$ 
candidates events that pass one of two level-3 $W$ or $Z$ triggers, 
and have an ``uncorrected'' electromagnetic cluster with
$E_T > 20$~GeV and an associated track with $p_T > 13$~GeV/c.

Candidate electrons are required to be in the fiducial region.
This requirement primarily removes EM clusters which
overlap with uninstrumented regions of the detector. To avoid azimuthal
cracks, $|x|$ is required to be less than 18~cm, and 
to avoid the crack between the $z > 0$ and $z < 0$ halves of the detector, 
$|z|$ is required to be greater than 12~cm.  
The transverse EM energy is required to be greater than 25~GeV,
and to have an associated track with $p_T >$~15~GeV/c.
The track must pass through all eight superlayers of the CTC, which 
improves the electron purity and limits the occurence of very hard 
bremsstrahlung.
No other track with $p_T >$ 1 GeV/c associated with the nominal vertex
may point at the electron towers. 
This criterion reduces the QCD dijet background in the $W$ sample.  It
also has the effect of removing the $W$ and $Z$ events which have
secondary tracks associated with the decay electrons.  These secondary
tracks can result from the conversion of hard bremsstrahlung photons or
through accidental overlap with tracks from the underlying event.
Both of these sources are included in the simulation.
Events are rejected when another track has an invariant mass
below 1 GeV when combined with the electron cluster.

A $\zee$ event can fake a $\wenu$ event if one of the
electrons passes through a crack in the calorimeter.  Most of these electrons 
are in the tracking volume.  An event is considered to be 
a $Z$ candidate if
there is a second track with $p_T > 10$~GeV/c which has opposite sign
to the electron track and points at either the $\theta = 90^\circ$ or 
$\theta = 30^\circ$ 
crack, or is extrapolated to $|x| > 21$~cm in the strip chamber.
$Z$ candidate events are removed from the $W$ sample.
For the $Z$ sample, the two electron tracks are required to have opposite
sign. 
The selection criteria described above are properly included 
in the Monte Carlo simulation~\cite{ag_thesis}.
The transverse mass in the region $65 < M_T < 100$~GeV/c$^2$ and
the invariant mass in the region $70 < M < 110$~GeV/c$^2$ are used for
extracting the $W$ mass and the $Z$ mass, respectively.
These transverse and invariant mass cuts apply only for mass fits and
are absent when we otherwise refer to the $W$ or $Z$ sample.
The final $W$ sample contains 42,588 events, of which 30,115 are
in the region 65 $< M_T <$ 100~GeV/c$^2$.
The final $Z$ sample contains 1,652 events, of which 1,559 are
in the region 70 $< M <$ 110~GeV/c$^2$.
The $E_T^e$, $E_T^\nu$, and $M_T$ after all cuts are shown in
Figure~\ref{f_wvar} for the $W$ sample. 

\begin{table}
\begin{center}
\begin{tabular}{|l|r|r|}
Criterion		&W events after cut&Z events after cut\\ \hline
Initial sample		& 108,455 & 19,527	\\
Z vertex requirement	& 101,103 & 16,724	\\
Fiducial requirements	& 74,475 & 9,493	\\
Tracks through all CTC superlayers 
			& 71,877 & 8,613	\\
$E_T^e >$ 25 GeV	& 67,007 & 6,687	\\ 
$E_T^\nu >$ 25 GeV 	& 55,960 & N/A 		\\
$|\bf u| <$ 20 GeV 	& 46,910 & N/A 		\\
$P_T^e >$ 15 GeV 		& 45,962 & 5,257 	\\
$N_{\rm tracks}$ in the electron towers = 1 	
			& 43,219 & 1,670 	\\
$M_{e,track} <$ 1 GeV 	& 43,198 & N/A	 	\\
Not a Z candidate 	& 42,588 & N/A	 	\\
Opposite sign		& N/A	 & 1,652	\\ \hline
Mass fit region 	& 30,115 & 1,559 	\\
\end{tabular}
\end{center}
\caption{Effect of selection cuts.}
\label{wenucuts}
\end{table}

\begin{figure}[p]
\epsfysize=6.0in
\epsffile[54 162 531 675]{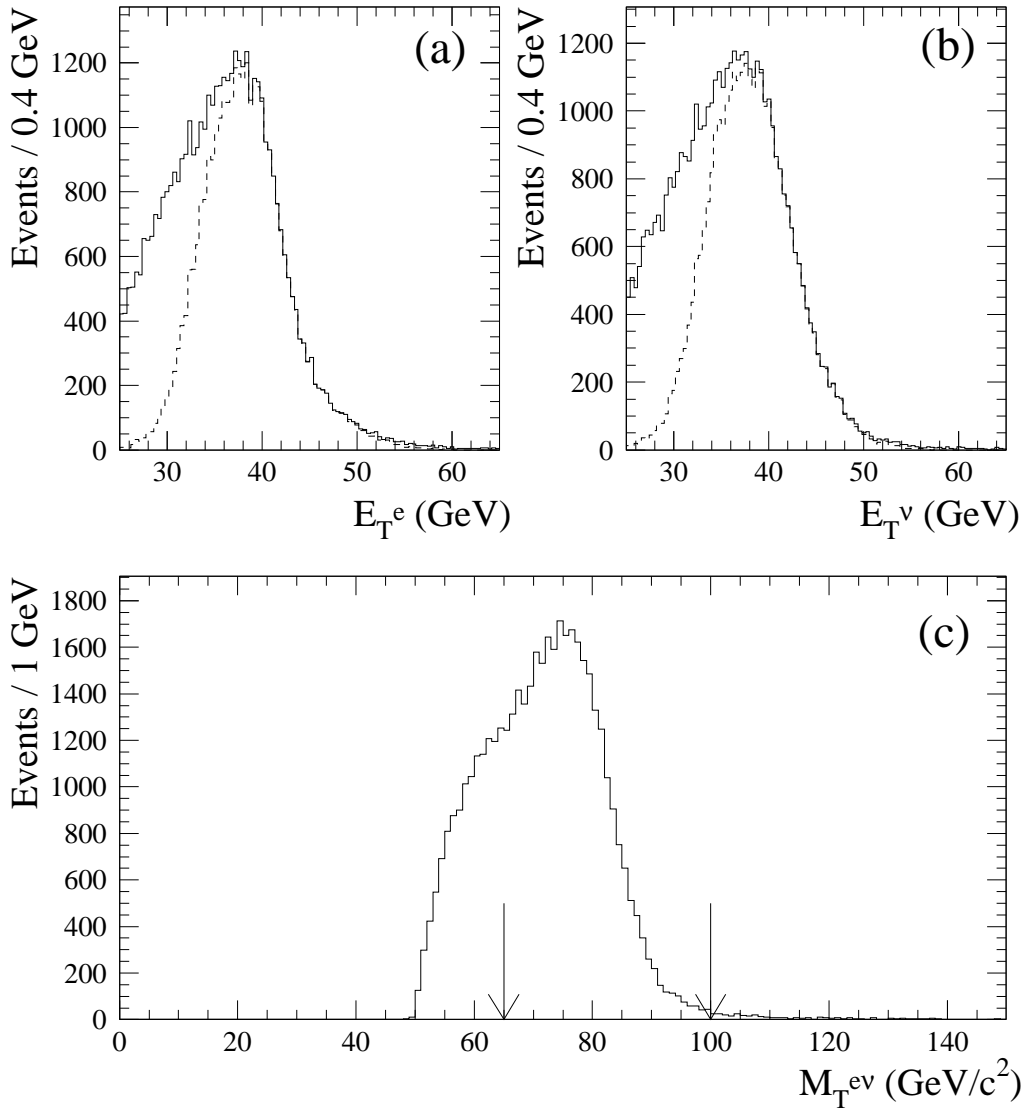}
\vspace*{0.5cm}
\caption{Kinematic quantities from the final $\wenu$ sample.
	$E_T$ distributions of (a) electrons and
	(b) neutrinos.
	The dashed curves show the events 
	in $65 < M_T < 100$ GeV, the fit region for the $W$ mass
	measurement.
	(c) Transverse Mass distribution. 
	The arrows indicate the region used in the $W$ mass fit.}
\label{f_wvar}
\end{figure}

\subsection{Electron Energy Scale and Resolution}
\label{escale}

All calibrations described above~\ref{rel_corr} are relative corrections 
designed to improve uniformity. The energy scale is extracted from the 
reconstruction of the $Z$ mass. The $Z$ Monte Carlo events are generated
in the manner described in Section~\ref{ctcscale}.
The Monte Carlo events are then processed through the detector 
simulation where the electron energy is smeared according to the
resolution:
\begin{equation}
 {\sigma_{E_T} \over E_T} = \sqrt{ {(13.5\%)^2 \over {E_T}} + \kappa^2}
\end{equation}
where all energies are in GeV, the stochastic term 13.5\% was measured in
the test beam, and the constant $\kappa$ includes such effects as shower
leakage and residuals from the uniformity corrections discussed in
Section 4.2. The parameter $\kappa$ is allowed to vary in the $Z$
mass fit. 
The other variable parameter in fitting the Monte Carlo events 
to the data is a scale factor, $S_E$.

For the fit, a binned maximum likelihood technique is used where the data
and Monte Carlo events for $M_Z$ are divided into 1~GeV/c$^2$ bins 
for the interval $70-110$~GeV/c$^2$. 
The results are:
\begin{equation}
  S_E(Z) = {M_Z^{\rm PDG} \over M_Z^{\rm CDF}} = 1.0000 \pm 0.0009
\label{escale_z}
\end{equation}
and 
\begin{equation}
\kappa = (1.53 \pm 0.27)\%,
\end{equation}
where the uncertainties come from the $Z$ statistics.
The fit results are shown in Figure~\ref{f_zmassfit}. 
The two parameters are largely uncorrelated.
The value of $S_E$ is equal to 1 by construction;
the initial value of $S_E$ was not 1, but 
we iterated the fit with the scale factor applied to the energy 
until the final scale factor becomes 1.
\begin{figure}[p]
\epsfysize=6.0in
\epsffile[54 162 531 675]{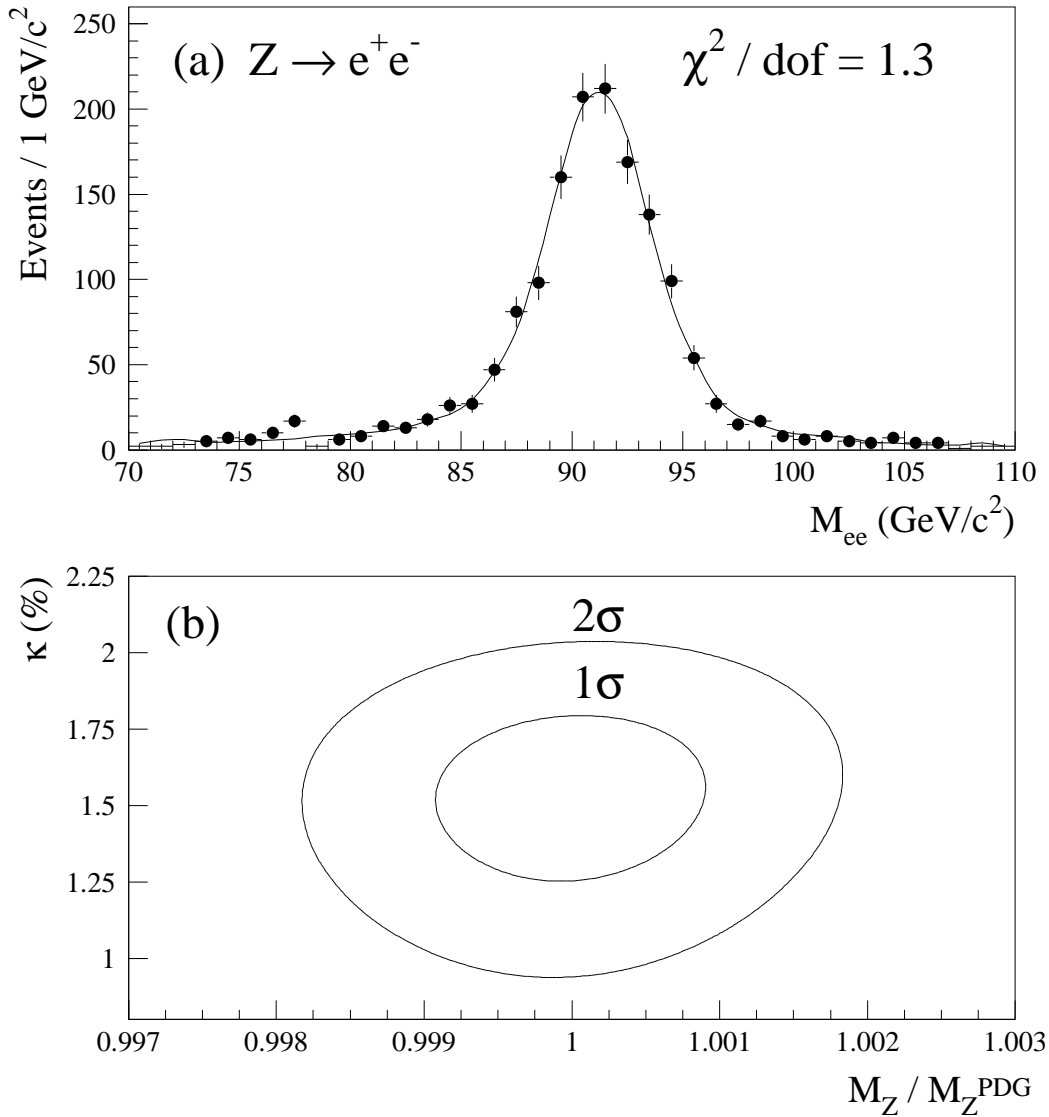}
\vspace{1cm}
\caption{Results of fit to $Z$ mass and energy resolution.
	(a) Invariant mass distribution. 
	The points are the data, and the solid line is the Monte 
	Carlo simulation (normalized to the data) with best fit.
	(b) Correlation between the scale factor ($S_E$) and the constant
	term ($\kappa$) in the resolution function.}
\label{f_zmassfit}
\end{figure}

A number of checks are performed to insure that these results are
robust and unbiased. For example, 1000 Monte Carlo subsamples
are created where each sample has the same size as the data, 
and are used to check that the likelihood procedure is unbiased
and that statistical uncertainties by the fit are produced correctly.
Moreover, compatible results are found when a much simpler technique 
is used, that is, comparing
the mean $M_Z$, in the interval $86-96$
GeV/c$^2$, between the data and the Monte Carlo events.
The Monte Carlo events include a 1\% QCD background
term. If the background term were omitted entirely, the energy scale and 
$\kappa$ would change by much less than their
statistical uncertainties; we conclude that the uncertainties in the
background have negligible contribution to the uncertainties in the fit
results. Finally a Kolmogorov-Smirnov (KS) statistic is used to
quantify how well the Monte Carlo events fit the data. 
The probability that a statistical fluctuation of the Monte Carlo parent
distribution would produce a worse agreement than the data is 19\%.
The likelihood fit is also checked by varying the
parameters in the KS fit to find a maximum probability. 
The result is $S_E = 1.0007 \pm 0.0010$, 
in good agreement with the likelihood method.

\subsection{Energy Nonlinearity Correction}
\label{enonlin}

The average $E_T$ for $Z$ decay electrons is about 4.5~GeV higher 
than those for $W$ decay. Since the energy calibration is done with 
the $Z$'s, any
nonlinearity in the energy response would translate to an 
incorrect energy scale at the $W$. 
The nonlinearity over a small range of $E_T$ can be expressed as
\begin{equation}
	{\Delta S_E \over S_E} = \xi \times \Delta E_T.
\end{equation}
The slope, $\xi$, could arise from several sources: energy loss
in the material of the solendoid, 
scintillator response versus shower depth, or shower
leakage into the hadronic part of the calorimeter.
The near equality of the $E/p$ scale factors for the $W$ and $Z$ samples
limits the slope to be less than about 0.0004 GeV$^{-1}$. The spread in
electron $E_T$ for each of the $W$ and $Z$ samples is larger than the
difference in the averages, so the most sensitive measure of $\xi$
is the variation of the mean $E/p$ between 0.9 and 1.1 
for both samples as a function of $E_T$. 
Their $E_T$ distributions and the residuals, $\langle E/p \rangle_{\rm
data} - \langle E/p \rangle_{\rm simulation}$, are shown in 
Figure~\ref{f_enonlin}.

\begin{figure}[p]
\epsfysize=6.0in
\epsffile[54 162 531 675]{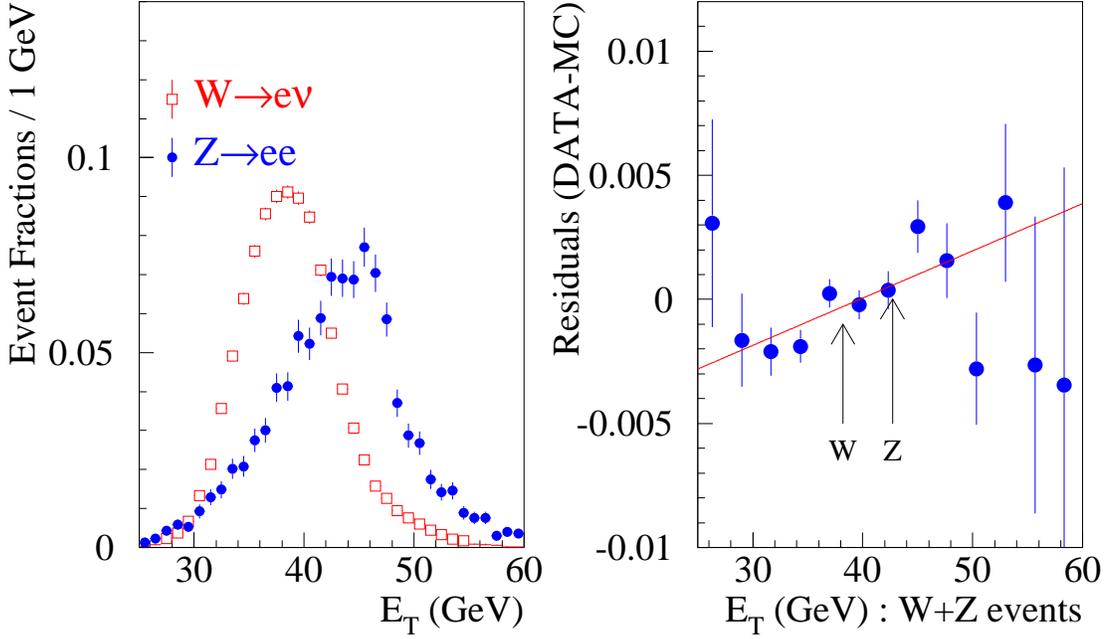}
\vspace{-4cm}
\caption{Left: The $E_T$ distributions of electrons from $W$ and $Z$ 
	decays.
	Right: Residual of data and Monte Carlo fit to $E/p$ versus 
	electron  $E_T$ for the W and Z samples. 
	The solid line is a linear fit with $\chi^2$/dof = 1.4.
	When the slope is forced to be zero, the $\chi^2$/dof 
	increases to 2.2.  The arrows represent the average $E_T$ values
	of the electrons for the $W$ and $Z$ samples.}
\label{f_enonlin}
\end{figure}

A linear fit to the $E/p$ residuals for the $W$ and $Z$ data
yields a slope of $(1.91 \pm 0.58) \times 10^{-4}$~GeV$^{-1}$ 
in $\langle E/p \rangle$. Correcting the relationship between 
$\langle E/p \rangle$ and the scale factor gives a slope
$\xi = - 0.00029 \pm 0.00013 ({\rm stat.}) 
\pm 0.00006 ({\rm syst.})$~GeV$^{-1}$, 
where the systematic uncertainty comes from 
backgrounds and the fitting procedure.
The electron $E_T$ is corrected by
\begin{equation}
	E_T \rightarrow E_T (1 - 0.00029 (E_T - 42.73~{\rm GeV}))
\end{equation}
before the final fit for the $W$ mass.
This correction shifts the fitted $W$ mass up by $(34 \pm 17)$~MeV/c$^2$.
The mean $E_T$ for the $Z$ sample is 42.73~GeV, so the energy scale is
unchanged at that point. 

\subsection{Check of Energy Scale and Momentum Resolution Using $E/p$}
\label{epscale}

The momentum scale was set with the $\zmumu$ mass
as discussed in Section~\ref{mumeas}, and the energy scale was set with the
$\zee$ mass as discussed in this section. In principle, the electron 
energy scale can be set by 
transferring the momentum scale from the $\Upsilon$(1s) or $J/\psi$ 
$\rightarrow \mu^+\mu^-$ mass as done in the Run IA analysis 
and equalizing $E/p$ for data and
simulation in $\wenu$ decays. This technique has great statistical power
and indeed was the preferred technique in previous CDF publications of
the $W$ mass~\cite{wmass_1a,trisch}.  However, systematic effects in 
tracking electrons are potentially much larger than for muons due to 
bremsstrahlung.  
To accurately simulate external bremsstrahlung effects~\cite{tsai},
the Monte Carlo program includes the magnitude and distribution of the  
material~(see Section~\ref{trkrecon}) traversed by electrons from the 
interaction region through the tracking volume, propagation
of the secondary electrons and photons,\footnote{The photons
are treated in the same manner as the electrons in the calorimeter 
simulation.} 
and a procedure handling the bias on the beam constrained momentum 
which is introduced through the non-zero impact parameters of electrons
that have undergone bremsstrahlung~\cite{ag_thesis}.

To fit to the $E/p$ distribution (see Figure~\ref{f_eoverpfit}) to 
determine the energy scale, 
the width of the $E/p$ distribution needs to be understood. 
It has a contribution from both the $E$ resolution and the $p$
resolution.  At the $W$ electron energies, the $p$ resolution
dominates.
When the $E/p$ distribution is fit to determine the energy scale, 
the $E$ resolution is fixed to the value determined
by the $Z$ data, and the $1/p_T$ resolution is allowed to vary.
As can be seen from Figure~\ref{f_ep_resolution}, the $E/p$ distribution
agrees well with the resolution values determined solely from the
$\zmumu$ data.  
However, there is an excess at the low $E/p$ tail region.
Studies of the transverse mass for data events
in this region show that the tail is due 
to mis-measured tracks in real $W$ events.
To account for this excess,  
the track parameters are smeared according to a second, wider Gaussian
term for 8\% of the Monte Carlo events.
The two Gaussians describe the overall $E/p$ distribution well.
However, adding the second Gaussian distribution 
does not significantly change the derived scale.

\begin{figure}[p]
\epsfysize=5.8in
\epsffile[54 162 531 675]{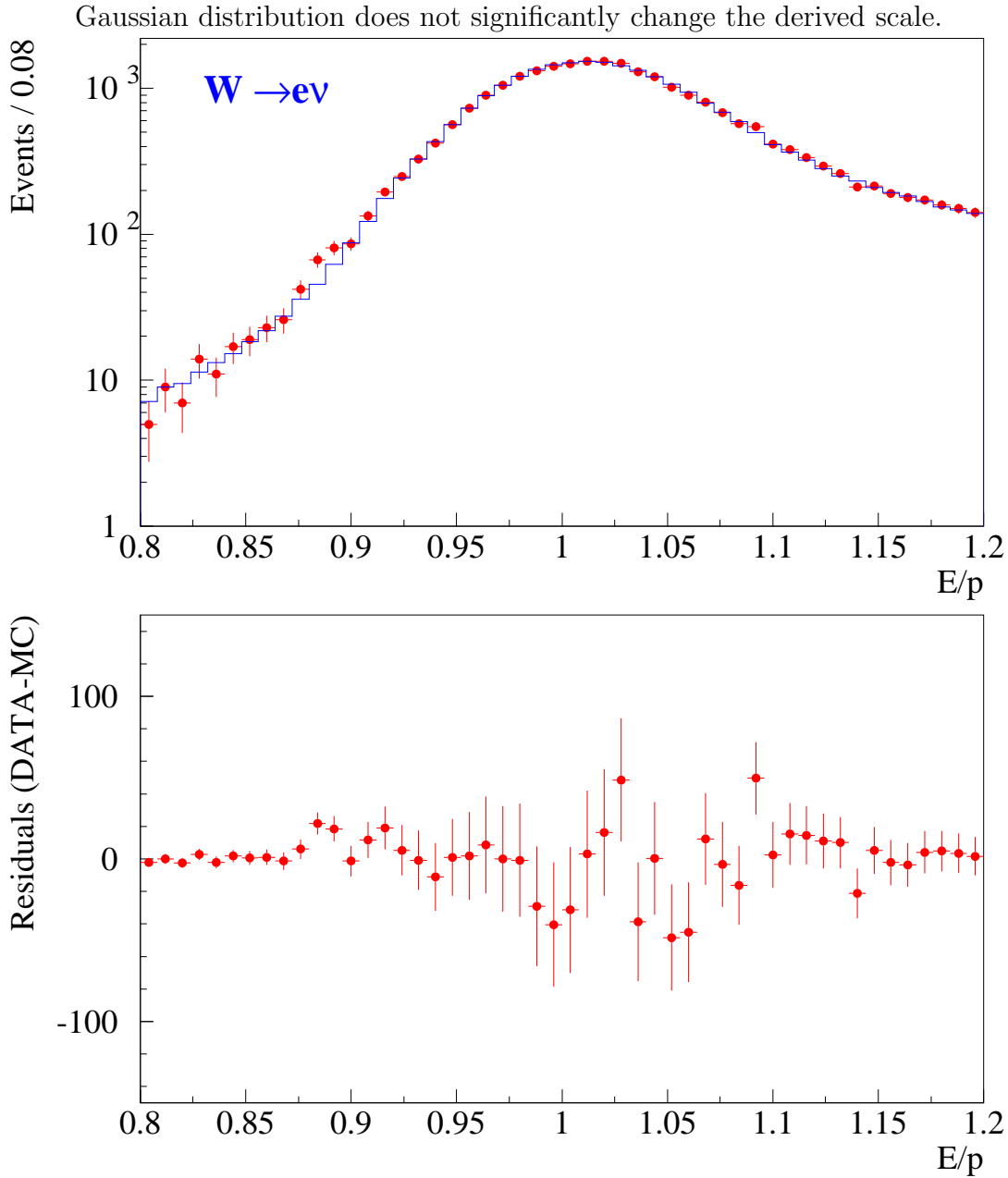}
\vspace{1.5cm}
\caption{Top: $E/p$ distribution for $W$ events (points) and 
	the best Monte Carlo fit.
	The solid histogram is the Monte Carlo fit normalized to data, 
	and the points are the data.
	The fit reproduces the shape very well as indicated
	by the $\chi^2$/dof =0.86.
	Bottom: The difference between the data and the best fit
	simulation.}
\label{f_eoverpfit}
\end{figure}

\begin{figure}[p]
        \centerline{\epsfysize 18cm
                    \epsffile{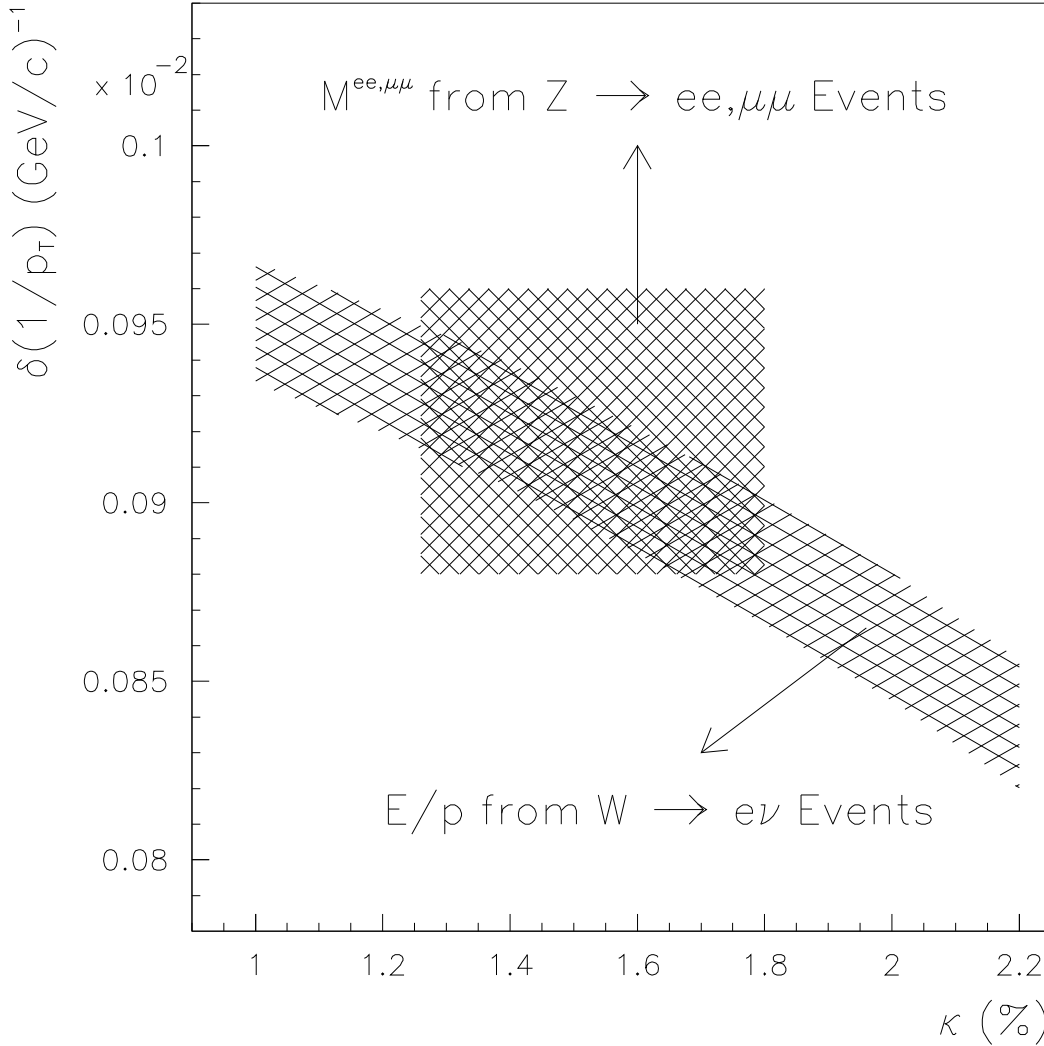}}
\vspace*{-2cm}
\caption{
The energy resolution $\kappa$ and tracking resolution 
$\delta (1/ p_T)$ as determined from fits to the $E/p$ 
distribution in $\wenu$
events, compared to the same resolutions determined from the $\zee$ and
$\zmumu$ data.}
\label{f_ep_resolution}
\end{figure}

The $E/p$ distribution is fit for an energy scale and tracking resolution
using a binned likehood method.  The method is similar to the one used to
fit the $Z$ mass.
The data are collected in 25 bins for the region 
$0.9 < E/p < 1.1$, containing 22,112 events as shown in 
Figure~\ref{f_eoverpfit}. 
The log likelihood is maximized with respect to
$S_E$ and the momentum resolution simultaneously. 
The energy scale factor is found to be 
\begin{eqnarray}
S_E(E/p) & = & 0.99633 \pm 0.00040{\rm (stat.)} \nonumber \\
         &   & \pm 0.00024(\kappa) \pm 0.00035(X_\circ) 
		\pm 0.00018(p_T {\rm ~scale}), \nonumber
\end{eqnarray}
where 0.00024 comes from the uncertainty in the calorimeter resolution,
0.00035 from the uncertainty in the radiation length measurement,
and 0.00018 comes from the uncertainty in the momentum scale which 
for this purpose is 
determined by the $\Upsilon$(1s) measurement (see Section~\ref{scalecheck}).
The result of the fit is shown in Figure~\ref{f_eoverpfit}. 
When we account for the nonlinearity of the calorimeter energy 
between $Z$ decay electrons and $W$ decay electrons as described in
Section~\ref{enonlin}, 
the scale factor becomes
\begin{eqnarray}
       S_E(E/p) = 0.99480 & \pm & 0.00040 {\rm ~(stat.)} \\ \nonumber 
                         & \pm & 0.00024 {\rm ~(\kappa)} 
                           \pm   0.00035 {\rm ~(X_\circ)} 
			   \pm   0.00018 {\rm ~(p_T~scale)} \\ \nonumber
			 & \pm & 0.00075 {\rm ~(CEM~nonlinearity).}
\label{esc_wep_xxx}
\end{eqnarray}
It is in poor agreement (3.9$\sigma$ discrepant)
with the energy scale determined from the $Z$ mass (Eq.~\ref{escale_z}).
When this scale factor is applied to the data, the $Z$ mass is
measured to be 0.52\% lower than the world-average value.

The $E/p$ distribution for the $Z$ sample is also used to extract
$S_E$.  The result is: 
\begin{eqnarray}
       S_E(E/p) = 0.99720 	& \pm & 0.00130 {\rm ~(stat.)} \\ \nonumber
                        	& \pm & 0.00024 {\rm ~(\kappa)} 
                        	  \pm   0.00035 {\rm ~(X_\circ)} 
				  \pm   0.00018 {\rm ~(p_T~scale).}
\label{esc_zep_xxx}
\end{eqnarray}
The systematic uncertainties with respect to $\kappa$, $X_\circ$, 
and momentum scale are common for the $W$ and $Z$ samples.
The difference between this scale value and the scale from 
the $Z$ mass is 2.0$\sigma$.  
When both the $W$ and $Z$ events are combined, 
the discrepancy is 5.3$\sigma$.

The disagreement between the energy scale determined from the $Z$ mass 
(Eq.~\ref{escale_z}) with that determined by the $E/p$ distribution 
(Eq.s~14 and 15) is significant; therefore
it would be incorrect to average the two. Moreover, the two techniques
applied to the $Z$ sample use the same energy measurements, thus 
hinting 
at a systematic problem between the tracking for muons and that for
electrons, or a systematic difference between the actual tracking and 
the tracking simulation. Another possibility is an incomplete modeling of
the calorimeter response to bremsstrahlung in the tracking volume.
Appendix A describes some possible causes. 

As a result of this disagreement, we choose to use conservative methods
for both the electron energy and muon momentum scale determination.
We use the $\zee$ mass instead of the $E/p$ distribution 
to set the electron energy scale 
since this is a direct calibration of the calorimeter measurement 
without reference to 
tracking or details of the bremsstrahlung process.
Although statistically much less precise, we use the $\zmumu$ mass
instead of the $\Upsilon$(1s) or $J/\psi$ mass
to set the muon momentum scale.

\subsection{Summary}

The electron energy scale is determined by normalizing the measured
$\zee$ mass to the world-average mass. The measurement is limited by the
finite statistics in the $Z$ peak which gives the uncertainty of 
72~MeV/c$^2$ on $M_W$. 
A small nonlinearity is observed, resulting in $\Delta M_W = (34 \pm
17)$~MeV/c$^2$.
Adding these uncertainties in quadrature, the total uncertainty on 
$M_W$ due to the energy scale determination is 75~MeV/c$^2$.
The energy resolution is measured from the width of the $\zee$ peak
in the same dataset:
${\sigma_{E_T} \over E_T} = \sqrt{ {(13.5\%)^2 \over {E_T}} + 
	(1.53 \pm 0.27)\%^2}.$
When the electron energy resolution is varied over this allowed range 
in the simulation, the measured $W$ mass changes by 25~MeV/c$^2$.


\section {Backgrounds}
\label{backgrounds}

Backgrounds in the $W$ samples come from the following processes:\\
\begin{tabular}{cl}
1.& $W \rightarrow \tau \nu \rightarrow\ell\nu\nu\nu$ \\
  & $W \rightarrow \tau \nu \rightarrow$ hadrons + $\nu\nu$\\
2.& $Z \rightarrow\ell^+\ell^- $ where the second charged lepton is not 
			detected\\
3.& Dijets (QCD) where jets mimic leptons \\
4.& cosmic rays\\
\end{tabular}
\\

\noindent
Contributions from $Z \rightarrow \tau^+ \tau^-$, $W^+W^-$, and $t\overline{t}$
are negligible.
In general, backgrounds have a lower average
transverse mass than $W \rightarrow \ell \nu$ decay, and, if not accounted for,
will lower the fitted mass. All the background distributions 
as shown in Figure~\ref{backmt} are included in the simulation.

\begin{figure}[p]
\vspace{-1.0in}
\epsfysize=3.0in
\epsffile[40 500 677 765]{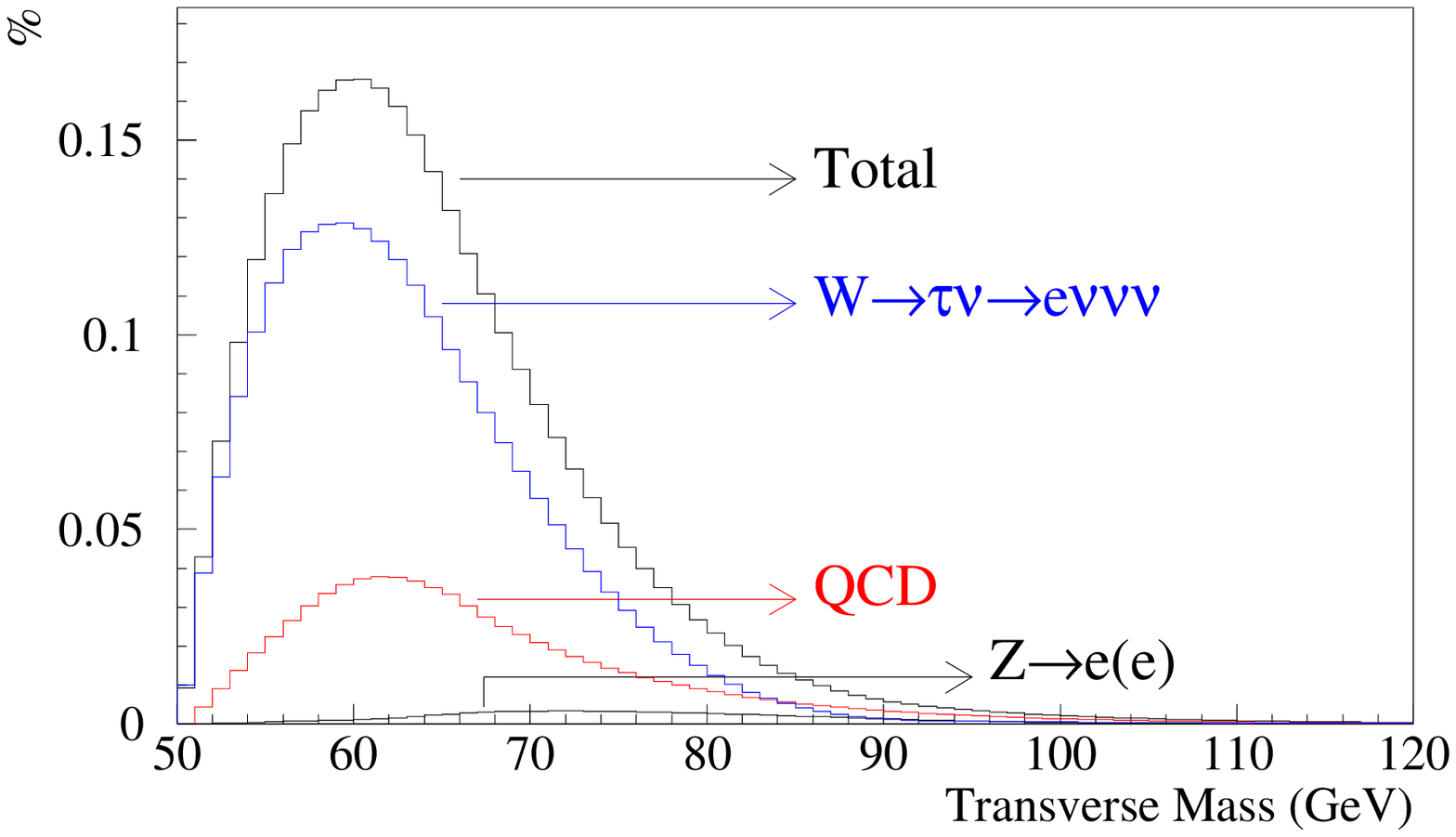}
\vspace{.2in}
\epsfysize=3.0in
\epsffile[40 500 677 765]{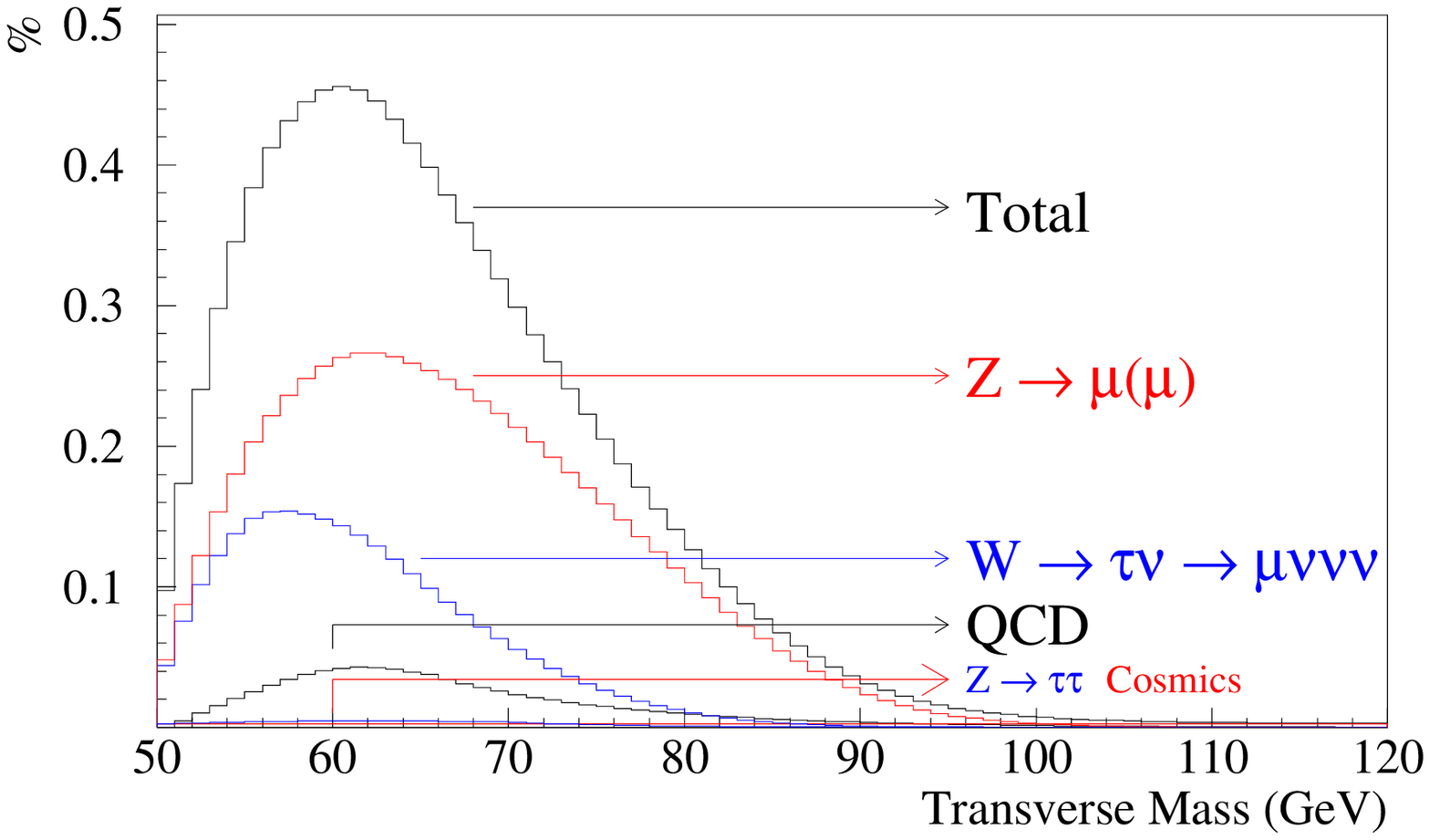}
\vspace{1.3in}
\caption{The fractions (\%) of backgrounds as a function of 
transverse mass distribution for the $\wenu$ 
sample (Top) and the $\wmunu$ sample (Bottom).
The smallest contributor, 
$W \rightarrow \tau\nu \rightarrow$ hadrons + $\nu\nu$, is not shown
in this figure.}
\label{backmt}
\end{figure}

\subsection{$W\rightarrow e\nu$ Backgrounds}

Few $W\rightarrow \tau\nu \rightarrow e\nu\nu\nu$ events pass the 
kinematic cuts since the electron $E_T$, the total neutrino 
$| {\bf E_T} |$, 
and $M_T$ are substantially lower than those in the $\wenu$ decay. 
$W\rightarrow\tau\nu\rightarrow e\nu\nu\nu$ events are estimated
to be 0.8\% of $\wenu$ events in the $W$ mass fitting region.  
This is the largest background in the $W\rightarrow e\nu$
sample, and is also the easiest to simulate.
We have also simulated the $W \rightarrow \tau\nu$ background where 
the $\tau$ decays hadronically.
We expect it to be $(0.054 \pm 0.005)$\% of the $W$ sample.
After $Z$ removal cuts, very few $\zee$ events can mimic $\wenu$
events. The Monte Carlo simulation predicts $(0.073 \pm 0.011)$\%
of the $W$ sample in the mass fitting region to originate from $\zee$.

Dijet events can pass the $W$ selection cuts if one of the jets 
mimics an electron and the other is mismeasured, creating $\met$.
Such events are refered to as ``QCD'' background.  The QCD background 
is estimated by 
selecting QCD candidates from the $W$ sample without $M_T$ and 
$| {\bf u}|$ cuts and plotting distributions of $|\bf u|$ and $M_T$ 
as shown in Figure~\ref{f_mt_vs_u} (a detailed description can be 
found in Reference~\cite{ag_thesis}).  The number of QCD events 
predicted in the signal region ``Region A'' (see the top figure) 
is given by 
\begin{eqnarray}
	N_{Region~A~(W)}
	& = & {N_{Region~A~(QCD)} \over N_{Region~B~(QCD)}} \times
	   N_{Region~B~(W)} \nonumber \\
	& = & 249 \pm 108, \nonumber
\end{eqnarray}
from which we find $119 \pm 56$ events or $(0.36 \pm 0.17)$\% of the $W$ events
are in the $W$ mass fitting region.
The kinematical distributions of the QCD events are derived
from the $\wenu$ sample with inverted electron quality cuts.

\begin{figure}[p]
\mbox{\hspace*{-0.2in} \epsfysize=6.2in \epsffile{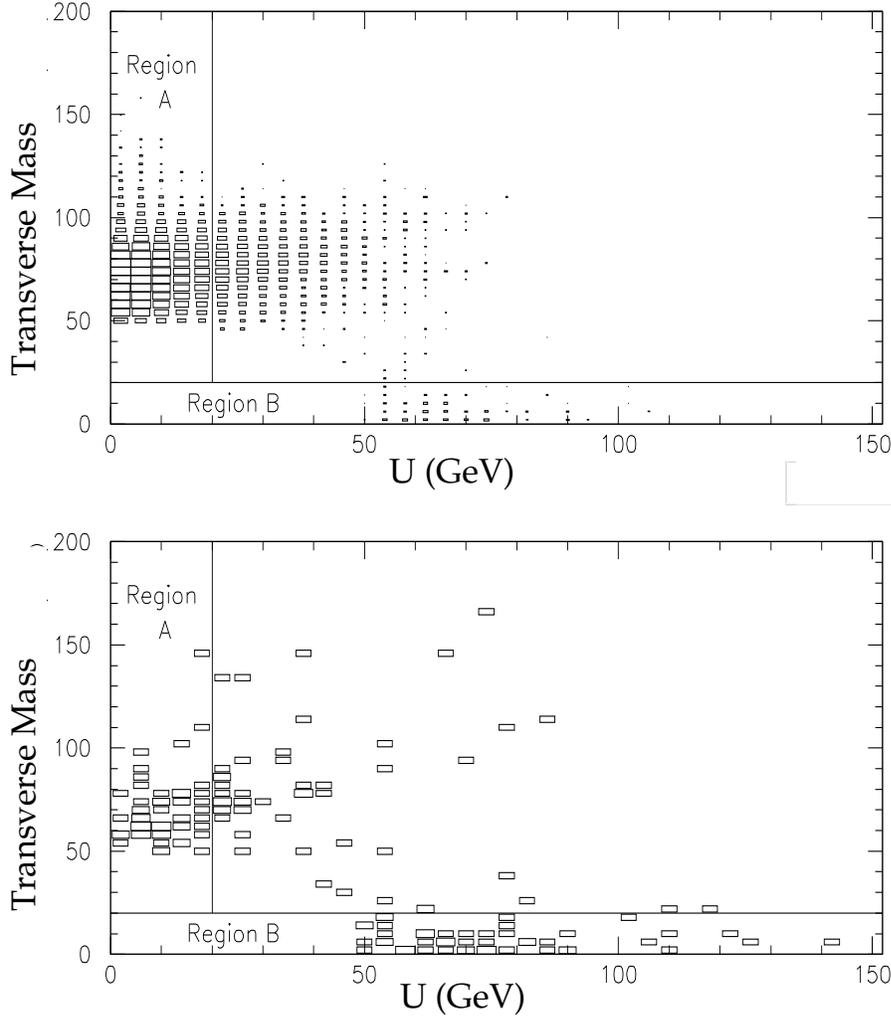}} 
\caption{$M_T$ (GeV/c$^2$) 
vs $|\bf u|$ distributions without $M_T$ and $|\bf u|$ cuts
for all $W$ data (Top), and a QCD subset of the $W$ data (Bottom).}
\label{f_mt_vs_u}
\end{figure}

\subsection{$W\rightarrow\mu\nu$ Backgrounds}

The largest background in the $W \rightarrow \mu\nu$ sample comes
from the $\zmumu$ process with one of the muons exiting at low
polar angle (outside of the CTC volume) which mimics 
a neutrino in the calorimeters.
The simulation predicts this background to be (3.6 $\pm$ 0.5)\%.
The uncertainty in the background estimate comes from two sources:
the uncertainty in the measured tracking efficiency at large $\eta$, 
and the choice of parton distribution functions.

The second largest background comes from 
the $W\rightarrow \tau\nu$ process where $\tau \rightarrow
\mu\nu\nu\nu$, which is 0.8\% of the $W$ sample.
The $W \rightarrow \tau\nu$ background where 
the $\tau$ decays hadronically is negligible.
Background from QCD is estimated by using the data in a similar manner
to the electron case.  The $\wmunu$ sample is estimated to contain 
$(0.4 \pm 0.2)$~\% of its events from the QCD process.
Cosmic rays can appear as two oppositely charged back-to-back tracks 
in $\phi$ when they cross the detector in time with ${\bar p}p$ 
collisions.
Most of them are removed by the $\wmunu$ selection criteria such as
the $Z$ removal cut or $|D_0| < 0.2$~cm (see Section~\ref{muonid}).
The number of cosmic rays remaining in the final sample 
is estimated by using events
which fail $|D_0| <$ 0.2~cm criteria, but which 
pass all the other selection criteria.
The expected number of cosmic ray events corresponds to 
$(0.10 \pm 0.05)$\% of the $W$ sample.

\subsection{Summary}

Table~\ref{t_background} summarizes the fraction of the background
events in the $W$ samples in the mass fitting region. 
The total backgrounds in the $\wenu$ and $\wmunu$ fit region are 
expected to be $(1.29 \pm 0.17)$\% and $(4.90 \pm 0.54)$\%, respectively.
Adding the backgrounds in the simulation leads to
shifts of $(80 \pm 5)$~MeV/c$^2$ and $(170 \pm 25)$~MeV/c$^2$ in 
the $\wenu$ and $\wmunu$ mass measurements, respectively.

\begin{table}
\begin{center}
\begin{tabular}{|l||c|c|}
Background source &
  $W\rightarrow e\nu$ sample 	& $W\rightarrow \mu\nu$ sample	\\ \hline
$W\rightarrow \tau\nu\rightarrow \ell\nu\nu\nu$ 
				&0.8\%		
				&0.8\% 			\\
$\wtaunu\rightarrow$ hadrons + $\nu\nu$	&$(0.054 \pm 0.005)$\%
				& $-$			\\
Lost $Z\rightarrow \ell \ell$	&$(0.073 \pm 0.011)$\%	
				&$(3.6 \pm 0.5)$\% 	\\
QCD				&$(0.36 \pm 0.17)$\%
				&$(0.4 \pm 0.2)$\% 	\\
Cosmic rays			& $-$	
				&$(0.10 \pm 0.05)$\% 	\\ \hline
Total				&$(1.29 \pm 0.17)$\%
				&$(4.90 \pm 0.54)$\%	\\
\end{tabular}
\end{center}
\caption{Backgrounds in the $\wenu$ and $\wmunu$ sample 
in the mass fitting region.}
\label{t_background}
\end{table}

\section {$W$ Production and Decay Model}
\label{wprod}

We use a Monte Carlo program to generate $W$ events according to a
relativistic Breit-Wigner distribution
and a leading-order ($p_T^W=0$) model of
quark-antiquark annihilation. The distribution in momentum of the quarks is
based on the MRS-R2 parton distribution functions (PDFs)~\cite{MRS_R2}. The
generated $W$ is Lorentz-boosted, in the center-of-mass frame of the
quark-antiquark pair, with a transverse momentum, {\ptw}. The \ptw\ spectrum
is derived from the \zee\ and \zmumu\ data and a theoretical prediction for
the ratio of $Z$ and $W$ \ptm\ spectra which is differential in the 
rapidity of the vector boson.  The Monte Carlo program also includes 
QED radiative effects~\cite{wgamma}.

\subsection{Parton Distribution Functions}
\label{SUBSECTION:PDF Uncertainty}

The uncertainty associated with PDFs is evaluated by varying 
the choice of PDF sets and by parametric modifications of PDFs.
Figure~\ref{f_w_charge_asym} shows the CDF data on the $W$ lepton charge
asymmetry~\cite{cdfwasym} 
which is sensitive to the ratio of $d$ to $u$ quark densities
$(d/u)$ at a given parton momentum fraction, $x$.  Of all modern 
PDFs, the two giving the best agreement, 
MRST~\cite{mrst} and CTEQ-5~\cite{cteq5}, 
are shown.\footnote{Predicted $W$ charge asymmetries
are calculated with the DYRAD NLO $W$ production program~\cite{DYRAD}.}
Unfortunately the agreement even with these PDFs is barely satisfactory.
Hence we follow reference~\cite{pdfmod} in making parametric
modifications to the MRS family of PDFs.  These modifications with
retuned parameters are listed in Table~\ref{t_PDFmod} and their
predictions are compared to the $W$ lepton charge asymmetry measurement
and the NMC $d/u$ data~\cite{NMC} in Figure~\ref{f_PDF_range}.
From the variation among the six reference PDFs, 
an uncertainty of 15 MeV/c$^2$ is taken which is common to the 
electron and muon analyses.

\begin{figure}[htbp]
\epsfysize=6.0in
\epsffile[54 162 531 675]{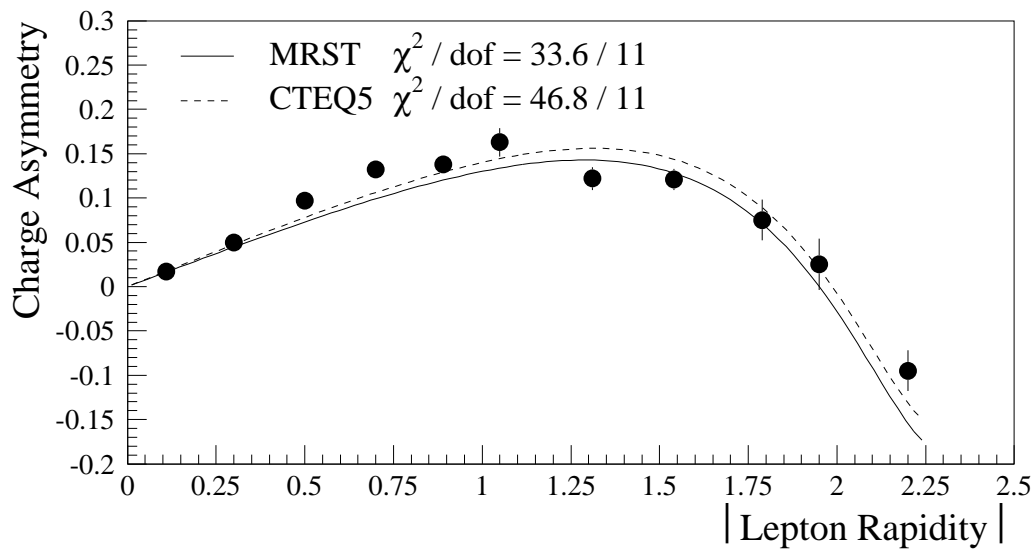}
\vspace{-7.0cm}
\caption{The CDF $W$ lepton charge asymmetry measurement compared
to predictions using the DYRAD calculations with MRST (solid) 
and CTEQ-5 (dashed) PDFs.}
\label{f_w_charge_asym}
\end{figure}

\begin{table}
\begin{center}
\begin{tabular}{|l|c|}
PDFs & Modification \\
\hline
MRST & $d/u \rightarrow d/u \times (1.07 - 0.07 e^{-8x})$ \\
MRS-R2 & $d/u \rightarrow d/u + 0.11 x \times (1 + x)$ \\
MRS-R1 & $d/u \rightarrow d/u \times (1.00 - 0.04 
e^{-\frac{1}{2}(\frac{(x-0.07)}{0.015})^2})$ \\
\end{tabular}
\end{center}
\caption{Reference PDFs and modifications.}
\label{t_PDFmod}
\end{table}

\begin{figure}[htbp]
\epsfysize=6.0in
\epsffile[54 162 531 675]{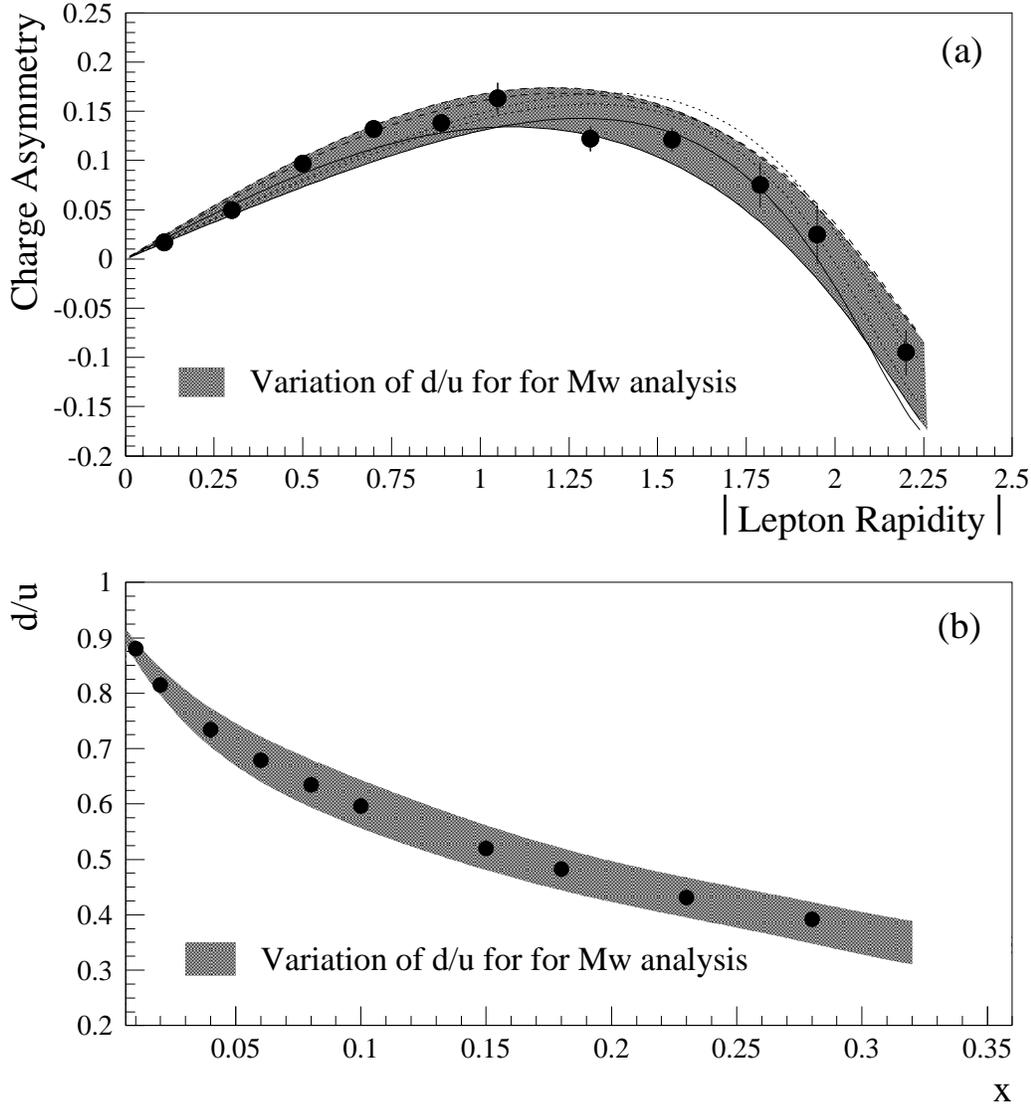}
\vspace*{0.5cm}
\caption{(a) The CDF measurement of the $W$ lepton charge asymmetry compared
with the six reference PDFs.  The upper and lower dotted curves are
MRS-R2 and MRS-R2 modified, the upper and lower dashed curves are
MRS-R1 modified and MRS-R1, and the upper and lower solid curves in
$|\eta|<1$ are MRS-T and MRS-T modified, respectively.
(b) The NMC $d/u$ data evolved to $Q^2 = M_W^2$.
The gray bands represent the range spanned by the six reference PDFs.}
\label{f_PDF_range}
\end{figure}

\subsection{$W$ Transverse Momentum Spectrum}
\label{SUBSECTION:PTW}

The spectrum of $W$ transverse momentum, $p_T^W$, is needed to simulate the
lineshape of transverse mass.  The $W$ mass measurement uses events at
low $p_T^W$ where the theoretical calculations are not reliable.
It would be difficult to extract $p_T^W$ from the $W$ data because
the neutrino momentum is not well measured.
However one can model $p_T^W$ through a measurement of $p_T^Z$,
which can be measured accurately using the charged leptons
from the $Z$ decays.
Theoretical calculations predict the cross-section ratio of $W$'s and
$Z$'s as a function of $p_T$ with 
small uncertainty since the production mechanisms are similar~\cite{reno}.
The measurement of $p_T^Z$ is combined with the theoretical calculations
of the ratio to derive $p_T^W$. This procedure is applied separately to 
the muon and electron samples, so the derived \ptw\ distributions are 
essentially independent although compatible.

For each $Z$ sample, a functional form for the $Z$ \ptm\ distribution is
assumed for input to a Monte Carlo generator. The lepton response is
modeled according to detector resolution and acceptance. The parameters
of the assumed functions are fit to give agreement with the observed
$Z$ \ptm\ distributions.  The observed $Z$ \ptm\ distributions are
shown in Figure~\ref{f_zpt} and are compared with the simulation 
which uses the best fit parameters for the input $p_T^Z$ distribution.

\begin{figure}[htbp]
\epsfysize=6.0in
\epsffile[54 162 531 675]{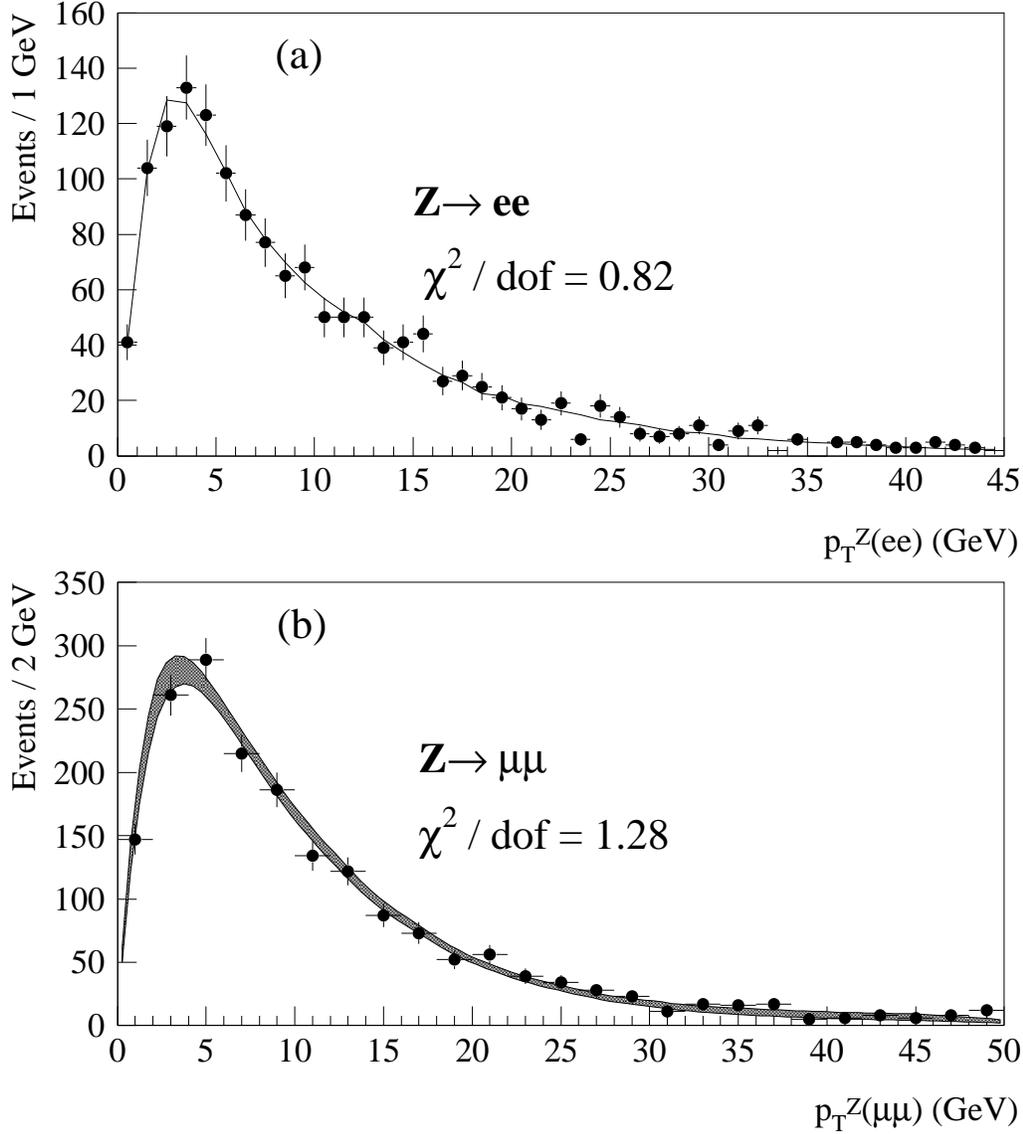}
\vspace{0.5cm}
\caption{The observed $Z$ \ptm\ distributions (points) 
for the (a) $\zee$ and (b) $\zmumu$ sample are compared
with the Monte Carlo simulation.
The solid line in (a) shows the best fit parameters for the
input $p_T^Z$ distribution, whereas the shaded band in (b) shows
the $1\sigma$ variation of the fit parameters.}
\label{f_zpt}
\end{figure}

Resummed calculations~\cite{resum,ptresum} are used for correcting 
the difference between the $W$ and $Z$ \ptm\ distributions, 
in terms of the ratio of the two distributions. 
As shown in Figure~\ref{f_pt_ratio} (a), (b) and (c), 
the ratio is between 0.9 and 1.0 over the $p_T$ range of interest.
Effects from the large ratio at $p_T \sim$ 0 is very small 
since $d\sigma / d(p_T) \rightarrow 0$ as $p_T \rightarrow 0$.
The variation of the ratio is studied by varying PDFs and nonperturbative
parameters in the resummed calculations, and by calculating it 
in two different resummed schemes, 
one in impact parameter space~\cite{resum} 
and the other in \ptm\ space~\cite{ptresum}.
There is a rapidity ($y^{boson}$) dependence to the \ptm\ distribution, 
illustrated in Figure~\ref{f_pt_ratio} (d) and (e).  
This rapidity dependence is taken into account when $p_T^W$ is 
derived from $p_T^Z$.
As indicated in Figure~\ref{f_pt_ratio}, the range of 
the possible ratio and rapidity dependence variation is about 2\%.

The extracted $p_T^W$ distribution for the muon channel 
at the generation level is shown in Figure~\ref{f_wpt}~(b).
The shaded band represents the total uncertainty on the 
$p_T^W$ distribution. 
The dominant uncertainty comes from the finite statistics of
the $Z$ sample.  The theoretical uncertainty in the \ptm\ ratio 
and rapidity dependence is small.
The fractional uncertainties on the $p_T^W$ distribution 
from the statistics and theoretical calculations are shown 
in Figure~\ref{f_wpt}~(a).

The uncertainty on the $W$ mass is evaluated by varying 
the $p_T^W$ distribution within the shaded band in 
Figure~\ref{f_wpt}~(a).
The finite statistics of the $Z$ sample contributes 
independent uncertainties of 15~MeV/c$^2$ and
20~MeV/c$^2$ for the $\wenu$ and $\wmunu$ channel.
The contribution of the theoretical uncertainty 
is 3~MeV/c$^2$ which is common for the electron and muon channel.

\begin{figure}[htbp]
        \centerline{\epsfysize 20cm
                    \epsffile{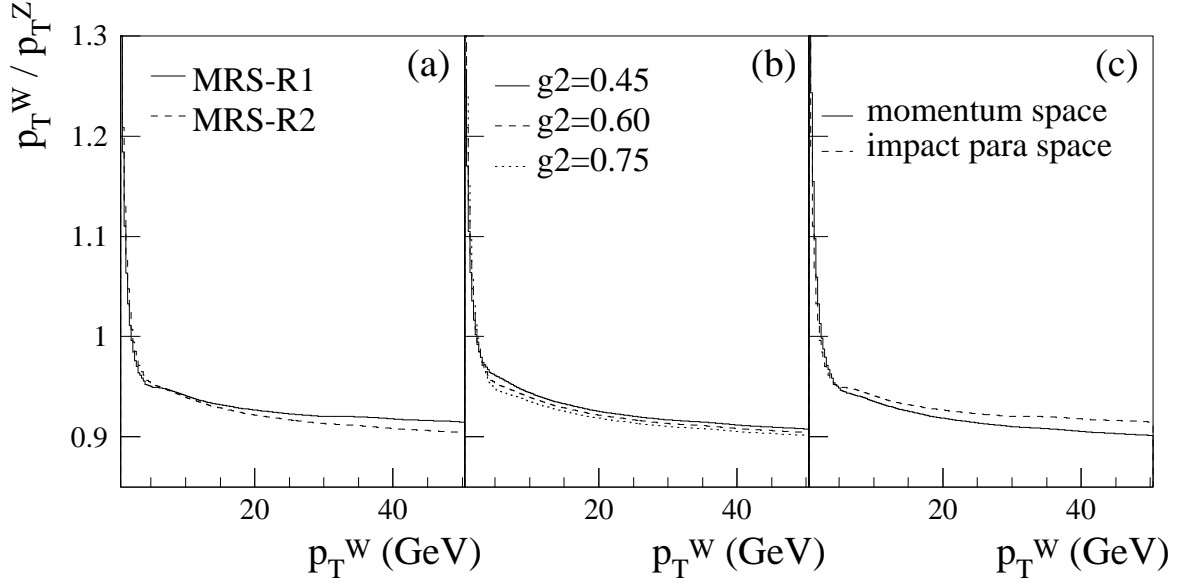}}
\vspace*{-10.5cm}
        \centerline{\epsfysize 18cm
                    \epsffile{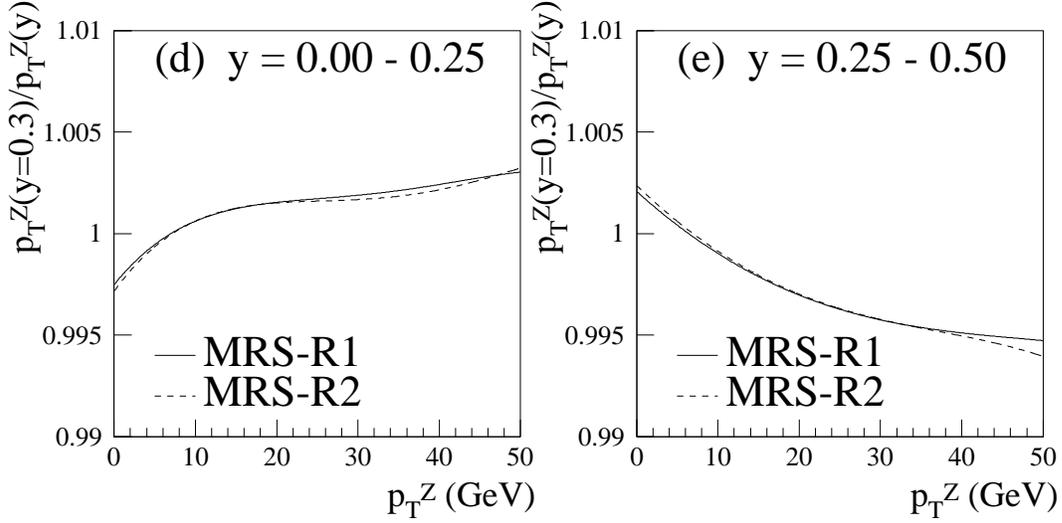}}
\vspace*{-8.5cm}
\caption{The ratios of the $p_T^W$ to $p_T^Z$ distribution from 
resummed calculations in impact parameter space showing (a) PDF
dependence, and (b) nonperturbative parameter dependence.
The ratios in impact parameter space and $p_T$ space are compared in
(c).
The ratio of $p_T^Z$ at $y^Z = 0.3$ to $p_T^Z$ for 
(d) $0 < y^Z < 0.25$, and (e) $0.25 < y^Z < 0.5$.}
\label{f_pt_ratio}
\end{figure}

\begin{figure}[htbp]
\epsfysize=6.0in
\epsffile[54 162 531 675]{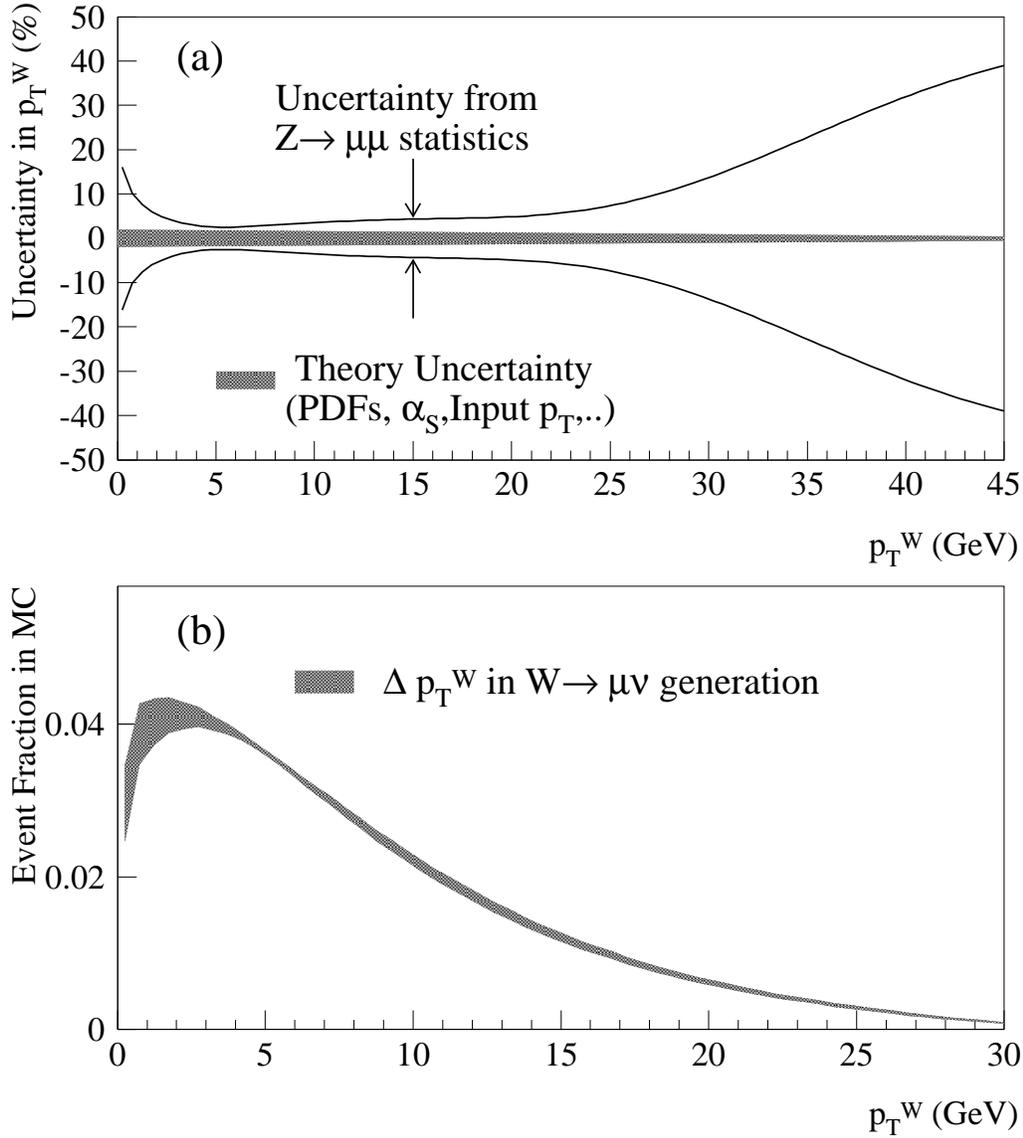}
\vspace{1.0cm}
\caption{
(a) The fractional uncertainties on $p_T^W$ 
as a function of $p_T^W$.  The solid lines show the uncertainty  
due to the $Z$ statistics and the shaded band the uncertainty 
due to the theoretical calculations.
(b) The $p_T^W$ distribution extracted from the $p_T^Z$ distribution 
and the theoretical calculations of $p_T^W / p_T^Z$ for the 
$\wmunu$ mass measurement.
The band represents the uncertainties.}
\label{f_wpt}
\end{figure}

\subsection{QCD Higher Order Effects}
 
The $W$ bosons are treated
as spin-one particles and decay via the weak interaction into a charged
lepton ($e$, $\mu$ or $\tau$) and a neutrino. 
The charged leptons are produced with an angular distribution determined
by the ${\cal{O}}({{\alpha}_s^2})$ calculation of~\cite{MIRKES_HO_ANGLE}
which, for $W^+$ bosons with a helicity of --1 with respect to the
proton direction, has the form :
\begin{eqnarray}
 \frac{d\sigma}{d\cos{\theta_{\rm CS}}} \propto 
    1 + a_1(p_T)\cos{\theta_{\rm CS}} + a_2(p_T)\cos^2{\theta_{\rm CS}}
   \label{EQUATION:ANG_DIST_W}
\end{eqnarray}
where $p_T$ is the transverse momentum of the $W$ and $\theta_{\rm CS}$
is the polar direction of the charged lepton with respect to the proton
direction in the Collins-Soper frame~\cite{COLLINS_SOPER_FRAME}. 
$a_1$ and $a_2$ are $p_T$ dependent parameters.
For $p_T$ = 0, $a_1 = 2$ and $a_2 = 1$ providing 
the angular distribution of a $W$ boson fully 
polarized along the proton direction.
For the \ptw\ values relevant to the $W$ mass analysis $(p_T^W < \sim
30)$, the change
in $W$ polarization as \ptw\ increases only causes a modest change in the
angular distribution of the decay leptons~\cite{MIRKES_HO_ANGLE}.
The uncertainty is negligible.

\subsection{QED Radiative Effects}

$W \gamma$ production and radiative $W$ decays  ($W\rightarrow\ell\nu\gamma$)
are simulated using the calculation by Berends and
Kleiss~\cite{wgamma,rgwgamma}. Most photons 
tend to be collinear with the lepton, often showering in the same calorimeter
towers as the lepton. For the electron channel, these photons are merged with
the electron cluster; for the muon channel, they reduce the muon momenta
by their energy. 
Radiative effects from collinear photons are thus expected
to be larger in the muon channel. Photons not collinear with the lepton are
included in  the calculation of ${\bf u}$ (see Figure~\ref{collide}), 
and have an effect that is  similar in both the electron and muon channels.

Shifts in the $W$ mass due to radiative
effects are estimated to be $(-65 \pm 20)$~MeV/c$^2$ 
and $(-168 \pm 10)$~MeV/c$^2$ for the
electron and muon channel, respectively.
Uncertainties of the radiative effects are estimated from
uncertainties in the theoretical calculation and in the calorimeter 
response to the photons. The Berends and Kleiss calculation~\cite{wgamma}
does not include all the radiative Feynman diagrams.  For example, 
it does not include initial state radiation ($t$- and
$u$-channel diagrams) and allows a maximum of one photon.
The effect arising from the missing diagrams is evaluated by
incorporating the PHOTOS package~\cite{photos} which allows
two photon emissions, and the calculation by U.~Baur
{\it et al.}~\cite{baur} who have recently developed a complete
$O(\alpha)$ Monte Carlo which incorporates the initial state QED 
radiation from the quark-lines and the interference between the initial
and final state radiation as well as including a correct treatment of 
the final state soft and virtual photonic corrections.
The effects on $M_W$ from the former case are less than 10~MeV/c$^2$
for the $\wenu$ channel and less than 5~MeV/c$^2$ for the
$\wmunu$ channel.
The effects on $M_W$ from the latter case are less than 20~MeV/c$^2$
for the $\wenu$ channel and  $\sim$10~MeV/c$^2$ for the
$\wmunu$ channel.
The uncertainty in the calorimeter response to the photons 
well-separated from the $W$ decay lepton, is evaluated by varying the photon
energy threshold,  the photon fiducial region, and the photon energy
resolution. The effect is 3~MeV/c$^2$ on the $W$ mass.

\subsection{Summary}

The uncertainty associated with PDFs is evaluated by varying the choice
of PDF sets.  It is estimated to be 15 MeV/c$^2$ 
which is common to the electron and muon analyses. 
The \ptw\ spectrum is derived from the \zee\ and \zmumu\ data and 
a theoretical prediction for the ratio of $Z$ and $W$ \ptm\ spectra 
differential in the rapidity of the vector boson.  The corresponding 
uncertainty in the $W$ mass is dominated by $Z$ statistics.
It is 15~MeV/c$^2$ for the $\wenu$ channel and 20~MeV/c$^2$ 
for the $\wmunu$ channel.
A common uncertainty of 3~MeV/c$^2$ comes from the theoretical 
prediction for the ratio.
The uncertainty in the $W$ mass due to QED radiative effects is 
estimated to be 20~MeV/c$^2$ to the $\wenu$ channel, and 
10~MeV/c$^2$ to the $\wmunu$ channel.

\section {Recoil Measurement and Model}
\label{recoil}

The transverse mass distribution used for the $W$ mass 
measurement is reconstructed using the $\bf E_T$ of the 
charged leptons (described in Section~\ref{mumeas} and \ref{emeas}) 
and the neutrinos.  The transverse energy of the 
neutrino is inferred from the charged lepton $\bf E_T$ and the recoil 
energy $\bf u$ (see Figure~\ref{collide}).
This section describes the reconstruction of $\bf u$, and 
an empirical model of the detector response to $\bf u$ which is 
implemented in the simulation. Since the $W$ and $Z$ share a common 
production mechanism and are close in mass, the recoil model is based 
mainly on $Z \rightarrow \ell^+\ell^-$ decays.

\subsection{Recoil Reconstruction}

The recoil vector $\bf u$ is calculated by summing over 
electromagnetic and hadronic calorimeter
towers within the detector range $|\eta| < 3.6$,
\begin{equation}
	{\bf u} = (u_x, u_y) 
		= \Sigma_{\rm towers} E \sin\theta (\cos\phi, \sin\phi).
\end{equation} 
Table~\ref{tableth} lists tower thresholds for online (Level-3)
reconstruction and this analysis. The thresholds for this analysis
correspond to 5 times the calorimeter noise level.

\begin{table}
\begin{center}
\begin{tabular}{|l|cc|}
Calorimeter & Online threshold (GeV) & Analysis threshold (GeV)\\ 
\hline
Central EM & 0.1 & 0.1 \\
Central Had. & 0.1 & 0.185 \\
Plug EM & 0.3 & 0.15 \\
Plug Had. & 0.5 & 0.445 \\
Forward EM & 0.5 & 0.2 \\
Forward Had. & 0.8 & 0.73 \\
\end{tabular}
\end{center}
\caption{Tower energy thresholds used to reconstruct $\bf u$ both 
in online and in this analysis.}
\label{tableth}
\end{table}

There are two contributions to the recoil vector $\bf u$.
The first contribution is the energy of the initial state 
gluons radiated from the quarks that produce the $W$ or $Z$ boson.
This energy balances the $p_T$ of the boson.  The second is the 
energy associated with multiple interactions and the remnants of
the protons and antiprotons that are involved in the $W$ or $Z$
production.  The latter energy is referred to as the underlying energy.
It is manifested in $\Sigma E_T$, where 
\begin{equation}
	\Sigma E_T
		= \Sigma_{\rm towers} E \sin\theta = \Sigma_{\rm towers}
		E_T.
\end{equation} 

The lepton energy should not be included in the $\bf u$ calculation,
and thus the towers containing energy deposited by the lepton are 
excluded in the sum.  This procedure removes two towers for muons, and 
two or three towers for electrons.
If the center of the electron shower is more than 10~cm away 
from the azimuthal center of the tower ($|x| > 10$~cm), 
there will be leakage in the
azimuthally adjacent towers which are also removed.
This procedure removes not only the lepton energy, but also
the underlying energy which needs to be added back to the sum.
The underlying energy
is estimated from the energy in calorimeter towers 
away from the lepton in the $W$ data. In the muon analysis, this 
energy is added back to the $\bf u$ calculation.  In the electron 
analysis, rather than correcting $\bf u$, the same amount of energy is 
removed from the Monte Carlo simulation.  

\subsection{Recoil Model}

For the purposes of modeling the response and resolution, 
it is natural to define $\bf u$ in terms of the components $u_1$ 
and $u_2$, anti-parallel and perpendicular to the boson direction, 
respectively.
The average value of $u_1$ is the average calorimeter response 
balancing the boson $p_T$, and the average value of $u_2$ is 
expected to be zero.
$u_1$ and $u_2$ are parameterized in the form
\begin{equation}
 \left( \begin{array}{c}
                        u_1 \\ u_2
                       \end{array} \right) =
            \left( \begin{array}{c}
                        f(p_T^{boson}) \\ 0
                       \end{array} \right) + 
            \left( \begin{array}{c}
                        G_1(\sigma_1) \\ G_2(\sigma_2)
                       \end{array} \right) 
\label{u_response}
\end{equation}
where $G_1(\sigma_1)$ and $G_2(\sigma_2)$ are Gaussian distributed
random variables of mean zero and widths $\sigma_1$ and $\sigma_2$, 
and the quadratic function $f(p_T^{boson})$ is the 
response function to the recoil energy.  
A detailed description can be found in Reference~\cite{ag_thesis}.

The resolutions $\sigma_1$ and $\sigma_2$ are expected to be dependent 
on $\sumet$.  For the minimum bias events which represent the underlying 
event in the $W$ and $Z$ sample, 
the resolutions $\langle \sigma_x \rangle$ and $\langle \sigma_y
\rangle$ are well parameterized with $\Sigma E_T$. 
A fit to the data, as shown in Figure~\ref{f_minbias_sumet}, gives 
\begin{equation}
	\sigma_{mbs}(\sumet) = 0.324 \times (\sumet)^{0.577}
\label{sumet_mbs}
\end{equation}
where $\sigma_{mbs}(\sumet)$ and \sumet\ are calculated in GeV.
For the $W$ and $Z$ events, a good description of the resolution 
requires additional parameters which accont for its boson $p_T$
dependence; the initial state gluons balancing the boson $p_T$ produce
jets which contribute to the resolution differently than the underlying
energy.
In order to allow this resolution difference, the widths are 
parameterized in the form
\begin{equation}
            \left( \begin{array}{c}
                        \sigma_1 \\ \sigma_2
                       \end{array} \right) =
		\sigma_{mbs}(\sumet) \times 
            \left( \begin{array}{c}
                        1 + s_1 \cdot (p_T^{boson})^2 \\ 
		        1 + s_2 \cdot (p_T^{boson})^2 
                       \end{array} \right)
\label{width_e}
\end{equation}
for the electron channel and 
\begin{equation}
            \left( \begin{array}{c}
                        \sigma_1 \\ \sigma_2
                       \end{array} \right) =
		\sigma_{mbs}(\sumet) \times 
            \left( \begin{array}{c}
                        \alpha_1 + \beta_1 \cdot p_T^{boson} \\ 
		        \alpha_2 + \beta_2 \cdot p_T^{boson} 
                       \end{array} \right)
\label{width_mu}
\end{equation}
for the muon channel, 
where $s_1$, $s_2$, $\alpha_1$, $\alpha_2$, $\beta_1$, and $\beta_2$
are free parameters. Although the two channels use different formulae, 
the fitted funtions are consistent with each other -- $\alpha_1$
and $\alpha_2$ are close to 1 and the difference between 
the linear term and the quadratic term is within the statistical 
uncertainty of the $Z$ sample.
The argument $\Sigma E_T$ in Eq.s~\ref{width_e} and \ref{width_mu} 
comes from the $\Sigma E_T$ distributions of the $W$ and $Z$ data.
The $\Sigma E_T$ distributions in various $p_T^Z$ bins are shown in 
Figure~\ref{f_sumet_pt}.  They are nicely fit to $\Gamma$-distributions
\begin{equation}
	\gamma(\sumet; a,b) = {a^b (\sumet)^{b-1} e^{-a (\sumet)} 
	\over \Gamma(b)}
\label{sumet_wz}
\end{equation}
where $a$ and $b$ are fit parameters, and $b$ is a linear function of
$p_T^{boson}$.  The term $a/\Gamma(b)$ normalizes the distribution.
Figure~\ref{f_sumet} shows the \sumet\ distributions 
and fits for the $Z$ and $W$ events.

\begin{figure}[thb]
        \centerline{\epsfysize 18cm
                    \epsffile{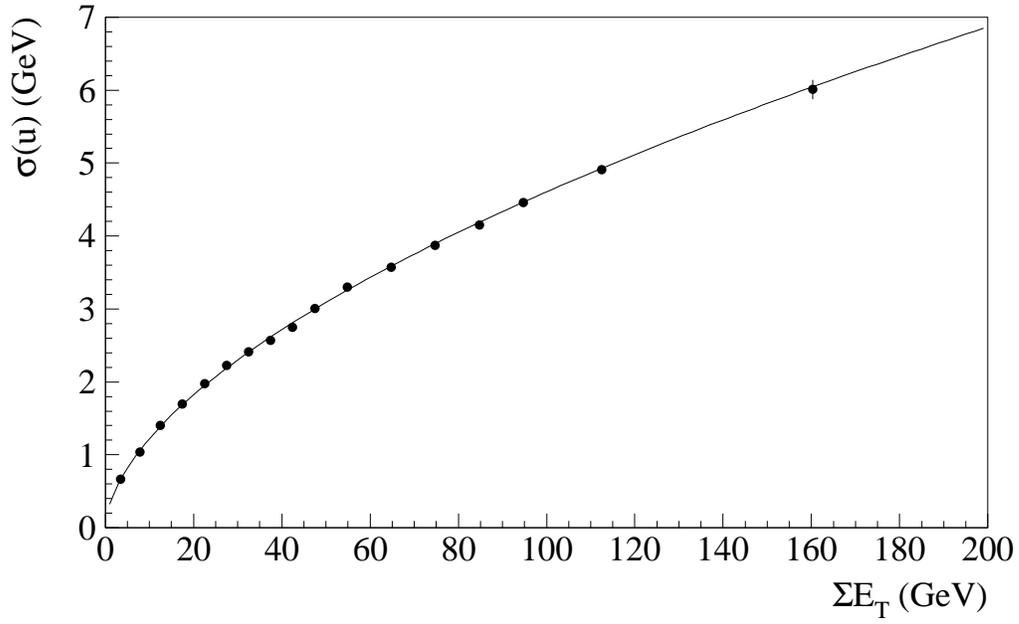}}
\vspace{-9.0cm}
\caption{The fit for the rms of the $u_x$ and $u_y$ distributions
	as a function of $\Sigma E_T$ using the minimum bias sample.}
\label{f_minbias_sumet}
\end{figure}

\begin{figure}[tbh]
	\epsfysize=6.0in
	\epsffile[54 162 531 675]{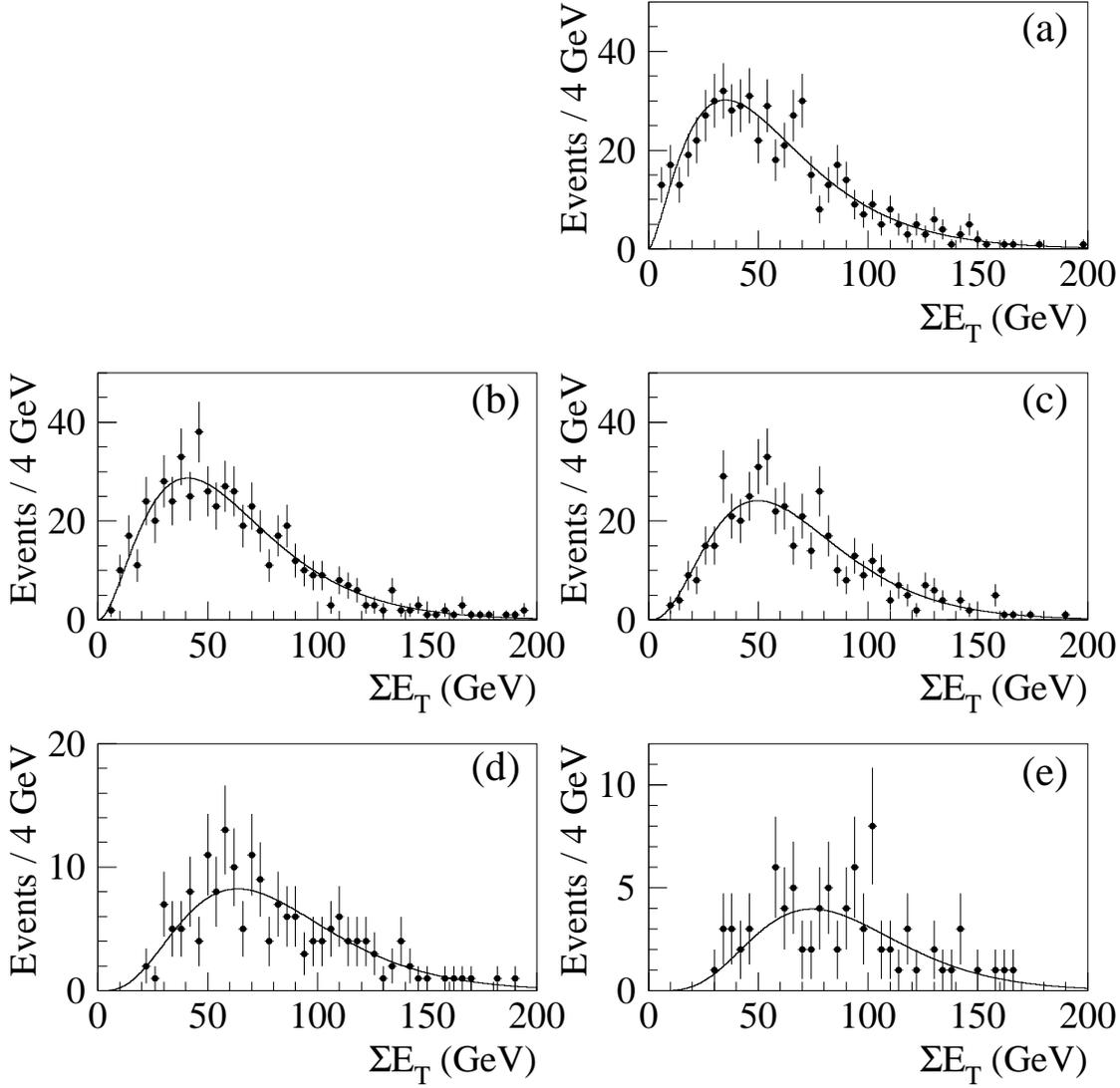}
\vspace{1.5cm}
\caption{The \sumet\ distributions in 5 different $p_T^Z$ bins
	for the $\zmumu$ data are shown:
	(a) for $p_T^Z < 5$~GeV,
	(b) for $ 5 < p_T^Z < 10$~GeV,
	(c) for $10 < p_T^Z < 20$~GeV,
	(d) for $20 < p_T^Z < 30$~GeV, and
	(e) for $30 < p_T^Z < 50$~GeV.}
\label{f_sumet_pt}
\end{figure}

\begin{figure}[tbh]
        \centerline{\epsfysize 19cm
                    \epsffile{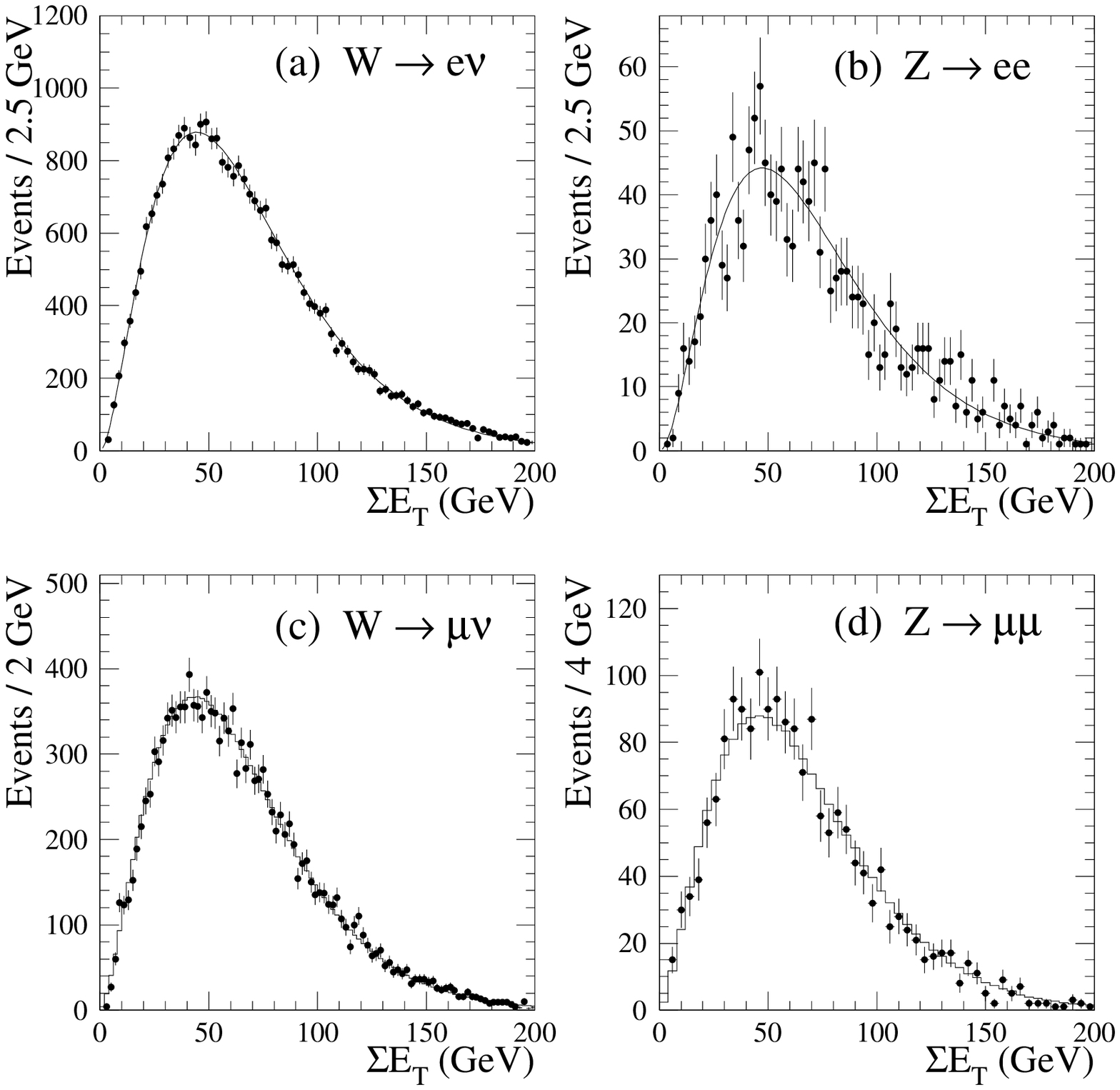}}
\vspace*{-2cm}
\caption{The \sumet\ distributions for 
	(a) the $\wenu$ sample, (b) the $\zee$ sample,
	(c) the $\wmunu$ sample, and (d) the $\zmumu$ sample.
	The solid lines are fits to the functions described
	in Eq.~\ref{sumet_wz}.}
\label{f_sumet}
\end{figure}

\begin{figure}[tbh]
        \centerline{\epsfysize 19cm
                    \epsffile{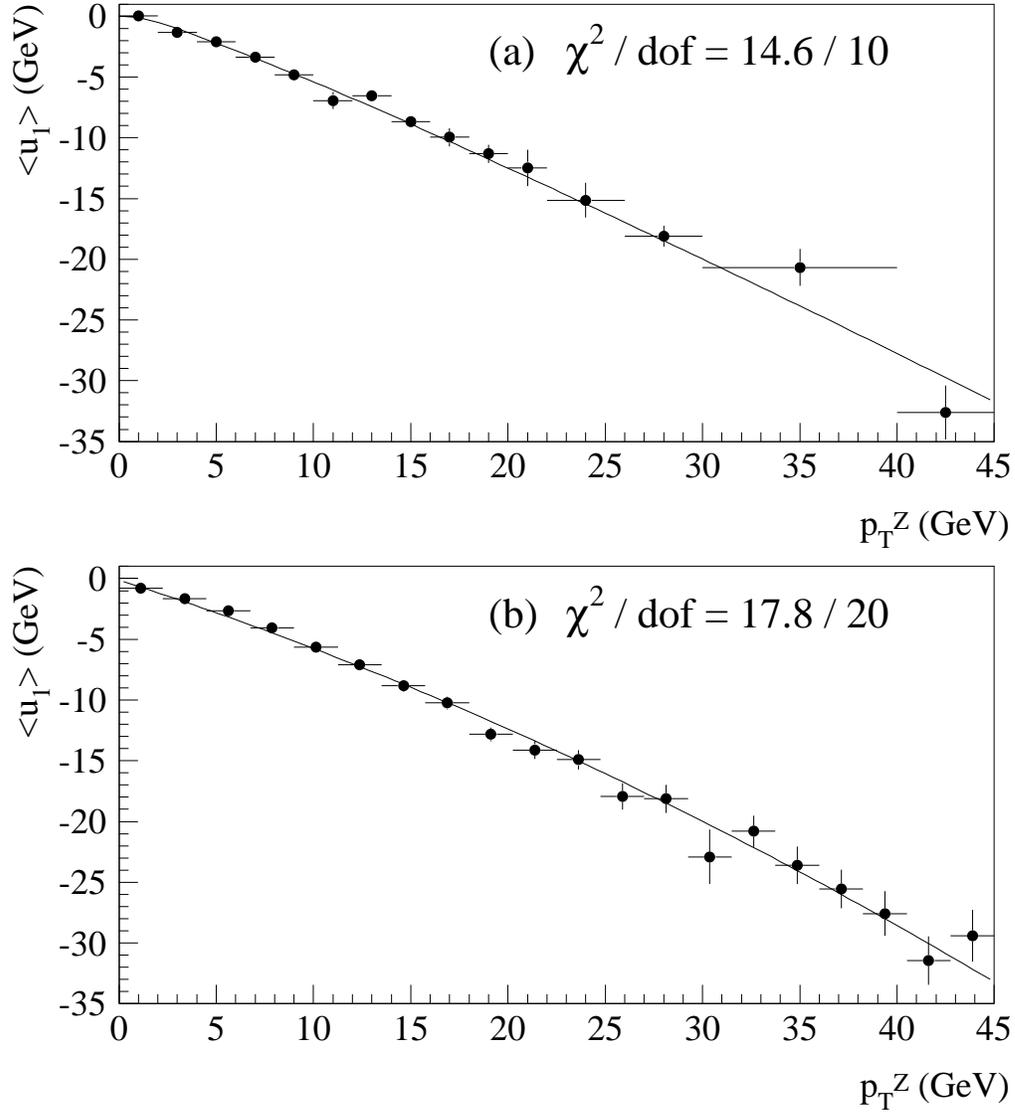}}
\vspace*{-2cm}
\caption{The $\langle u_1 \rangle$ versus $p_T^Z$ (solid lines) 
	as derived from $Z$ sample
        fits for (a) the electron channel and (b) the muon channel.
        The fits are compared with the data points.}
\label{f_u1_vs_zpt}
\end{figure}

\begin{figure}[thb]
	\epsfysize=6.0in
	\epsffile[54 162 531 675]{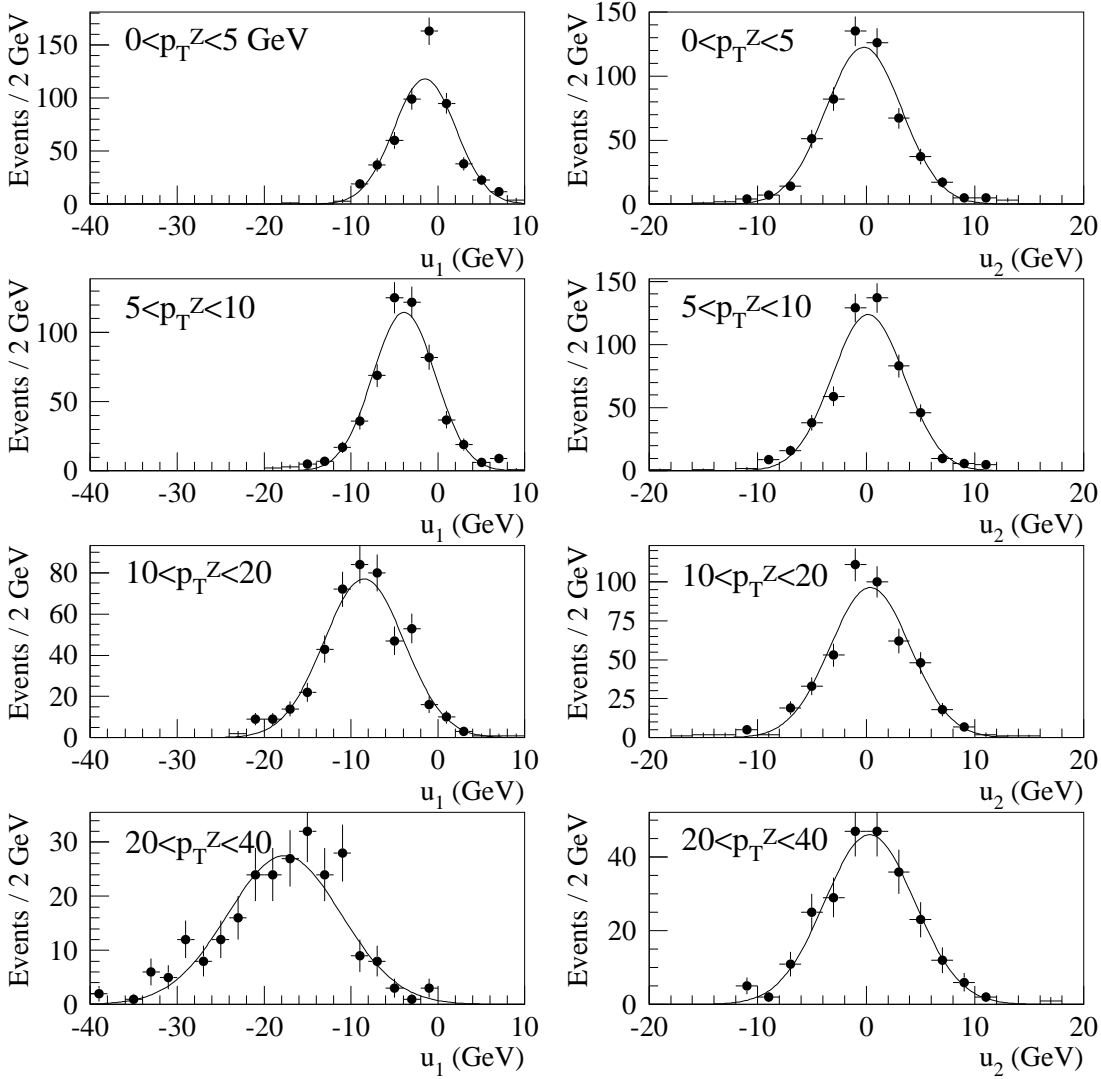}
\vspace{1.5cm}
\caption{The fit of the $u_1$ and $u_2$ distributions in ranges of
	$Z$ \ptm\ in the muon sample, 
	illustrating the adequacy of assuming Gaussian 
	resolution(solid lines).}
\label{FIGURE:UGAUS}
\end{figure}

The $Z$ data provide  $u_1$, $u_2$, $\sumet$, and the $p_T$ of the $Z$.
The parameters in 
Eqs.~\ref{u_response},~\ref{sumet_wz},~\ref{width_e}, and \ref{width_mu} 
are derived by fitting to these variables.
Figure~\ref{f_u1_vs_zpt} compares $\langle u_1 \rangle$ as a function 
of $p_T^Z$ from the $Z$ data with the fit functions 
$f(p_T^Z)$ described in Eq.~\ref{u_response}.
The validity of a Gaussian parameterization in Eq.~\ref{u_response}
is illustrated in Figure~\ref{FIGURE:UGAUS}. 
The parameterization of the recoil response model is further 
cross-checked by distributions of $u_1$, $u_2$, and $|\bf u|$.
As shown in Figure~\ref{f_model_uave}, they all agree well.
The $u$ resolutions in the $\zmumu$ data are shown as a function of
$p_T^Z$ in Figure~\ref{f_model_sigma}, where the data is
compared with the recoil model with (the solid histograms) and
without (the dashed histograms) including the effect of gluons 
against the $W$.
As expected, the resolution gets worse in $u_1$ as the jet
structure of the recoil becomes apparent, increasing $\Sigma E_T$ in the
$u_1$ direction.

While the $Z$ sample, where the boson \ptm\ is well understood, allows 
the unfolding of response and resolution, the $W$ samples do not allow 
these effects to be separately understood.  However, the $W$ samples 
can be used to optimize the model parameters for the $W$ data
while preserving a good description of the $Z$ data.
This is demonstrated in Figure~\ref{FIGURE:CON_Z_W}.
The ultimate recoil model includes the $\bf |u|$ and \uperp\ (the
component of $\bf u$ perpendicular to the lepton direction)
distributions from the $W$ data in the fit.

\begin{figure}[tbh]
        \centerline{\epsfysize 18cm
                    \epsffile{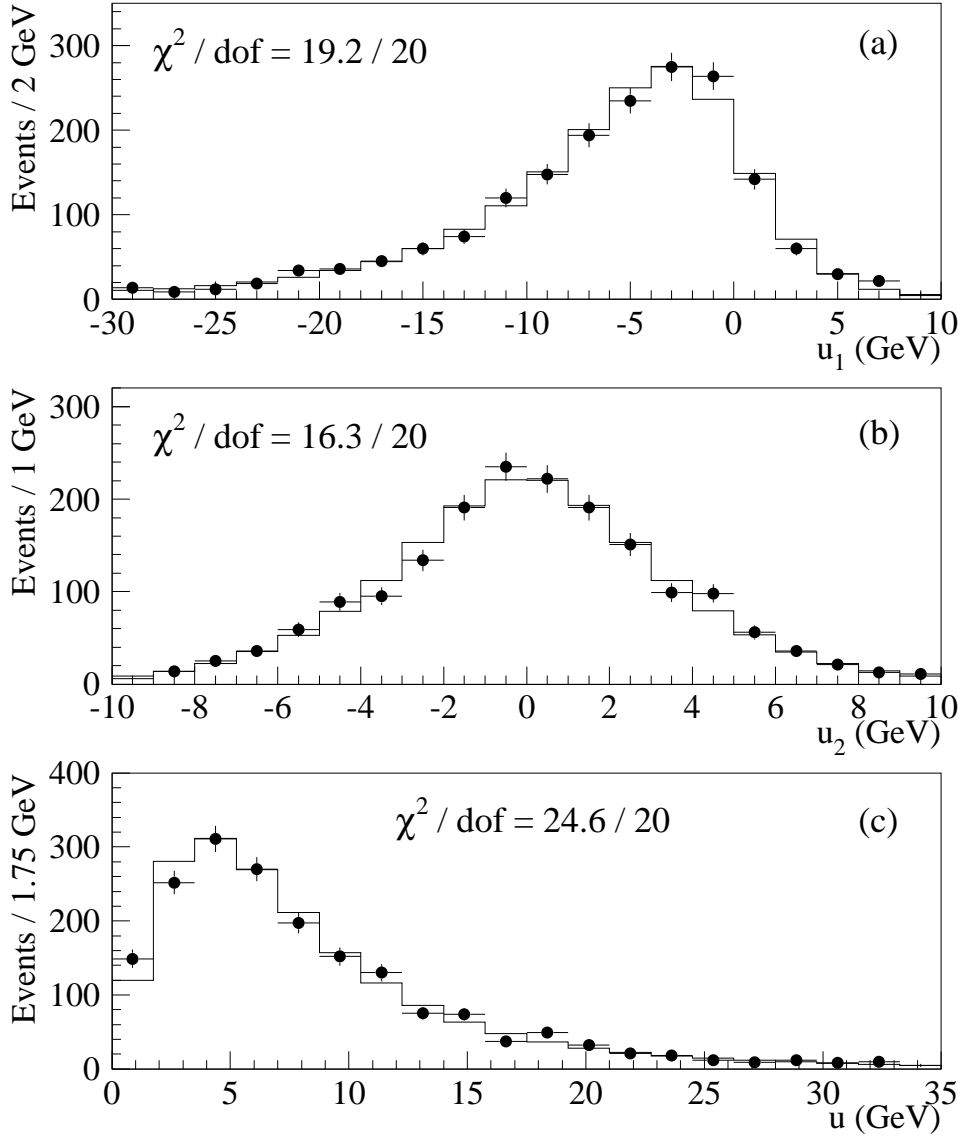}}
\caption{(a) $u_1$, (b) $u_2$, and (c) $|\bf u|$ distributions for
        the $\zmumu$ data.  The histograms are the simulation using the
        recoil model parameters.}
\label{f_model_uave}
\end{figure}

\begin{figure}[tbh]
\vspace*{-1.5cm}
        \centerline{\epsfysize 20cm
                    \epsffile{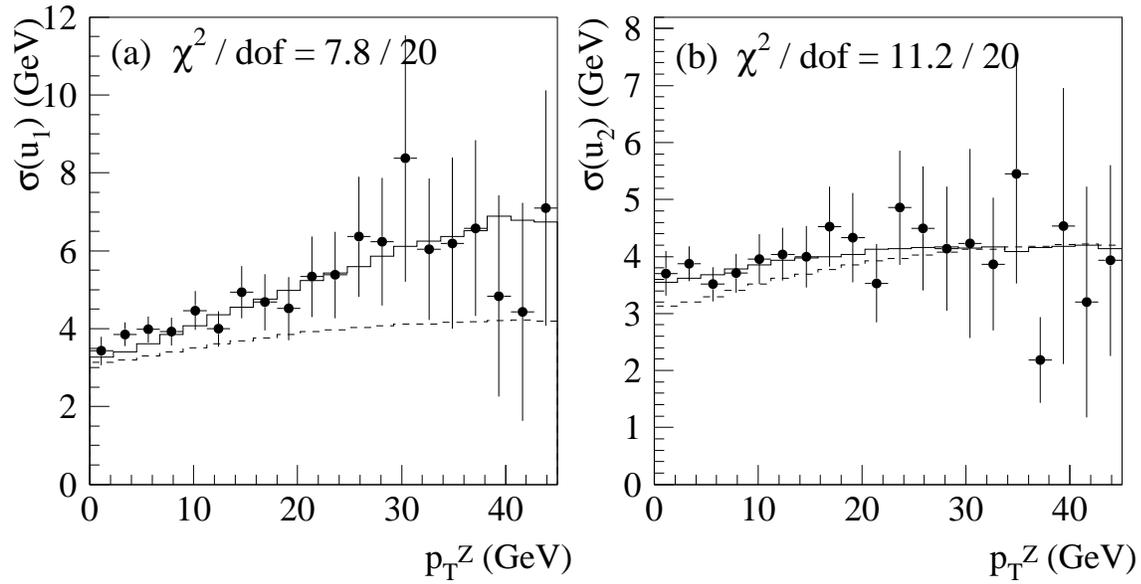}}
\vspace{-10.0cm}
\caption{$\sigma(u_1)$ and $\sigma(u_2)$ as a function of
	$p_T^Z$ for the $\zmumu$ sample.
        The points are the data, and the solid histograms are the
        simulation using the recoil model parameters.
	The dashed histograms show $\sigma_{mbs}(\Sigma E_T)$,
	the resolutions of the underlying energy.}
\label{f_model_sigma}
\end{figure}

\begin{figure}[tbh]
	\epsfysize=6.0in
	\epsffile[54 162 531 675]{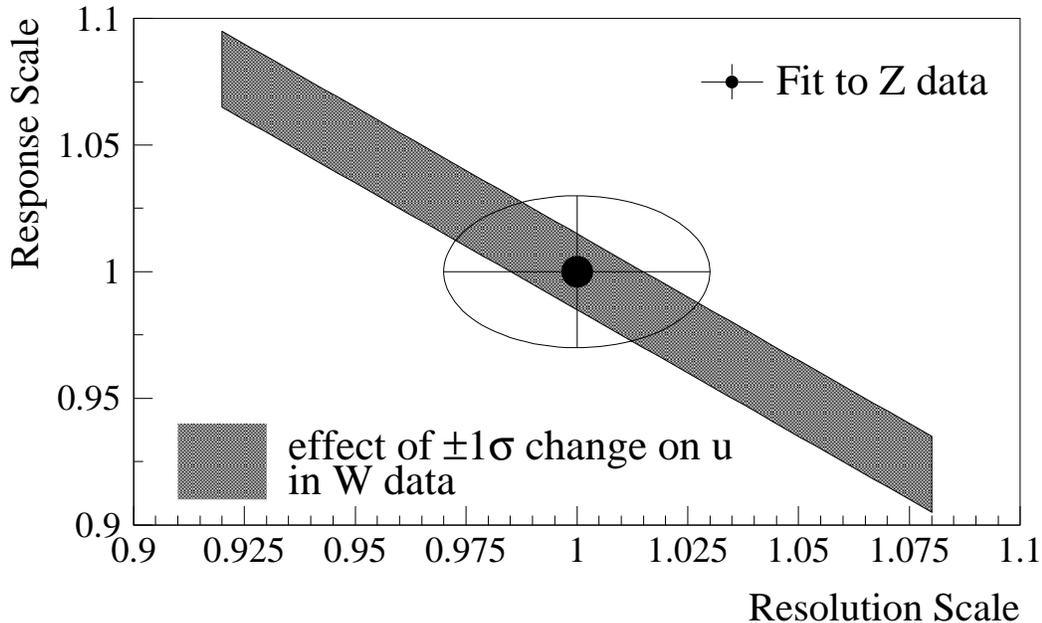}
\vspace{-6.5cm}
\caption{The muon $Z$ fits separately constrain resolution and
	response, as shown by the ellipse, while the $W$ data 
	gives a further correlated constraint, as shown by the band.
	This is obtained from the Monte Carlo studies.}
\label{FIGURE:CON_Z_W}
\end{figure}

\subsection{Comparison of Data and Simulation in the $W$ Samples}

This section compares the data with the simulation which uses
the best fit parameters of the modeling.
The $W$ data is more naturally described in terms of components 
\upara\ and $\uperp$ of recoil defined with respect to the charged 
lepton direction -- the component along the lepton direction and 
the component perpendicular to the lepton direction, respectively
(see Figure~\ref{upicture}).\footnote{When
$| {\bf u} | << E_T^\ell$, the transverse mass becomes 
	$M_T^W \approx 2 E_T^\ell + \upara$.} 
The $|\bf u|$ and \upara\ distributions and residuals are shown in
Figure~\ref{FIGURE:U_UPARA_E} and Figure~\ref{FIGURE:U_UPARA_MU}.  
The \uperp\ distribution is shown in Figure~\ref{FIGURE:UPERP_MU}. 
The means for \uperp\ are consistent with
zero and the other $\bf u$ projection numbers are listed in
Table~\ref{TABLE:UNUMB}.
The models reproduce the basic characteristics well.

\begin{figure}[p]
\vspace*{12cm}
\epsfysize=3.0in
\epsffile[125 146 500 345]{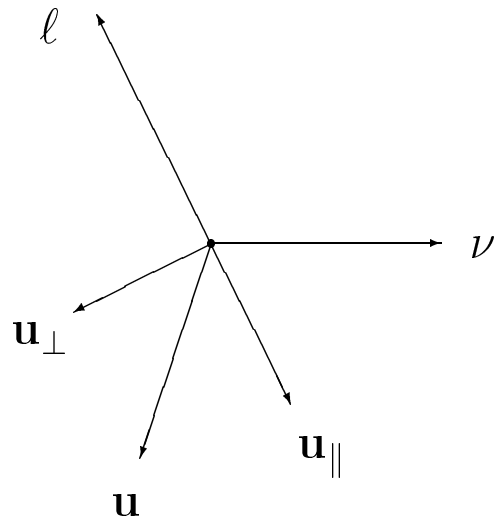}
\vspace*{-8cm}
\caption{Kinematics of leptons from the $W$ decay and the transverse
	energy vector recoiling against the $W$, 
	as viewed in the plane transverse to the
        antiproton-proton beams. 
	$\upara$ is the component of ${\bf u}$ along the lepton 
	direction and $\uperp$ the component of  ${\bf u}$ 
	perpendicular to the lepton direction.}
\label{upicture}
\end{figure}

\begin{figure}[tbh]
\epsfysize=6.0in
\epsffile[54 162 531 675]{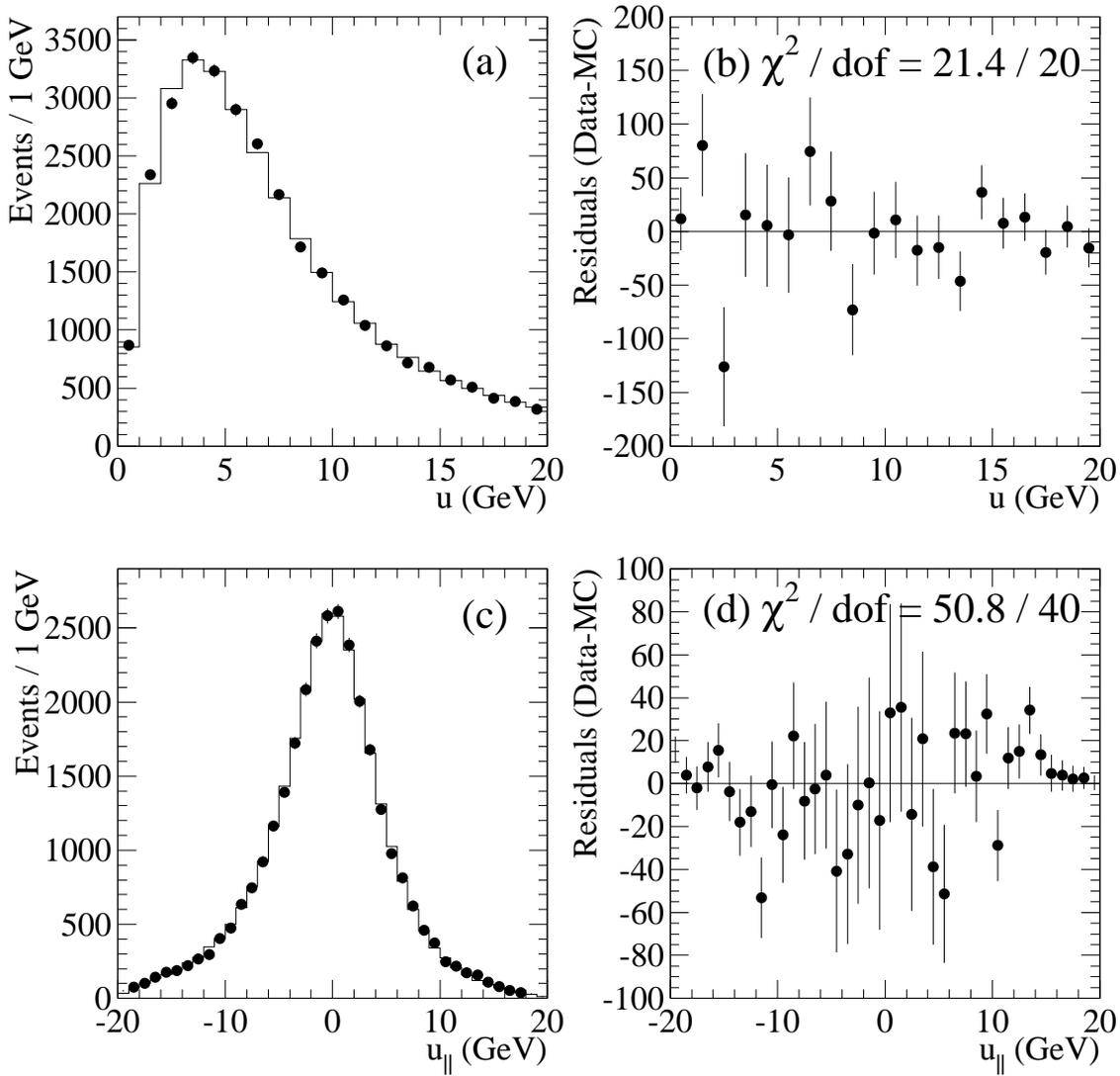}
\vspace*{0.5cm}
\caption{The (a) $|\bf u|$ and (c) \upara\ distribution 
	distribution for the $\wenu$ sample.
	The points (histograms) are the data (simulation). 
	The differences between the data and the simulation
	are shown in (b) and (d).}
\label{FIGURE:U_UPARA_E}
\end{figure}

\begin{figure}[tbh]
\epsfysize=6.0in
\epsffile[54 162 531 675]{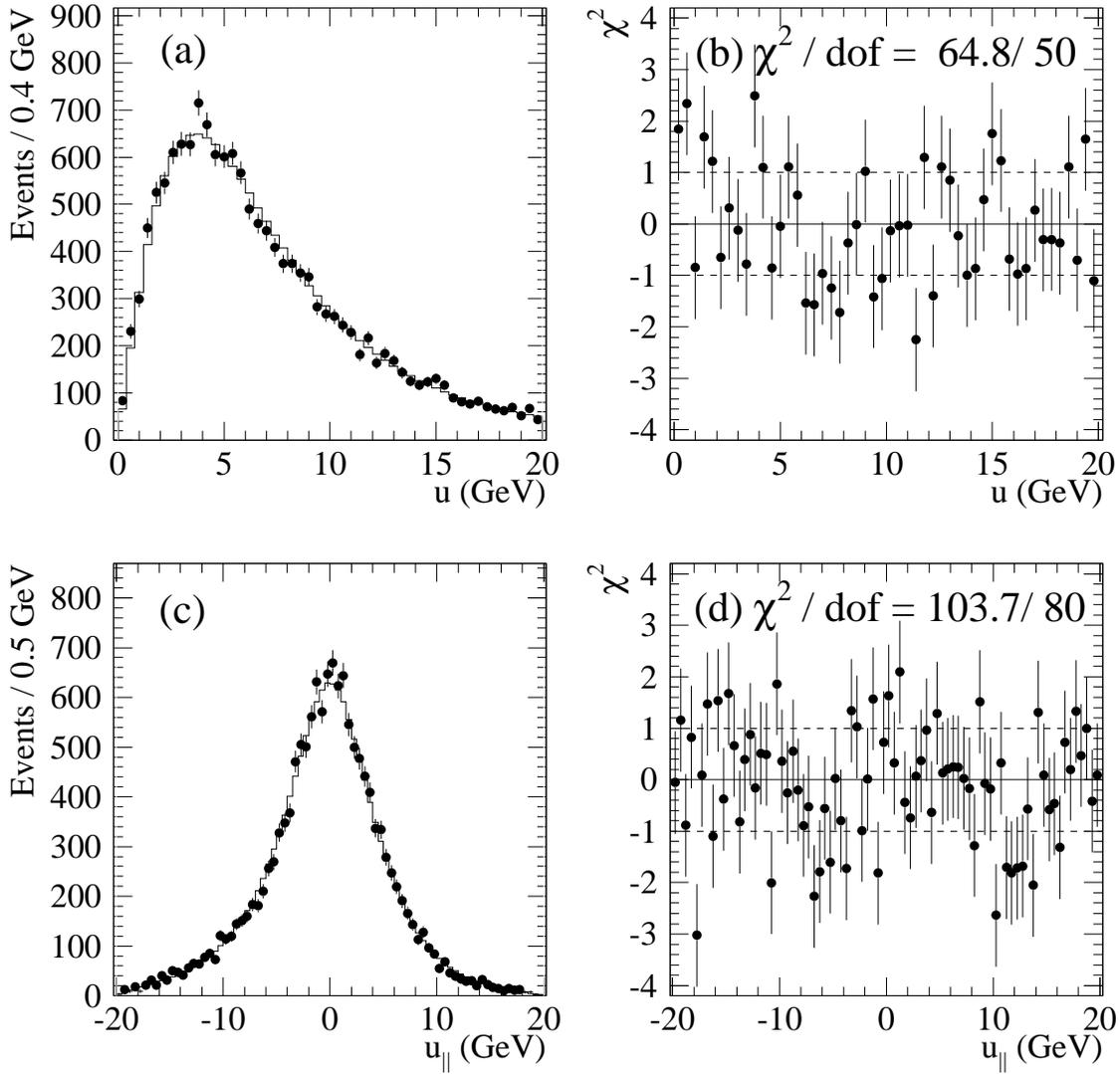}
\vspace*{0.5cm}
\caption{The (a) $|\bf u|$ and (c) \upara\ distribution 
	for the $\wmunu$ sample.
	The points (histograms) are the data (simulation). 
	The differences between the data and the simulation
	normalized by the statistical uncertainty are shown 
	in (b) and (d).}
\vspace*{0.5cm}
\label{FIGURE:U_UPARA_MU}
\end{figure}

\begin{figure}[tbh]
\epsfysize=6.0in
\epsffile[54 162 531 675]{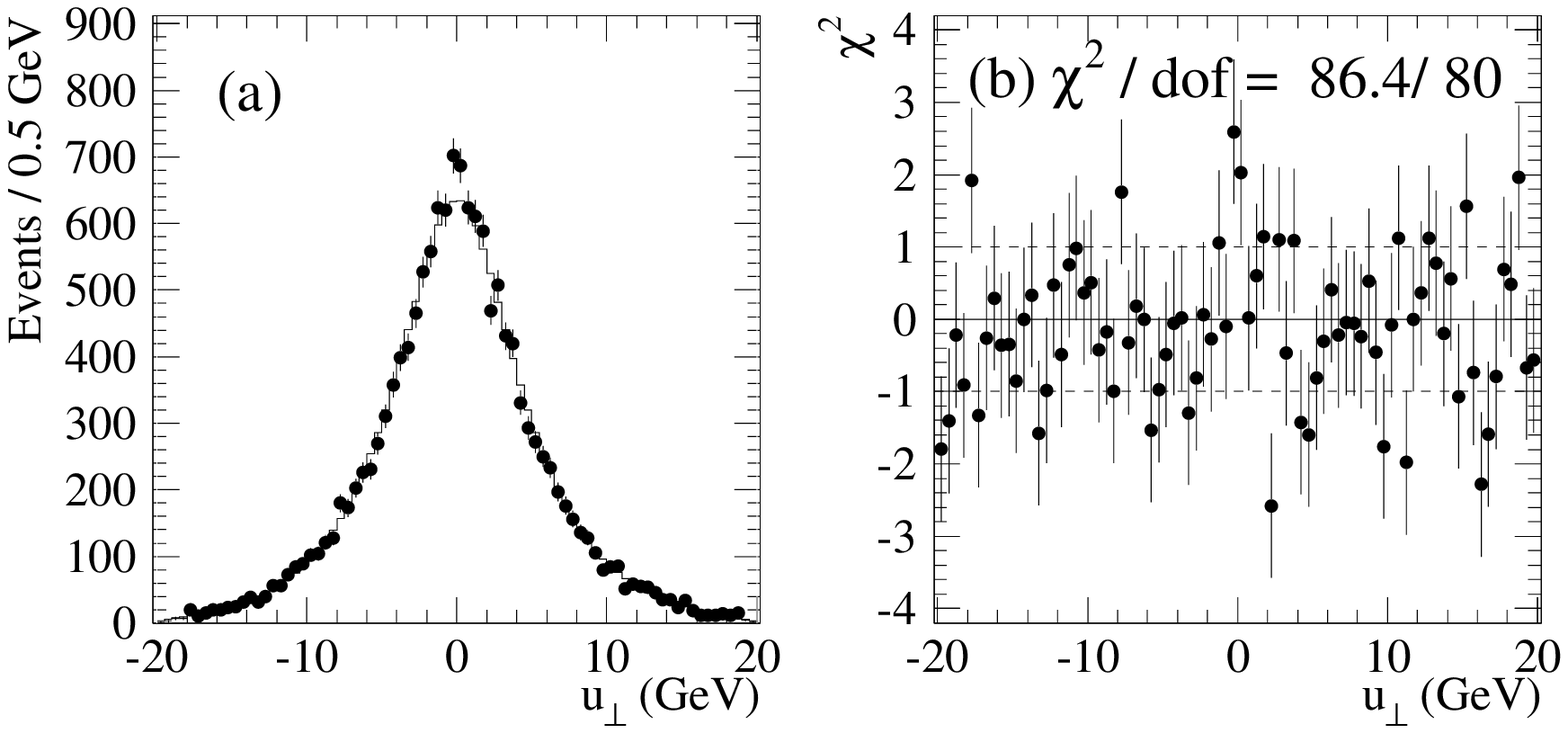}
\vspace*{-6cm}
\caption{(a) The \uperp\ distribution for the $\wmunu$ sample.
	The points (histogram) are the data (simulation).
	(b) The difference between the data and the simulation
	normalized by the statistical uncertainty.}
\label{FIGURE:UPERP_MU}
\end{figure}

\begin{table}
\begin{center}
\begin{tabular}{|c|c|r|r|}
Quantity          & Mode &   Data~~~~~~~~~ & Simulation~    \\ \hline
$\sigma^{\rm rms}(\uperp)$ &e$\nu$   &  $5.684\pm 0.034$ GeV& $ 5.765$ GeV \\
$\sigma^{\rm rms}(\uperp)$ &$\mu\nu$ &  $5.640\pm 0.065$ GeV& $ 5.672$ GeV \\
\hline
$\sigma^{\rm rms}(\upara)$ &e$\nu$   &  $5.877\pm 0.024$ GeV& $ 5.827$ GeV \\
$\sigma^{\rm rms}(\upara)$ &$\mu\nu$ &  $5.732\pm 0.069$ GeV& $ 5.750$ GeV \\
\hline
$\langle \upara \rangle$   &e$\nu$   & $-0.573\pm 0.034$ GeV& $-0.639$ GeV \\
$\langle \upara \rangle$   &$\mu\nu$ & $-0.436\pm 0.048$ GeV& $-0.422$ GeV \\
\end{tabular}
\end{center}
\caption{Widths and means for recoil response projections for data 
	and simulation. The simulation includes the $W$ constraint
	and background bias.  Uncertainties shown here are
	only statistical, and do not include systematic uncertainties
	due to $p_T^W$ and the recoil model.}
\label{TABLE:UNUMB}
\end{table}

One can further examine whether or not the model describes correlations
among variables. The distributions in \upara\ are examined
in four bins of $|{\bf u}|$, shown for the electron analysis in
Figure~\ref{FIGURE:UPAR_U_E} and
for the muon analysis in Figure~\ref{FIGURE:UPAR_U_MU}.
The correlation of \upara\ and transverse mass is illustrated in
Figure~\ref{FIGURE:UPAR_MT_E}
and the trend of $\langle \upara \rangle$ with azimuthal angle between 
the lepton and $\bf u$ is shown in Figure~\ref{FIGURE:UPAR_DPHI_MU}.
As indicated in these figures, the simulation well represents the data.

\begin{figure}[tbh]
\epsfysize=6.0in
\epsffile[54 162 531 675]{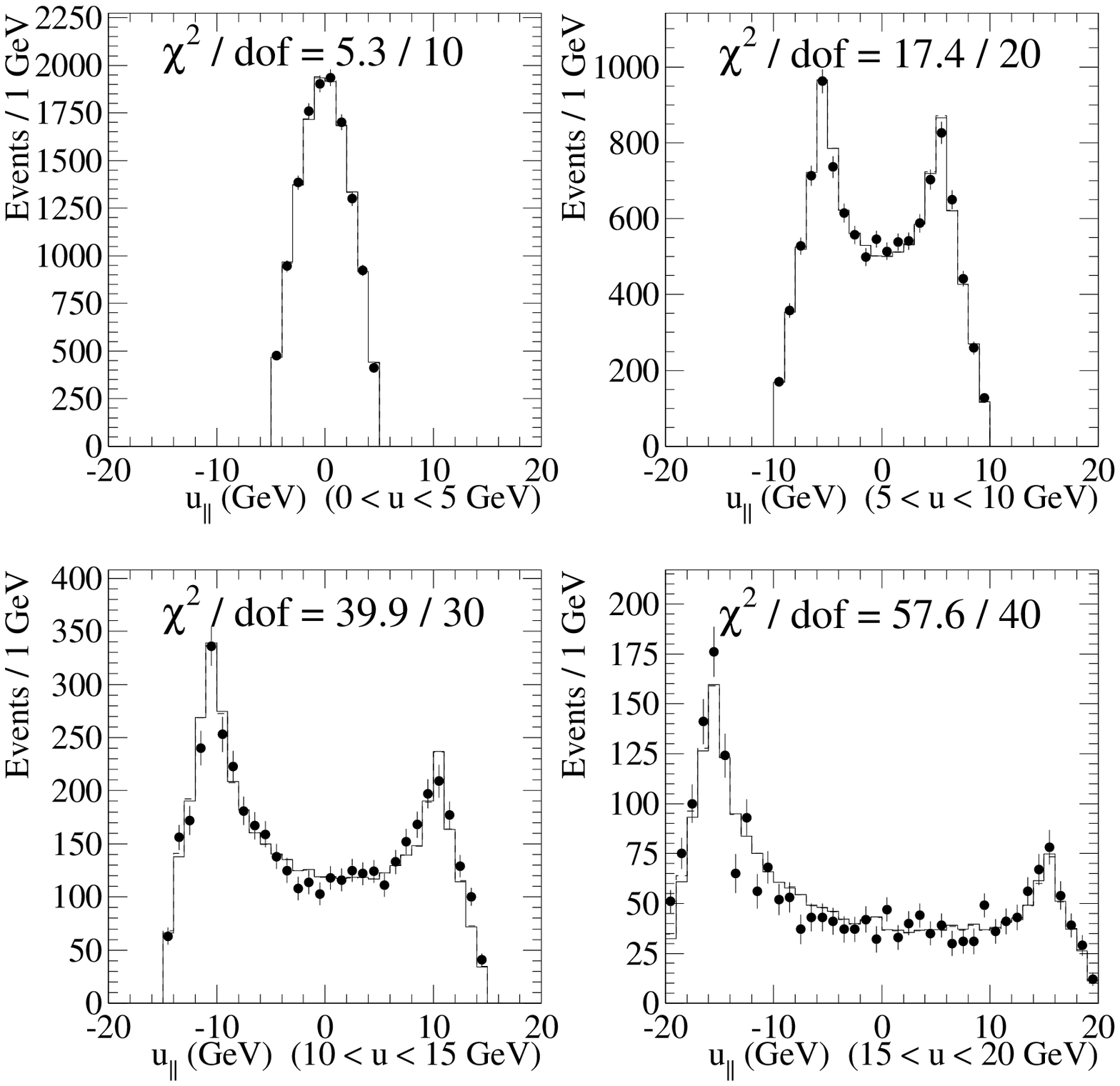}
\caption{The $\upara$ distributions for the $\wenu$ 
	sample in four bins of $|{\bf u}|$. The points are the data, and
	the histograms the simulation.}
\label{FIGURE:UPAR_U_E}
\end{figure}

\begin{figure}[tbh]
\epsfysize=6.0in
\epsffile[54 162 531 675]{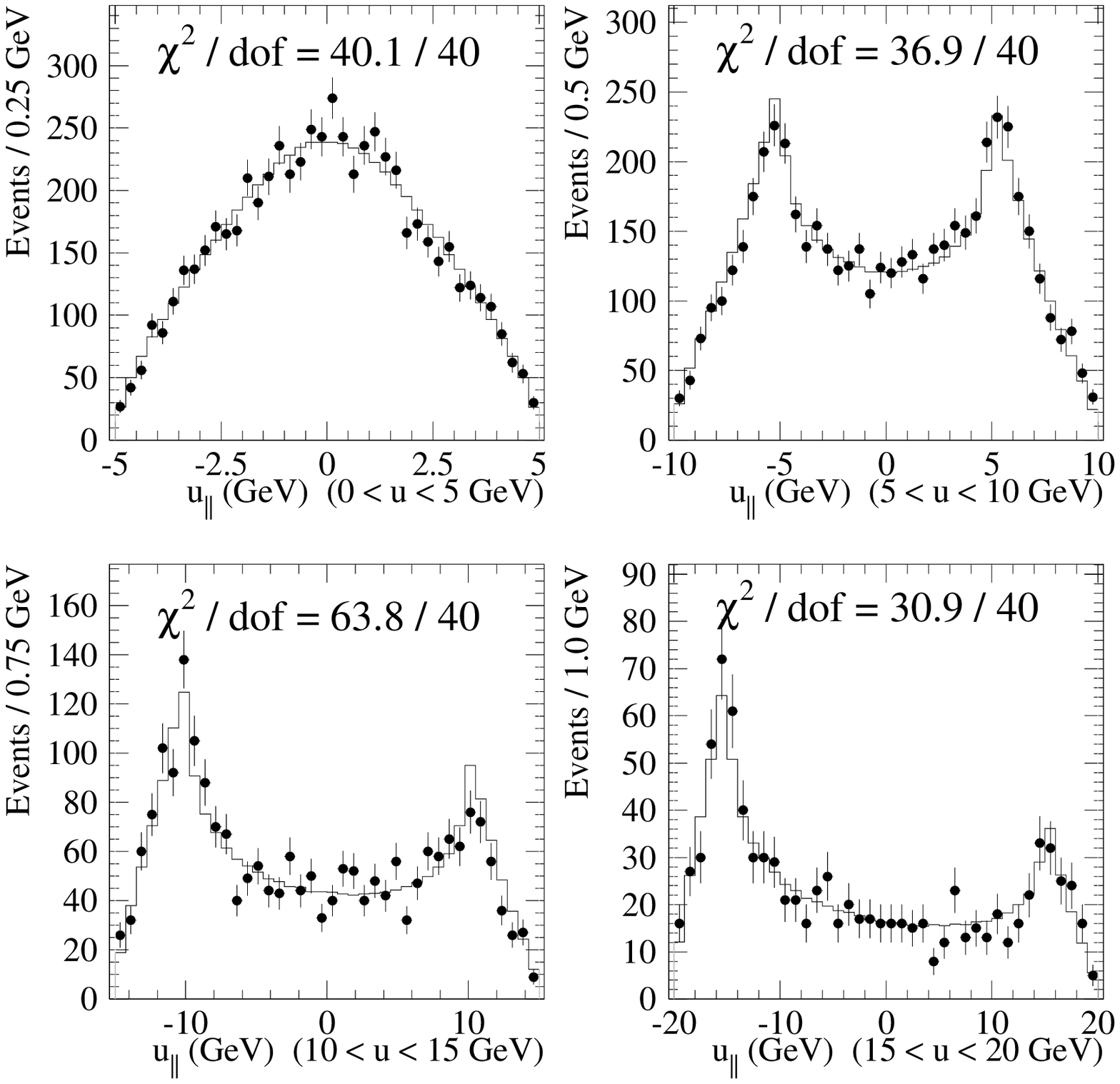}
\caption{The $\upara$ distributions for the $\wmunu$ 
	sample in four bins of $|{\bf u}|$. The points are the data, and
	the histograms the simulation.}
\label{FIGURE:UPAR_U_MU}
\end{figure}

\begin{figure}[tbh]
\epsfysize=6.0in
\epsffile[54 162 531 675]{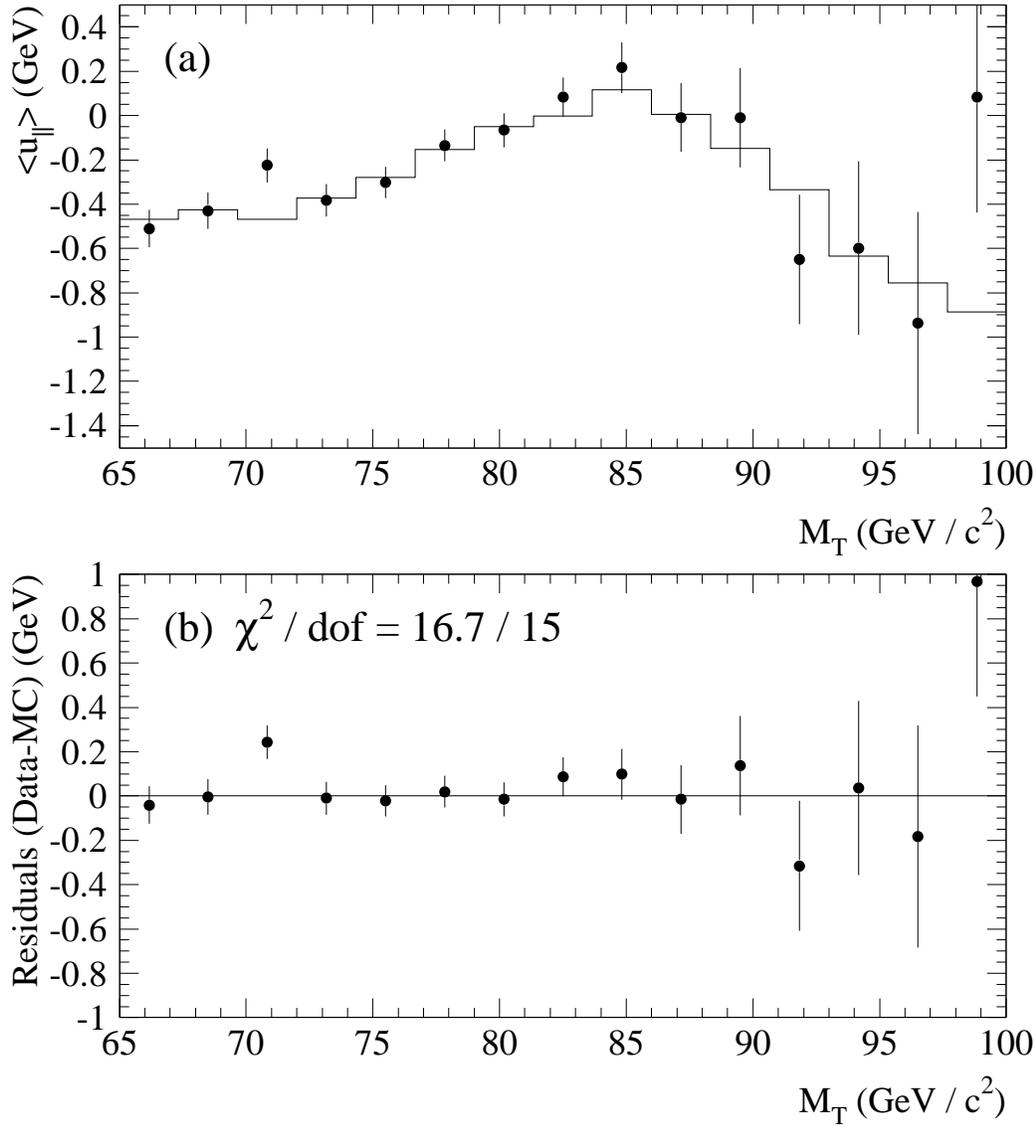}
\vspace*{1.0cm}
\caption{(a) The average value of ${\rm u_{\parallel}}$
        as a function of $M_T$ for the $\wenu$ sample.
        The points are the data, and the solid histogram is
        for the simulation.
        (b) Residuals between the data and the simulation.}
\label{FIGURE:UPAR_MT_E}
\end{figure}

\begin{figure}[tbh]
\vspace*{-1.5cm}
        \centerline{\epsfysize 18cm
                    \epsffile{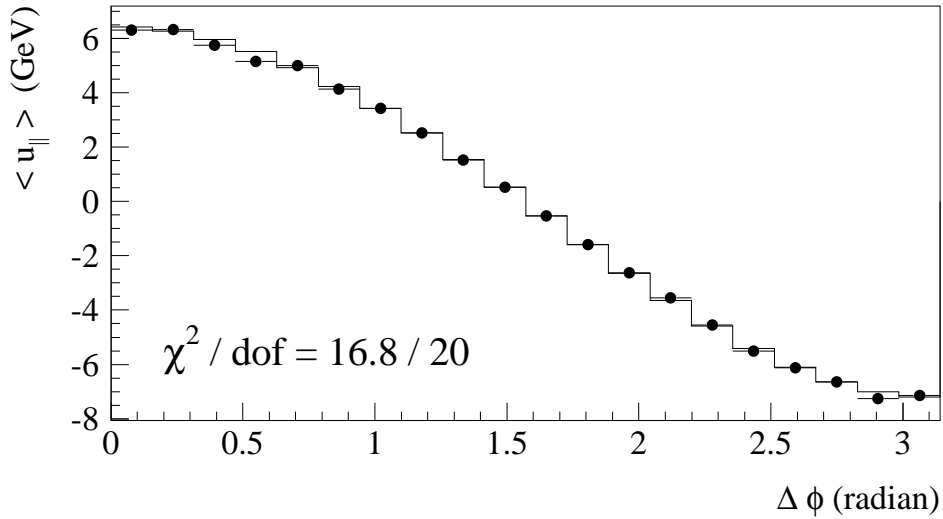}}
\vspace{-9cm}
\caption{$\langle \upara \rangle$ as a function of azimuthal angle
        between the lepton and $\bf u$ for the $\wmunu$ sample.
        The points are the data and the histogram is the simulation.}
\label{FIGURE:UPAR_DPHI_MU}
\end{figure}

\subsection{Uncertanties on $M_W$}

The uncertainty on the $W$ mass is evaluated by varying 
the model parameters within their uncertainties.
The size of the parameter uncertainties is taken from the 
$Z$ statistics and does not include the reduction produced by
including the $W$ data in the model.
For each set of model parameters a set of transverse mass templates
are produced which are fit to the transverse mass distributions of
the data and a standard Monte Carlo template.
The rms of $M_W$ values obtained from the fit to the Monte Carlo
template is 37~MeV/c$^2$ for the electron channel and
35~MeV/c$^2$ for the muon channel.

\subsection{Summary}

The detector response to the recoil energy against the $W$ is 
modeled primarily using the $Z \rightarrow \ell^+\ell^-$ data.
The $W$ data are used to optimize the model.
The model is empirical in the sense that its form is justified by the
data and its parameters determined from the data.
The modeling procedure is applied separately to the muon and
electron samples, so the uncertainties on the $W$ mass 
due to the recoil model are essentially independent.
The parametrizations are compatible in the two channels.

The uncertainty on the $W$ mass is evaluated by producing
a set of transverse mass templates with the model parameters 
allowed within their uncertainties, and 
fitting to the transverse mass distributions of
the data and a standard Monte Carlo template.
It is 37~MeV/c$^2$ for the electron channel and
35~MeV/c$^2$ for the muon channel.

\section {Results and Conclusions}
\label{results}

This section summarizes the $W$ mass results. Cross-checks which
support the results are discussed.  The 
results of the two lepton channels are combined with previous CDF
measurements. The combined result is compared with other 
measurements and with global fits to all precise electroweak
measurements which predict a $W$ mass as a function of 
the Higgs boson mass.

\subsection{Fitting Procedure}
\label{fitting}

The $W$ mass is obtained from a binned maximum likelihood fit to the 
transverse mass spectrum. This spectrum cannot be predicted analytically
and must be simulated using a Monte Carlo program which produces 
the shape of the transverse mass distribution as a function of $M_W$. 
This program incorporates 
all the experimental effects relevant to the analysis, including 
$W$ production and decay mechanisms as described in Section~\ref{wprod}, 
the detector acceptance for the charged leptons from the $W$ decay,
the detector responses and resolutions of the leptons 
as described in Sections~\ref{mumeas} and \ref{emeas}, 
and the detector response and resolution of the recoil energy 
against the $W$ as described in Section~\ref{recoil}.
The Monte Carlo program generates $M_T$ distributions used
as templates for discrete values
of $M_W$. The width of the $W$, $\Gamma_W$, is taken as the Standard Model 
value~\cite{rosner} for that $W$ mass.\footnote{$\Gamma_W$ is precisely
predicted in terms of the masses and coupling strengths of the gauge
bosons. The leptonic partial width $\Gamma(W\rightarrow \ell\nu)$ can be
expressed as $G_F M_W^3 / 6\sqrt{2\pi}(1 + \delta_{\rm SM})$ where
$\delta_{\rm SM}$ is the radiative correction to the Born-level
calculation.  Dividing the partial width by the branching ratio,
Br$(W \rightarrow \ell\nu) = 1 / (3 + 6 (1 + \alpha_s(M_W)/\pi
+ {\cal{O}}(\alpha_s^2)))$, gives the SM prediction for $\Gamma_W$.}
The transverse mass distribution templates also include 
the background contributions.  
The mass fit compares the data transverse
mass distribution to the templates.

The transverse mass fitting procedure is tested by using large Monte
Carlo samples and
by generating pseudo-samples of the size of the data and extracting a
mass value for each dataset. We investigated the bias in
the fit and confirmed the statistical errors returned by the fits. 
The results are illustrated for the muon fit in Figure~\ref{f_fiterror}. 
No biases are observed in the fitting procedure 
and the fit errors returned by the simulation datasets and the variation 
in returned mass values are consistent with
the statistical uncertainties of the fits to the data.

\begin{figure}[tbh]
    \centerline{\epsfysize 19cm
                    \epsffile{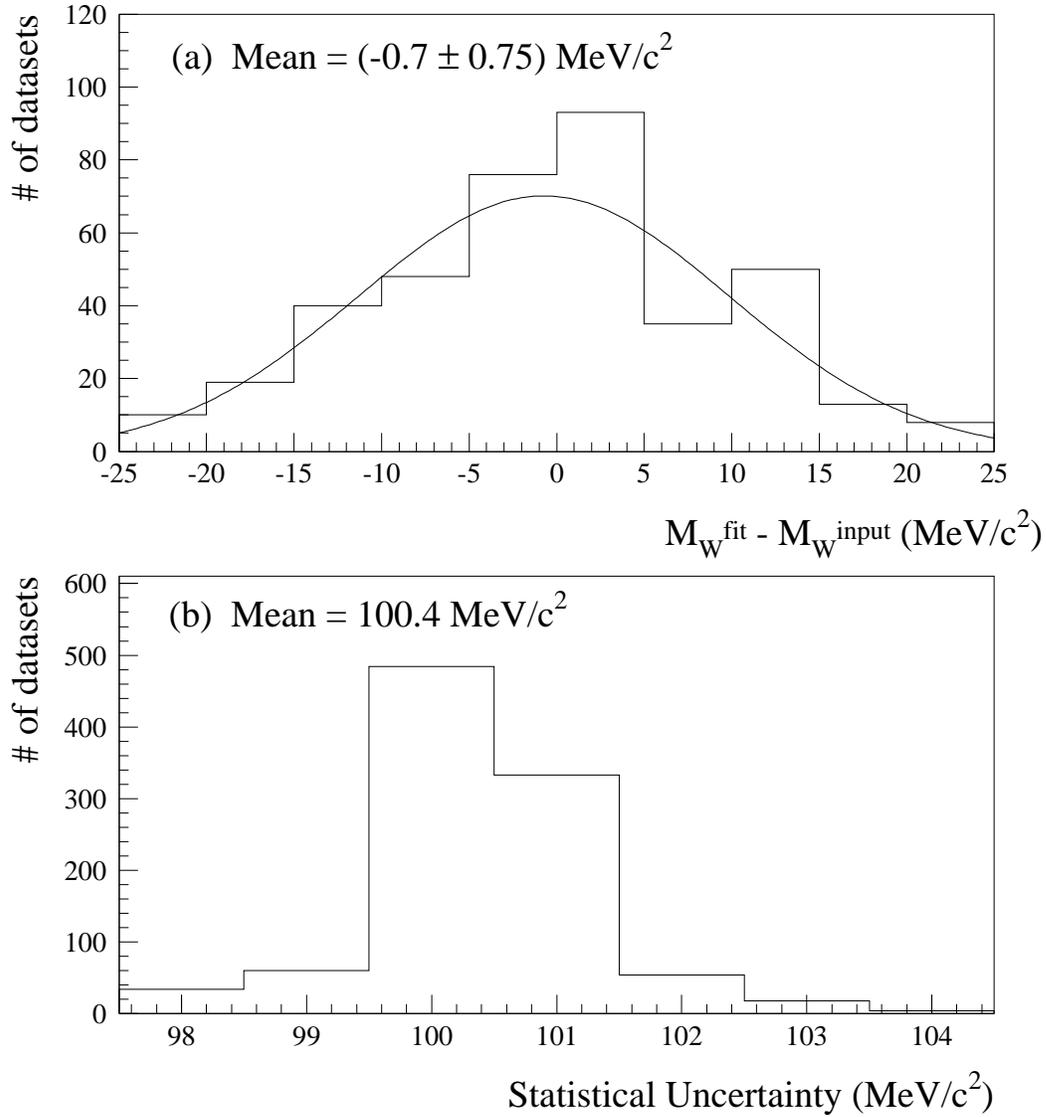}}
\vspace*{-2cm}
\caption{(a) Difference between the input $M_W$ values and the 
	returned values by fits to Monte Carlo pseudo-samples. 
	Each sample is 100 times the size of the $\wmunu$ data.
	(b) The (statistical) error returned by fitting 1000 Monte Carlo
	pseudo datasets of the same size as the $\wmunu$ data.}
\label{f_fiterror}
\end{figure}

\subsection{The $W$ Mass Measurement}
\label{sysunc}

The fit results yield the measurements of the $W$ mass 
in the electron and muon channels.  They are:
$$M_W^e   = 80.473 \pm 0.065 ~{\rm (stat.)} 
		   \pm 0.092 ~{\rm (syst.)} ~{\rm GeV/c^2}$$
and
$$M_W^\mu = 80.465 \pm 0.100 ~{\rm (stat.)} 
		   \pm 0.103 ~{\rm (syst.)} ~{\rm GeV/c^2}.$$
The negative log likelihood distribution for the muon sample is 
shown in Figure~\ref{f_likelihood} as a function of $M_W$.  
A similar distribution is obtained for the electron sample.
The transverse mass distributions for the $\wenu$ and 
$\wmunu$ samples are compared to the simulation with the best fits
in Figures~\ref{f_mtefit} and \ref{f_mtmufit}. 
The fit curves give $\chi^2$/dof of
32.4/35 and 60.6/70 for the electron and muon samples, respectively.
If we extend the region of comparison from 
$65 < M_T < 100$~GeV/c$^2$ to $50 < M_T < 120$~GeV/c$^2$,
the curves give $\chi^2$/dof of 82.6/70 and 147/131, and 
Kolmogornov-Smirnov (KS) probabilities of 16\% and 21\%.

\begin{figure}[p]
        \centerline{\epsfysize 16cm
                    \epsffile{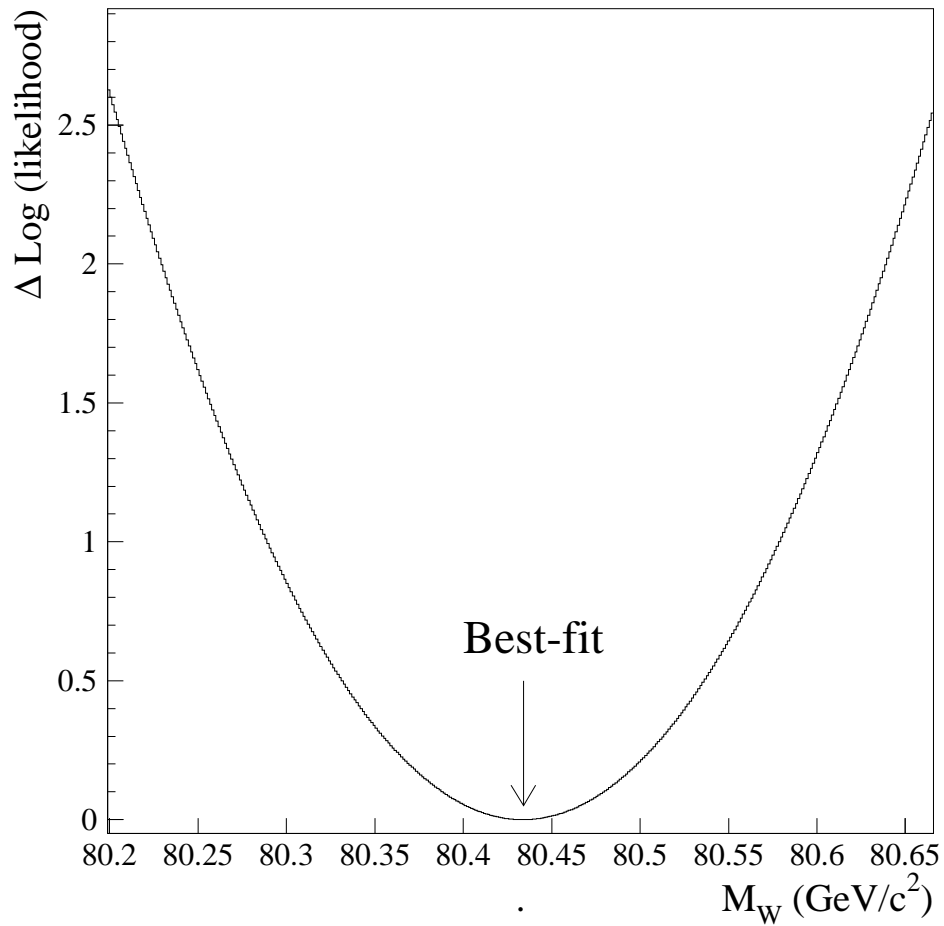}}
\vspace*{-1.5cm}
\caption{The deviation of the negative log likelihood from the minimum
for the $\wmunu$ sample.  The $W$ width is fixed at the Standard Model
value in the fit.}
\label{f_likelihood}
\end{figure}

\begin{figure}[p]
        \centerline{\epsfysize 18cm
                    \epsffile{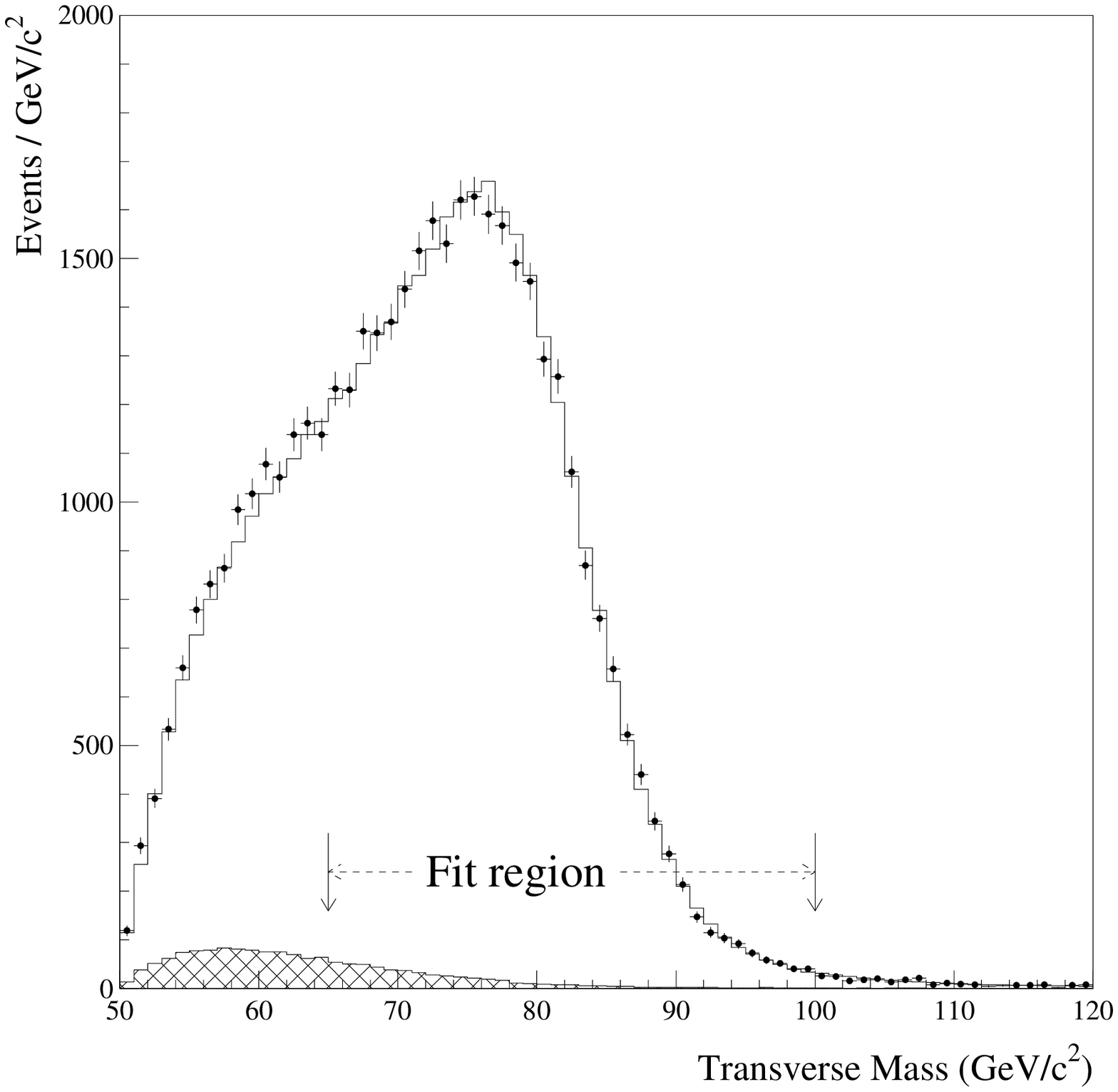}}
\vspace*{-1.5cm}
\caption{$W$ Transverse mass distributions compared to the best fit
	for the $\wenu$ channel.}
\label{f_mtefit}
\end{figure}

\begin{figure}[p]
        \centerline{\epsfysize 18cm
                    \epsffile{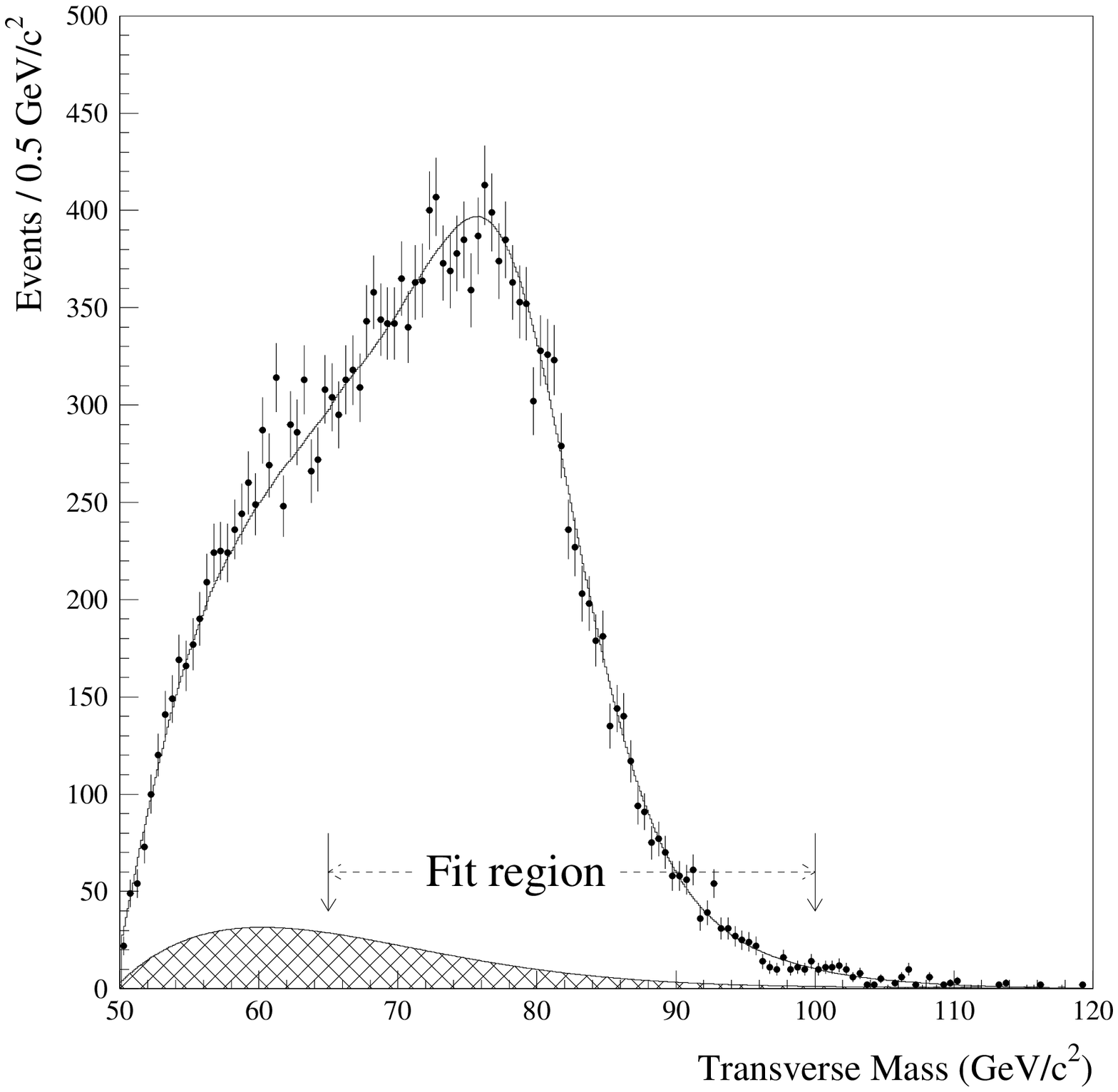}}
\vspace*{-1.5cm}
\caption{$W$ Transverse mass distributions compared to the best fit
	for the $\wmunu$ channel.}
\label{f_mtmufit}
\end{figure}

A summary of all systematic uncertainties is given in Table~8.1.  
They are estimated by measuring the subsequent shifts in
$M_W$ when each source is varied by its uncertainty in the
Monte Carlo simulation.
The largest uncertainties come from the finite statistics of the $Z$
samples.  The $Z$ statistics are the predominant source of
the uncertainties on lepton scale, lepton resolution, the $p_T^W$ model,
as well as the recoil model.  As muon and electron analyses use the muon and 
electron $Z$ sample separately, the statistical effects are independent.  
The theoretical uncertainty in the $p_T^W$ distribution gives a small 
common-contribution.  The uncertainty 
due to the choice of PDFs is evaluated for the muon acceptance and is 
essentailly the same for the electron
acceptance. We take the PDF uncertainties to be identical and common
for the two channels.  
Although the
QED corrections are rather different for electrons and muons, there is 
common as well as independent uncertainty.

The total common uncertainty for the two
lepton channels is 16~MeV/c$^2$, due almost entirely to the common
determination of the parton distribution function contribution.
Accounting for the correlations, the combined value is:
	$$M_W = 80.470 \pm 0.089 ~{\rm GeV/c^2}.$$

\begin{table}
\begin{center}
\begin{tabular}{|l|c|c|c|}
Source of uncertainty 	& $\wenu$ & $\wmunu$ 	 & common \\ \hline
Lepton scale 		& 75 	  & 85 		 & 	  \\
Lepton resolution 	& 25  	  & 20 		 & 	  \\
PDFs 			& 15  	  & 15 		 & 15 	  \\
$P_T^W$ 		& 15  	  & 20 		 & 3      \\
Recoil 			& 37  	  & 35 		 &        \\
Higher order QED 	& 20  	  & 10 		 & 5 	  \\
Trigger \& Lepton ID bias & $-$   & 15$\oplus$10 & 	  \\
Backgrounds 		& 5   	  & 25 		 & 	  \\ \hline
Total			& 92 	  & 103 	 & 16 	  \\
\end{tabular}
\caption{
\label{t_uncertainty}
Systematic uncertainties in the $W$ mass measurement in MeV/c$^2$.}
\end{center}
\end{table}

\begin{figure}[p]
\vspace*{-2cm}
 \centerline{\epsfysize 12cm
	\epsffile{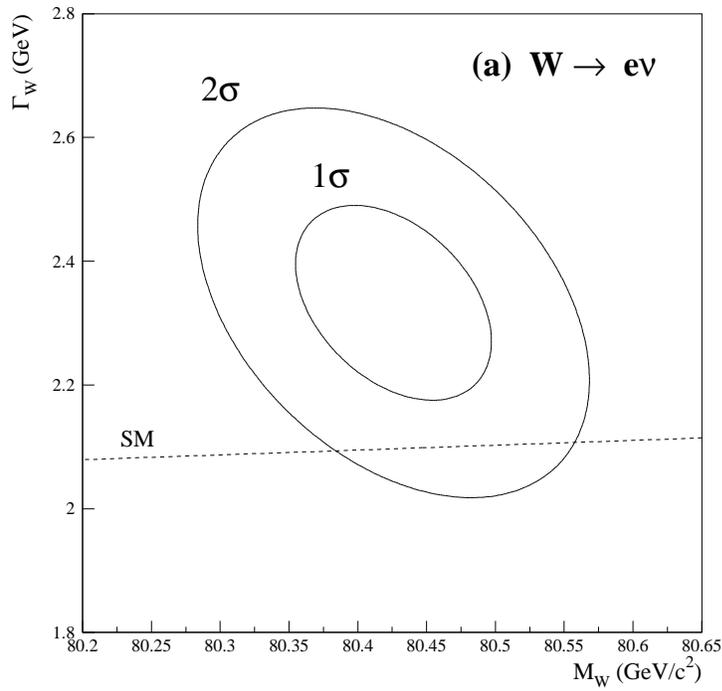}}
\vspace*{-2cm}
 \centerline{\epsfysize 12cm
	\epsffile{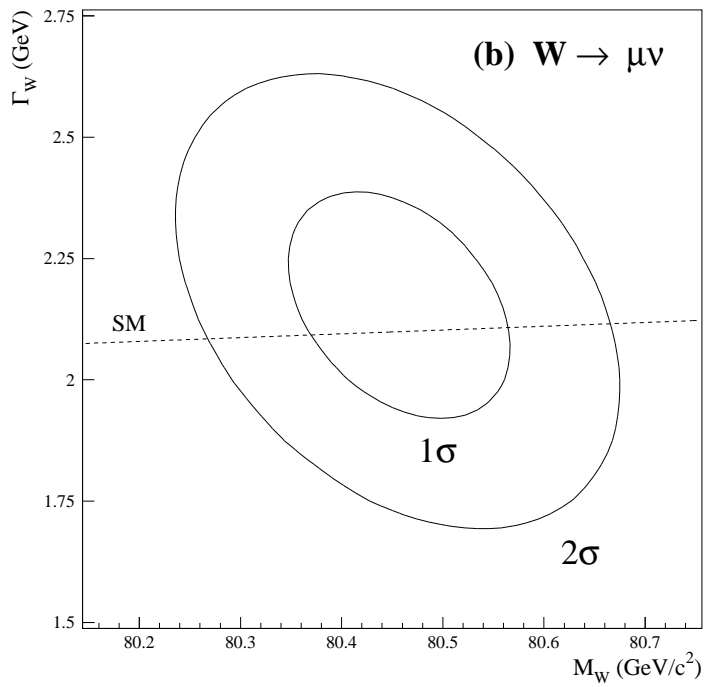}}
\vspace*{-1cm}
\caption{The 1-$\sigma$ and 2-$\sigma$ contours in $\Gamma_W$
	versus $M_W$ of the transverse mass fit when the width
	is floated for 
	(a) the $\wenu$ channel and (b) the $\wmunu$ channel.
	The dashed lines are the predicted $\Gamma_W$ as a function
	of $M_W$.}
\label{f_widthfit}
\end{figure}

\subsection{Cross-Checks of the $W$ Mass Measurement}

The reliability of the measurement can be checked by fitting lepton $p_T$ 
instead of transverse mass, by subdividing the $W$ samples, and by 
removing the constraint on the $W$ width as a function of mass.

The $W$ width, $\Gamma_W$, can be extracted from the transverse mass
distributions by fitting either in the region near the Jacobean edge 
or in the high-$M_T$ region.
The CDF experiment measured $\Gamma_W$ to be $2.04 \pm 0.14$~GeV
using $100 < M_T < 200$~GeV/c$^2$~\cite{wwidth}. 
By generating $M_T$ templates at discrete values of $M_W$ and 
$\Gamma_W$, and allowing them to vary in the fit, one can measure
both $M_W$ and $\Gamma_W$ simultaneously from the region near the 
Jacobean edge.  Since $\Gamma_W$ provides similar effects to 
the input $p_T^W$ and the detector resolution of $\bf u$ in this region,
the measurement of $\Gamma_W$ 
provides a check on the recoil and $p_T^W$ models.
Figure~\ref{f_widthfit} shows the 1-$\sigma$ and 2-$\sigma$ contours of
the fitted $W$ width versus $W$ mass.
The widths are consistent with the Standard Model:
it is almost identical to the SM value for the muon channel, 
and about 1.5~$\sigma$ away for the electron channel.
The fitted $W$ mass differs by 60~MeV/c$^2$ for the electron channel
and 10~MeV/c$^2$ for the muon channel from the values with $\Gamma_W$
fixed.  
We do not derive measurements of the width from these fits due to
the large systematics variations which come from changing resolutions and
modeling. 

The transverse momentum spectra of the leptons as shown in
Figures~\ref{f_etfit_wenu} and
\ref{f_etfit_wmunu} also contain $W$ mass information. 
$W$ mass values obtained from maximum likelihood fits  
are consistent with the values from the transverse mass fit.  
The distributions from the simulation 
with the best fits are compared with the data in the figures.

The $W$ mass results are cross-checked by making various selection criteria
on the data and Monte Carlo simulation, and refitting for the $W$ mass.
The events are divided into positively and negatively
charged lepton samples. For the electon sample 
the charge difference listed in Table~\ref{t_crosscheck} involves
statistical uncertainty only and corrreponds
to the mass difference of $123 \pm 130$~MeV/c$^2$
between the $W^+$ and the $W^-$. For the muon sample 
the table entries include the tracking alignment uncertainty of 
$50$ MeV/c$^2$.
The mass difference of $136 \pm205$ MeV/c$^2$ is observed between
the $W^+$ and the $W^-$.
The electron and muon results are combined to give 
a mass difference of $127 \pm110$ MeV/c$^2$.

The samples are also partitioned into four bins of 
$|{\bf u}|$ as shown in Figures~\ref{f_mt_wenu_ubin} and 
\ref{f_mt_wmunu_ubin}.  
The Monte Carlo simulation reproduces the data very well in all the
$|{\bf u}|$ bins, indicating that the $W$ $p_T$ and recoil energy
are well modeled in the simulation.
When the events are partitioned into 
$p_T^\mu > 35$ GeV/c and $p_T^\mu < 35$ GeV/c samples, 
the $M_T$ shapes between the two samples (see
Figure~\ref{f_mt_wmunu_pt}) are dramatically different.
Yet there is good agreement between the data and simulation.

The extracted $W$ masses described above are summarized 
in Table~\ref{t_crosscheck}.

\begin{figure}[p]
\epsfysize=6.0in
\epsffile[54 162 531 675]{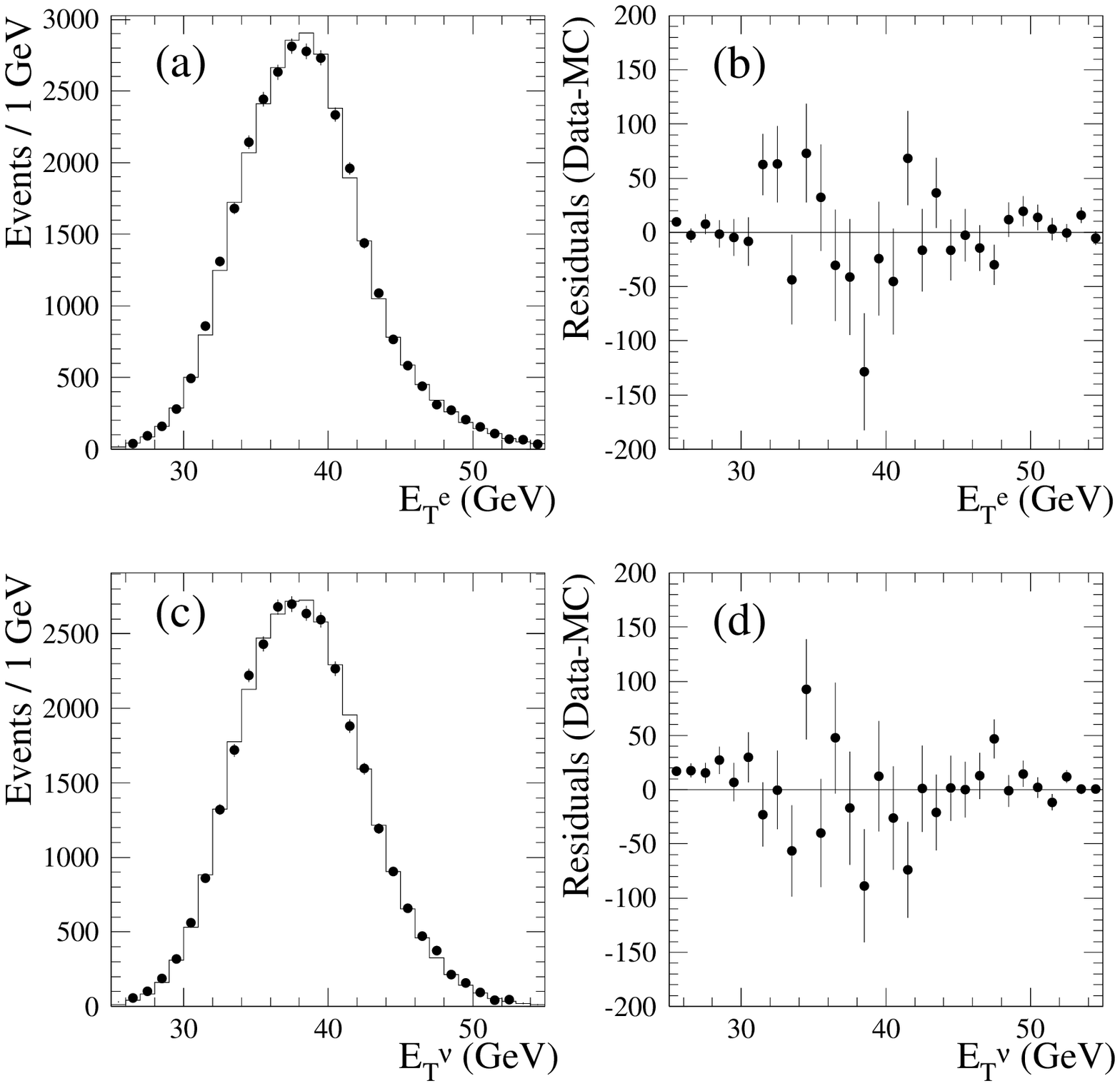}
\caption{$E_T$ distributions of (a) electrons and 
	(c) neutrinos in the $\wenu$ channel.
	The points are the data and the histograms the best fit 
	simulation.	
	The differences between the data and simulation are shown in
	(b) and (d).}
\label{f_etfit_wenu}
\end{figure}

\begin{figure}[p]
\epsfysize=6.0in
\epsffile[54 162 531 675]{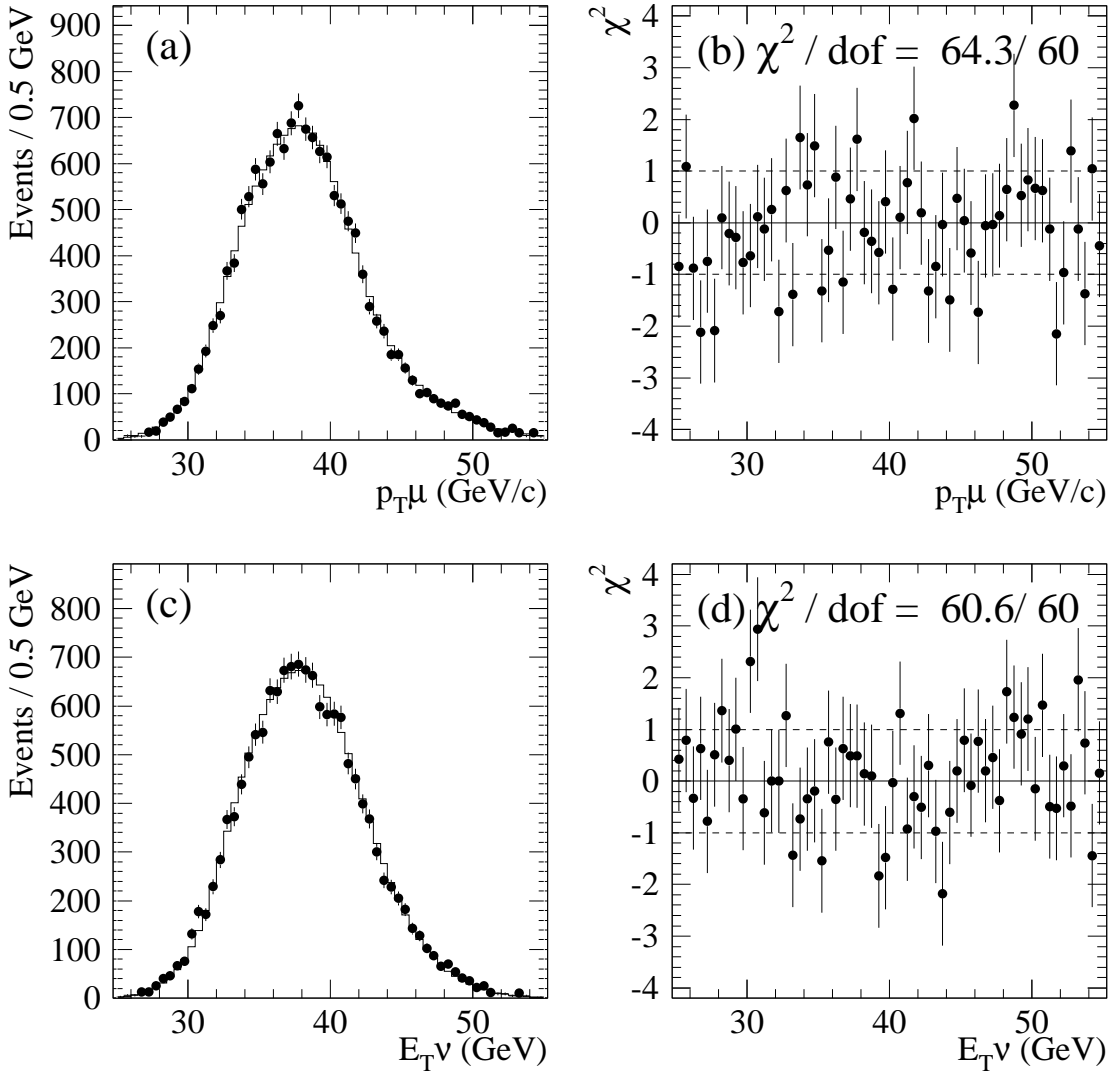}
\caption{(a) $p_T$ distribution of muons and 
	(c) $E_T$ distribution of neutrinos in the $\wmunu$ channel.
	The points are the data and the histograms the best fit 
	simulation.	
	(b) and (d) The difference between the data and simulation
	normalized by the statistical uncertainty.} 
\label{f_etfit_wmunu}
\end{figure}

\begin{figure}[p]
\vspace{-5.0cm}
\epsfysize=6.0in
\epsffile[54 162 531 675]{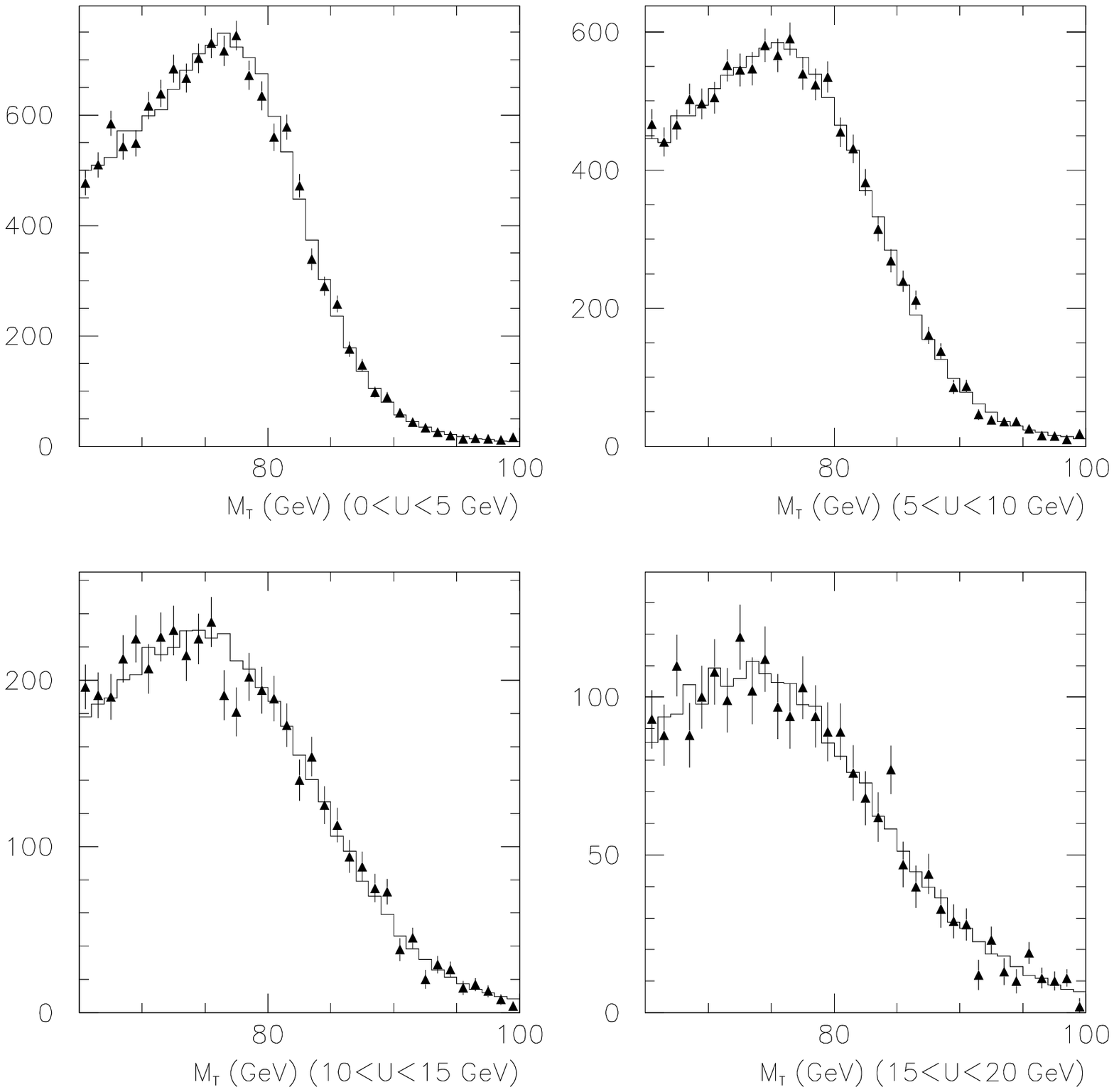}
\vspace{4.5cm}
\caption{Transverse mass distributions in bins of $|{\bf u}|$ 
	for the $\wenu$ data (triangles) and the best fit simulation
	(histograms).  The four $|{\bf u}|$ bins are,
	$0 < |{\bf u}| < 5$~GeV (Top Left),
	$5 < |{\bf u}| < 10$~GeV (Top Right),
	$10 < |{\bf u}| < 15$~GeV (Bottom Left), and 
	$15 < |{\bf u}| < 20$~GeV (Bottom Right)}
\label{f_mt_wenu_ubin}
\end{figure}

\begin{figure}[p]
\epsfysize=6.0in
\epsffile[54 162 531 675]{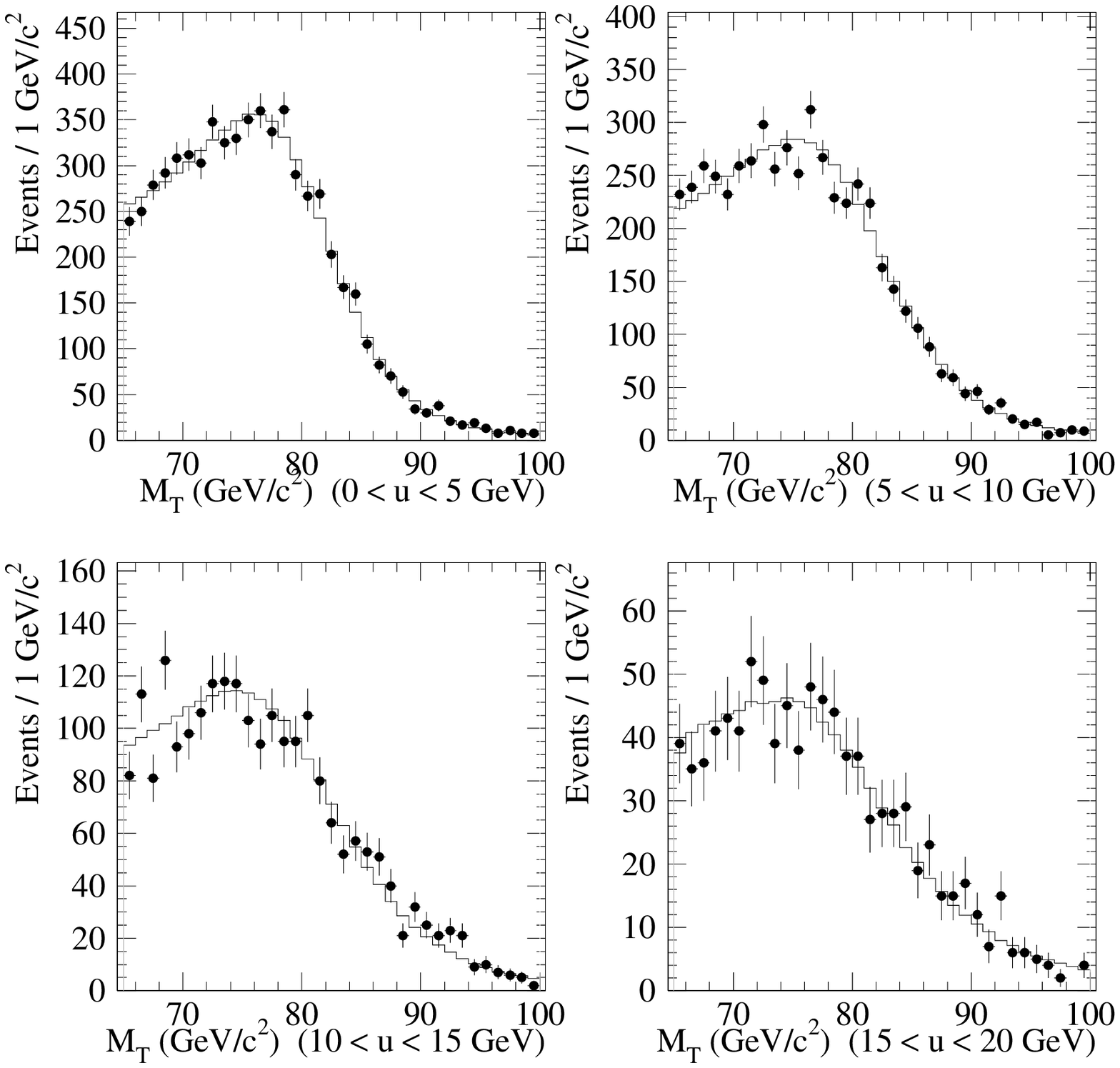}
\vspace{1cm}
\caption{Transverse mass distributions in bins of $|{\bf u}|$ 
	for the $\wmunu$ data (points) and the best fit simulation
	(histograms).  The four $|{\bf u}|$ bins are,
	$0 < |{\bf u}| < 5$~GeV (Top Left),
	$5 < |{\bf u}| < 10$~GeV (Top Right),
	$10 < |{\bf u}| < 15$~GeV (Bottom Left), and 
	$15 < |{\bf u}| < 20$~GeV (Bottom Right)}
\label{f_mt_wmunu_ubin}
\end{figure}

\begin{figure}[p]
\epsfysize=6.0in
\epsffile[54 162 531 675]{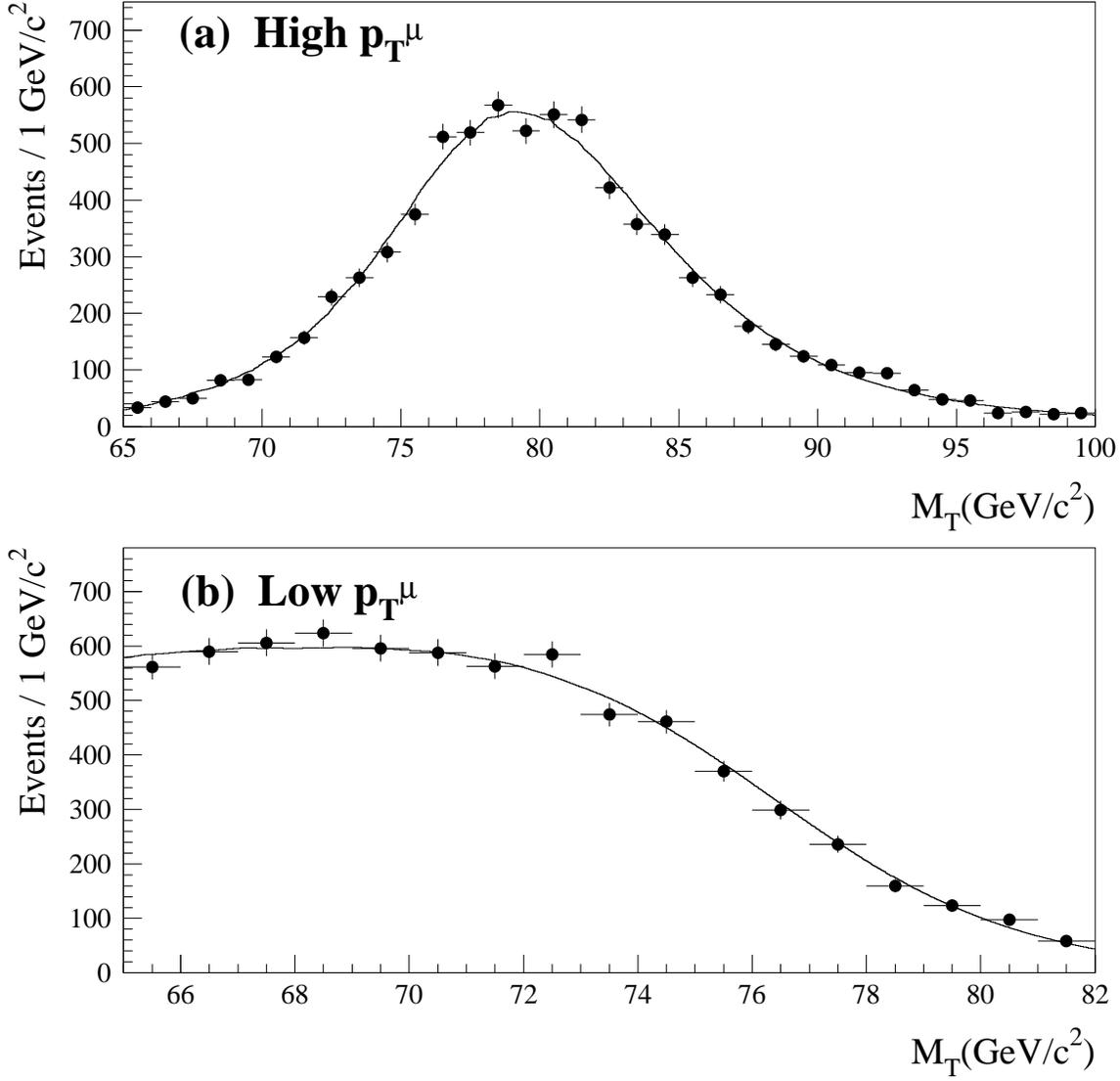}
\vspace*{2.0cm}
\caption{Transverse mass distributions for (a) low $p_T$ 
	and (b) high $p_T$ muons in the $\wmunu$ data (squares) and
	simulation (lines).}
\label{f_mt_wmunu_pt}
\end{figure}

\begin{table}
\begin{center}
\begin{tabular}{|c|c||c|c|}
Fitting & Selection & $\Delta M_W$ (MeV/c$^2$) (e$\nu$) &
		      $\Delta M_W$ (MeV/c$^2$) ($\mu\nu$) \\ \hline
$E_T^e$, $p_T^\mu$ 	& -- 	& $-80 \pm 60$ & $-19 \pm 132$ \\
$E_T^\nu$ 		& --	& $+76 \pm 60$ & $-20 \pm 127$ \\
$M_T$	& $\ell^+$		   & $ +62 \pm  90$ & $ +67 \pm 145$ \\
$M_T$	& $\ell^-$		   & $ -61 \pm  90$ & $ -69 \pm 145$  \\
$M_T$	& $ 0 < |{\bf u}| < 5$ GeV & $  -1 \pm  86$ & $ -41 \pm 135$ \\
$M_T$	& $ 5 < |{\bf u}| <10$ GeV & $ -36 \pm 110$ & $-164 \pm 169$ \\
$M_T$	& $10 < |{\bf u}| <15$ GeV & $+161 \pm 204$ & $+484 \pm 301$ \\
$M_T$	& $15 < |{\bf u}| <20$ GeV & $-348 \pm 385$ & $+534 \pm 450$ \\
\end{tabular}
\caption{
\label{t_crosscheck}
Difference from the nominal value of
        extracted $M_W$ values from lepton transverse momentum fits
	and from various subsample transverse mass fits.}
\end{center}
\end{table}

\subsection{Combined $W$ Mass}

The issue of combining the present results with previous CDF
measurements~\cite{wmass_1a} merits some additional discussion since the
lepton energy and momentum scales were determined differently.  In
particular, in our the previous analyses the
electron scale was determined with the $E/p$ method.
In the present work that procedure is shown to result in a $Z$ mass
discrepant by $(0.52 \pm0.13)$\%; in the Run IA analysis, the discrepancy 
was $(0.28 \pm0.24)$\%.  
The statistics of Run~IA are insufficient to distinguish the two cases 
-- that the $E/p$ method worked well or was systematically off as 
indicated in the Run~IB result. Moreover, the experimental conditions differ 
for the two runs. For example, the aging and rate effects in the CTC due to
higher luminosity are more pronounced for the present work.  For these
reasons and because the underlying cause for the $E/p$ discrepancy
remains unresolved, we believe that applying a correction factor to the
Run IA result is not warranted.  We prefer to
average the results as published with the stated errors.  Thus
the combined CDF result is:
		$$M_W = 80.433 \pm 0.079 ~{\rm GeV/c^2}.$$
This value is precise to 0.1\% and corresponds to a total integrated
luminosity of $\sim$105~pb$^{-1}$.

\subsection{Comparison with Other Results}

The present results are compared with other published results in
Table~\ref{t_mwcomparison}~\cite{spps,d0,lepdr,lepewwg}.
The agreement is excellent.
The direct measurement of the $W$ mass is an important test of the 
Standard Model.  The $W$ mass is indirectly 
predicted precisely by including loop corrections involving the top
quark and Higgs boson. The corresponding implication for the Higgs
boson mass is shown in Figure~\ref{f_mw_vs_mhiggs}.  Our result agrees
well with the Standard Model, and when combined with all other
electroweak results~\cite{lepewwg} prefers a light Higgs boson.

\begin{table}
\begin{center}
\begin{tabular}{|l|l|}
UA2 	& $80.360 \pm 0.370$ GeV/c$^2$\\ \hline
CDF 	& $80.433 \pm 0.079$ GeV/c$^2$\\ 
D0 	& $80.474 \pm 0.093$ GeV/c$^2$\\ \hline
ALEPH 	& $80.418 \pm 0.076$ GeV/c$^2$ upto $\sqrt s$ = 189 GeV\\
	& $80.423 \pm 0.123$ GeV/c$^2$ upto $\sqrt s$ = 183 GeV\\
DELPHI 	& $80.270 \pm 0.144$ GeV/c$^2$ upto $\sqrt s$ = 183 GeV\\
L3 	& $80.610 \pm 0.150$ GeV/c$^2$ upto $\sqrt s$ = 183 GeV\\
OPAL 	& $80.432 \pm 0.080$ GeV/c$^2$ upto $\sqrt s$ = 189 GeV (preliminary)\\
	& $80.380 \pm 0.130$ GeV/c$^2$ upto $\sqrt s$ = 183 GeV\\ \hline
Indirect Meas. & $80.381 \pm 0.026$ GeV/c$^2$ \\
\end{tabular}
\caption{
\label{t_mwcomparison}
Measurements of the $W$ mass. CDF and D0 measurements have
a common error mostly due to Parton Distribution Functions.
The LEP II measurements have common errors including the LEP beam 
energy. 
The indirect measurement includes the LEP and SLC $Z$ pole
measurements, the $\nu$N measurement, and the Tevatron Top mass
measurements.}
\end{center}
\end{table}

\subsection{Conclusions}

We have measured the $W$ mass to be $M_W = 80.470 \pm 0.089~{\rm
GeV/c^2}$ using data with an integrated luminosity of $\sim$85~pb$^{-1}$
collected from 1994 to 1995. 
When combined with previously published CDF data, we obtain
 $M_W = 80.433 \pm 0.079~{\rm GeV/c^2}$.
 
\begin{figure}[p]
\epsfysize=6.0in
\epsffile[54 162 531 675]{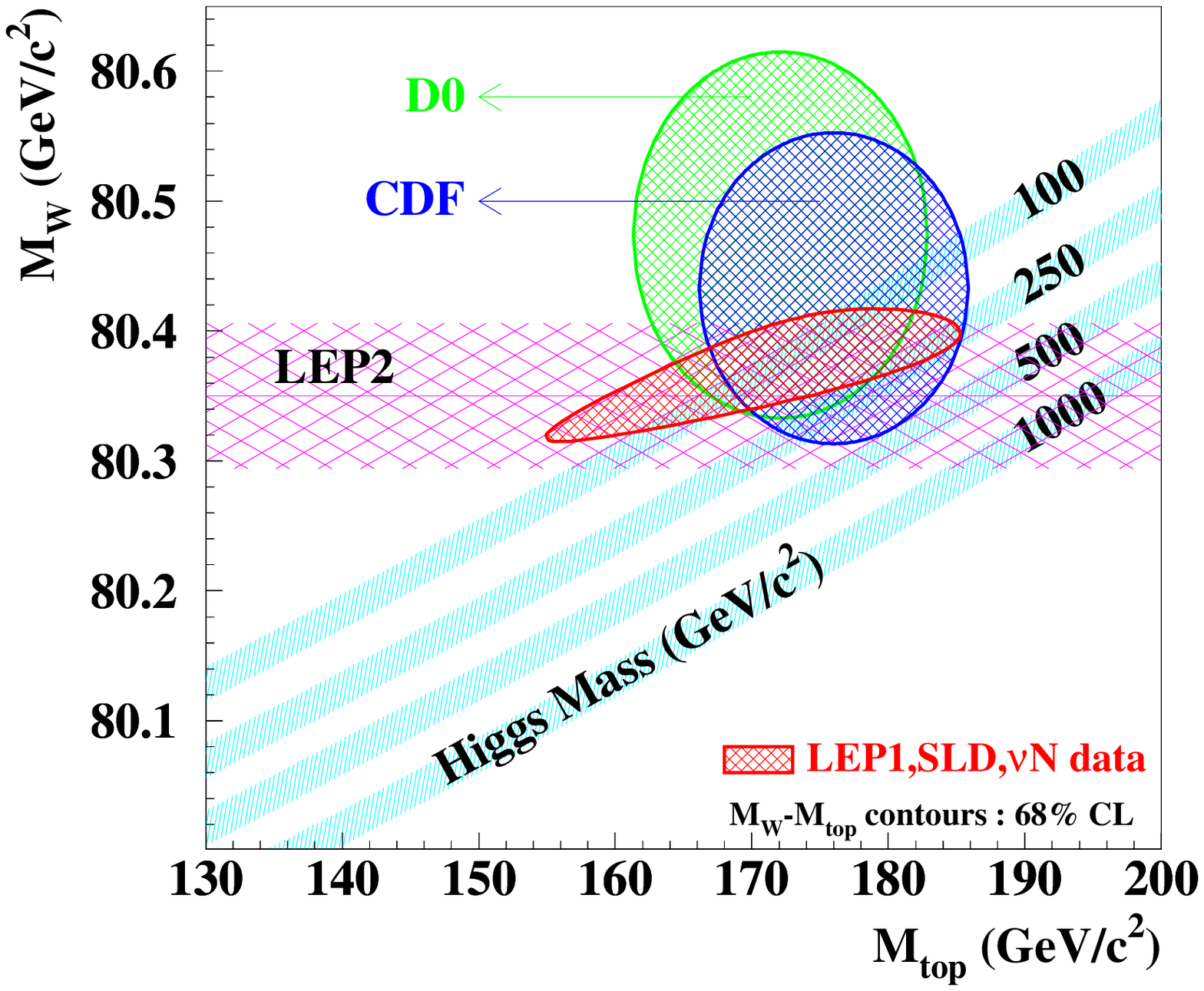}
\caption{The direct measurements of the $W$ and top quark mass from CDF and
	\dzero\ experiments, the direct measurement of the $W$ mass
	from LEP~II experiments, and the indirect $W$ and top mass
	measurement from LEP, SLC, and Tevatron neutrino experiments.
	The curves are from a calculation of the dependence of the $W$
	mass on the top mass in the Standard Model
	using several Higgs boson masses.  The band on each curve
	is the uncertainty obtained by folding in quadrature
	uncertainties on $\alpha(M_Z^2)$, $M_Z$, and $\alpha_s(M_Z^2)$.
	The uncertainty is dominated by the hadronic contribution to
	$\alpha(M_Z^2)$, 
	$\Delta \alpha_{had} = 0.028 \pm 0.0007$ (Ref.48).}
\label{f_mw_vs_mhiggs}
\end{figure}

\vspace*{1cm}

\begin{center}
{\large\bf Acknowledgments}
\end{center}

     We thank the Fermilab staff and the technical staffs of the
participating institutions for their vital contributions.  This work was
supported by the U.S. Department of Energy and National Science Foundation;
the Italian Istituto Nazionale di Fisica Nucleare; the Ministry of Education,
Science, Sports and Culture of Japan; the Natural Sciences and Engineering 
Research Council of Canada; the National Science Council of the Republic of 
China; the Swiss National Science Foundation; the A. P. Sloan Foundation; the
Bundesministerium fuer Bildung und Forschung, Germany; and the Korea Science 
and Engineering Foundation.
We also thank Ulrich Baur and Keith Ellis in support and advice.

\appendix
\section {Discussion of Discrepancy between $M_Z$ and $E/p$ Methods}
\label{epchecs}

The calorimeter energy scale for the $W$ mass measurement in this paper
is set using the invariant mass distribution of {\zee} events.  
Ideally, the $E/p$ distribution would be used to set the energy scale 
where the momentum scale is determined by the 
$\Upsilon \rightarrow \mumu$ data.  The $E/p$ distribution 
has a smaller statistical uncertainty than the method of using 
the {\zee} mass because it makes use of the higher statistics of 
the $W$ and $\Upsilon$ samples.  
The $E/p$ method, however, gives a significantly
different result than the {\zee} mass method. 

The \zee\ mass method gives the energy scale of 1 by construction
(see Section~\ref{escale}) :
	$$S_E = {M_Z^{\rm PDG} \over M_Z^{\rm CDF}} 
				= 1.0000 \pm 0.0009.$$
The $E/p$ distribution for the $\wenu$ data
does not agree with the simulation with the energy scale given by
the $Z$ mass method.  The best fit between the data
and the simulation requires an energy scale, 
		$$S_E = 0.99613 \pm 0.00040 {\rm ~(stat.)}.$$
Including the non-linearity correction described in 
Section~\ref{enonlin} 
the energy scale becomes 
\begin{eqnarray*}
       S_E = 0.9948 & \pm & 0.00040 {\rm ~(stat.)} \\
                         & \pm & 0.00024 {\rm ~(\kappa)} \\
                         & \pm & 0.00035 {\rm ~(X_\circ)} \\
			 & \pm & 0.00018 {\rm ~(p_T~scale)} \\
			 & \pm & 0.00075 {\rm ~(CEM~Non-linearity)}
\end{eqnarray*}
where the uncertainty on the momentum scale comes from 
the $\Upsilon$ mass measurement 
(see Section~\ref{scalecheck}).
The difference between the $M_Z$ result and the $E/p$ result is 
\begin{equation}
\label{eq:scale_disc}
		\frac{1.0000-0.9948}
		     {\sqrt{
			 	0.0009^{2}+
			 	0.0010^{2}
				}} = 3.9
\end{equation}
standard deviations.  This is unlikely to be a statistical fluctuation.
A Kolmogorov-Smirnov statistic is calculated for the comparison of the 
data to the Monte Carlo.  The probability that a statistical fluctuation
would produce a worse agreement in the integrated distributions is
$5.5\times 10^{-6}$. 

This Appendix discusses checks given by various data samples, and 
possible explanations of the discrepancy between
$E/p$ and $M_Z$ methods.  

\subsection{Checks on $E$ and $p$ Scales}

The energy scale, $S_E$, is checked using various data samples.
The $\zee$ sample is used for extracting the $E$ scale from $E/p$.
The $J/\psi \rightarrow \mumu$ and $\zmumu$ samples are used
for extracting the $p$ scale.
The momenta of electron tracks for 
the $\psi \rightarrow \ee$, $\Upsilon \rightarrow \ee$, 
   and $\zee$ samples are used for setting the $p$ scale 
(see Figure~\ref{f_mass_ee}). 
The results are summarized in Table~\ref{escale_summary} and
Figure~\ref{f_e_check}.
While all the results are consistent with each other, 
the central values are closer to 1 when the $E/p$ scale is determined
using the $\zee$ sample instead of the $\wenu$ sample, 
or when the $p$ scale is determined using electron tracks instead of 
muon tracks.
Problems in the electron non-linearity correction or 
differences between the electron and muon tracks beyond our simulation
could cause this.  However our results are not statistically significant
enough to be conclusive.

\begin{figure}
\centerline{\epsfysize 18cm
                    \epsffile{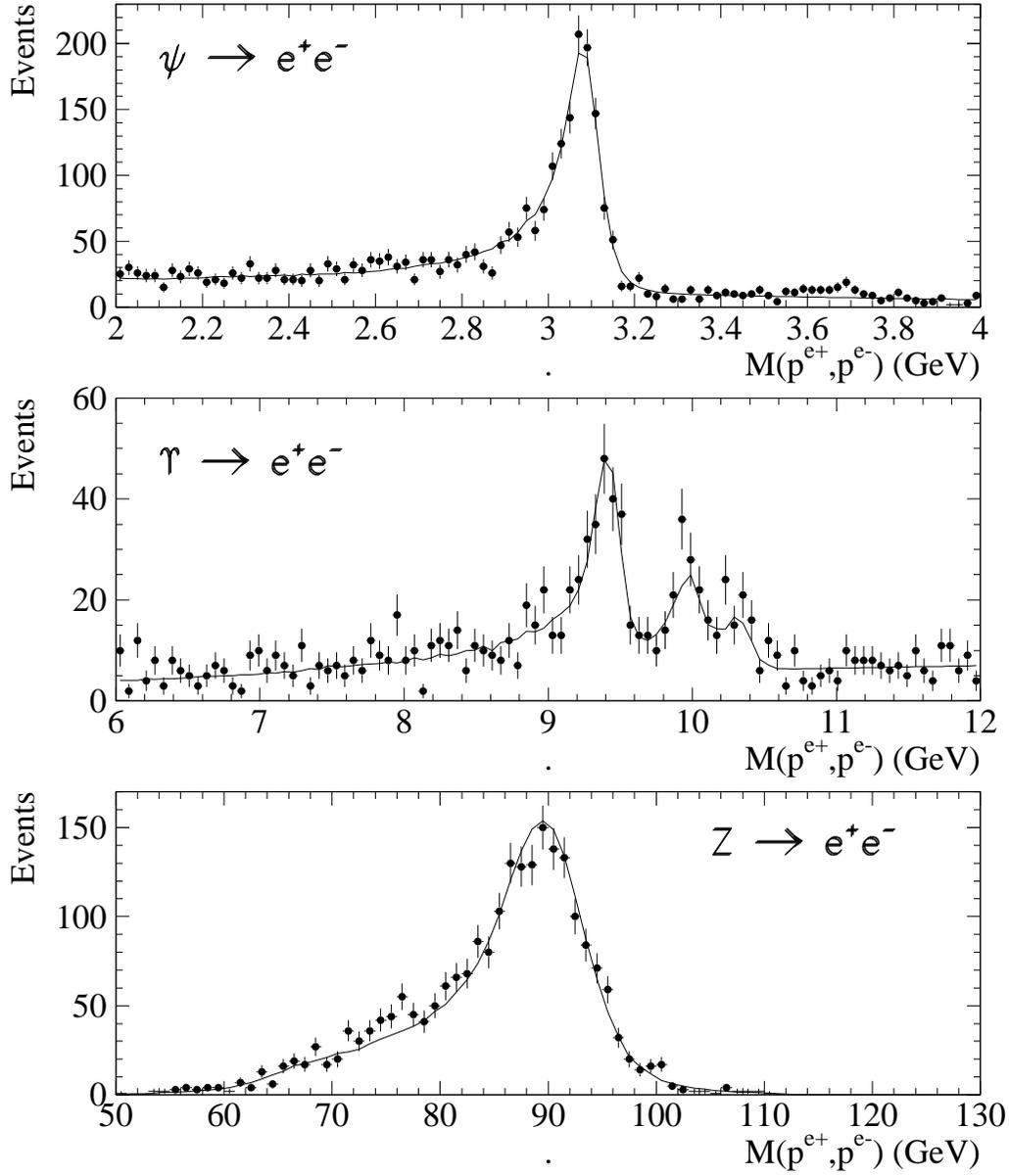}}
\caption{Invariant mass distributions of electrons using their momenta
	for $\psi \rightarrow \ee$, $\Upsilon \rightarrow \ee$, and
	$\zee$ data samples. The solid lines are the best fits from
	the Monte Carlo simulation.}
\label{f_mass_ee}
\end{figure}

\begin{table}
\begin{center}
\begin{tabular}{|c|c|c|c|c|}
\#& Data Sample for $p$ scale   & Data Sample for $E/p$ scale   
		& $S_E$ & Dev. from 1 \\ 
\hline\hline
1& $\Upsilon \rightarrow \mumu$ 	     & $\wenu$ 
 & $0.9948 \pm 0.0010 \pm 0.0002$     & $-3.9\sigma^*$ \\ \hline
2& $\Upsilon \rightarrow \mumu$ 	     & $\zee$  
 & $0.9972 \pm 0.0014 \pm 0.0002$     & $-2.0\sigma$ \\ \hline
3& $J/\psi \rightarrow \mumu$ 	     & $\wenu$  
 & $0.9947 \pm 0.0010 \pm 0.0004$     & $-3.8\sigma^*$ \\ \hline
4& $Z \rightarrow \mumu$ 	     & $\wenu$  
 & $0.9952 \pm 0.0010 \pm 0.0011$     & $-2.8\sigma^*$ \\ \hline
5& $Z \rightarrow \ee$(tracks)   & $\wenu$ 
 & $0.9955 \pm 0.0010 \pm 0.0026$     & $-1.5\sigma^*$ \\ \hline
6& $\Upsilon \rightarrow \ee$(tracks)& $\wenu$ 
 & $0.9970 \pm 0.0010 \pm 0.0020$     & $-1.2\sigma^*$ \\ \hline
7& $J/\psi \rightarrow \ee$(tracks)  & $\wenu$ 
 & $0.9959 \pm 0.0010 \pm 0.0015$     & $-2.0\sigma^*$ \\
\end{tabular}
\end{center}
\caption{Required energy scales for various data samples.
The errors on $S_E$ come from the $E/p$ scale (first) and
the $p$ scale (second). *: the deviation from 1 includes the $Z$
statistical uncertainty ($\pm0.0009$).}
\label{escale_summary}
\end{table}

\vspace{-2cm}
\begin{figure}
\centerline{\epsfysize 18cm
                    \epsffile{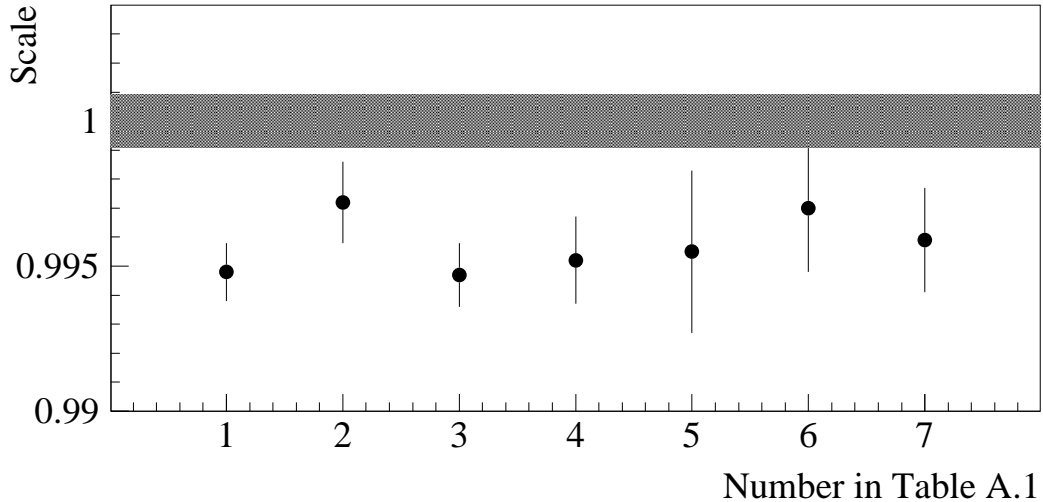}}
\vspace{-9.0cm}
\caption{Required energy scales for various data samples.
The shaded area represents the energy scale determined by the $\zee$
mass.}
\label{f_e_check}
\end{figure}

\subsection{Momentum Non-Linearity}

A non-linearity in the $p_T$ measurement could produce a discrepancy 
between the two methods.
The average $p_T$ of $\Upsilon$ ($\psi$) decay 
muons is $\sim 5.0$ GeV/c ($\sim 3.5$ GeV/c),
while the average $p_T$ of W and Z decay electrons is $\sim 40$ GeV/c.
Figure~\ref{f_p_non_linearity} shows the difference between the measured
mass and the expected mass as a function of the sum of $1/p_T$ of 
the two muons in $\Upsilon$ and $\psi$ decays. 
$W$ and $Z$ events occur on the far left of the plot. 
No significant momentum non-linearity is observed.\footnote{
Without the new CTC calibration and alignment for this analysis, 
there appears to be a small non-linearity in momentum measurement 
(0.1\% non-linearity from 2~GeV to 50~GeV).  
This went away with the CTC calibration and alignment.  The change has not been
fully understood.}

\subsection{Differences between the Electron and Muon Tracks}
\label{section:scale_disc_pt}

In the $E/p$ method, the electron momentum scale is determined 
from the muon momenta.
In many ways, electron tracks are different from those of muons.
They are produced with different internal bremsstrahlung.
The external bremsstrahlung is also different, resulting in 
different momenta.
Furthermore the external bremsstrahlung causes the tracks to have
a non-zero impact parameter, which introduces a bias on the 
beam-constrained momentum.
The simulation should take into account all the differences between 
electrons and muons,\footnote{Note that no material effects are
included for the muons from the $W$ and $Z$ decays because they
are negligible} when the momentum scale determined by muons
is transferred to the electron momentum.
However, mishandling any of these differences 
in the simulation may cause a difference between the electron 
momentum scale and the muon momentum scale, causing
a discrepancy between the $Z$ mass and $E/p$ methods.
In principle, the electron momentum scale can be checked using
electron tracks.  However, as shown in 
Table~\ref{escale_summary},  the uncertainties are too large to 
allow us to have concrete conclusions.

This section describes the differences between electron tracks 
and muon tracks, how the simulation treats them, and the size of
possible biases.

\begin{description}

\item [Internal Bremsstrahlung Distribution:] ``Internal'' photons
are photons which are produced at the vertex in a radiative $\wenu\gamma$ event
(or ${\zee}\gamma$ event).  For Monte Carlo events with no external photons, we
find that the average $E/p$ between $0.9$ and $1.1$ is $1.0039$.  Part of this
shift above 1, 0.0014, is from cut biases, and the internal
bremsstrahlung shifts the peak by 0.0025.  
The distribution we are using would have to be wrong by $\sim$100\% for 
our fitted energy scale to come out shifted enough to account for the 
discrepancy between the energy scale from $M_Z$ and $E/p$.

\begin{itemize}

\item The generator that is used for $E/p$ simulation 
in these studies (PHOTOS~\cite{photos} in two-photon mode) has
been compared to the calculation by Berends and Kleiss of
Reference~\cite{wgamma}, and the two generators give similar energy-angle
distributions. 

\item Laporta and Odorico~\cite{LandO} argue that inclusion of multiple photon
radiation from the final state electron may change the energy loss distribution
of the electron relative to a single photon calculation, such as Berends and
Kleiss.  Reference~\cite{LandO} contains an algorithm to calculate the effect
of a cascade of final state photons.  By construction, this algorithm reduces
to Berends and Kleiss for the case of single photon emission.  
Their algorithm is implemented for $W$ decays.
The Laporta and Odorico case has the mean $E/p$ between $0.9$ and 
$1.1$ lower by $0.00033$.  This is not insignificant, but
it is not nearly large enough to account for the discrepancy 
between the $M_Z$ and $E/p$ methods.  The statistical error on the Monte 
Carlo for this  calculation is 0.00015. 

\item Baur, Keller, and Wackeroth~\cite{baur} have done a calculation of the
${\wenu\gamma}$ process which includes radiation from the $W$ propagator.
We have
received their calculation in the form of a Monte Carlo~\cite{baurper}.  The
Monte Carlo can implement their calculation, and it can also implement Berends
and Kleiss.  We run separately in each mode and implement some simple 
model of CEM
clustering of the photons and measurement resolutions.  
We find that~\cite{baur}
produces a value for the mean of $E/p$ between $0.9$ and $1.1$ 
that is $0.00023$ lower than the Berends and Kleiss result. 

\end{itemize}

\item [External Bremsstrahlung Distribution:] The formula we
are using for the photon energy distribution was calculated in
1974 by Tsai~\cite{tsai}. 
This formula is still referenced in papers written today,
but it is possible that the formula is unexpectedly
breaking down at high energies. Evidence that it is not is given by the SLAC
measurement of the Landau-Pomeranchuk-Migdal effect described below~\cite{lpm}.
They measured the rate and energy distribution of bremsstrahlung of $25$ GeV
electrons incident on different targets.  For all the targets, they measured
some level of bremsstrahlung suppression at low photon energies, as expected,
but at higher photon energies, their measured distributions agreed well with
the expectation from~\cite{tsai}.  

\item[Low Energy Bremsstrahlung Cutoff:] 
Since the number of
external photons diverges as $1/E$, we only consider external photons above a
certain energy.  In particular, we only simulate photons above $y=0.1\%$, where
$y$ is the fraction of the electron energy taken up by the photon. However, we
can integrate the total fraction of the electron energy that is carried by
photons below the cutoff.  The total fraction is $y=0.1\%\times 0.085$, where
$0.085$ is an approximation of the effective number of radiation lengths seen
by the electrons, including the CTC gas and wires.  We expect this to affect
the energy scale by less than $0.0001$, which is a negligible amount.  As a
simple check we have increased the cutoff and we do not see any significant
change in the fitted energy scale.  A similar argument holds for the
internal photons. 

\item [Beam Constraint Biasing $E/p$:] 
The beam constraint can bias tracks that have undergone external 
radiation (bremsstrahlung)
before the CTC active volume.  Bremsstrahlung causes the tracks to 
have a non-zero impact parameter which biases the beam-constrained 
momentum.  The simulation follows the same procedure, and so we expect
this bias to be reproduced. 
Two possibilities are considered.

\begin{itemize}

\item {\bf The radial distribution of material may be wrong.}\\
The average radius of external radiation (including half the CTC gas) 
occurs at $22.21$ cm in the simulation.
The bias depends on $r^{2}$, and so 
the location of the material might be sensitive to the scale.
As a check the simulation is run with all the material before 
the CTC gas placed in the beampipe,
or with all placed in the CTC inner can.  The material is scaled 
so that $\langle X_\circ \rangle$ is the same for both cases.  $f_{tail}$ 
for the beampipe case is higher than the CTC case by about $1\%$ of
itself.  The average $E/p$ from $0.9$ to $1.1$ is higher in 
the beampipe case than the CTC case by $0.0003$.  
Both of these changes are small.
Considering that these are extreme cases for variations in the
possible distributions of the material, the expected changes are
negligible.

\item  {\bf In the simulation, the correlation between curvature and impact
parameter mismeasurement may not be correct.}\\
This would cause the Monte Carlo
to produce the wrong bias from the beam constraint.  However, in the Monte
Carlo, we use CTC wire hit patterns from the real $W$ data to derive a
covariance matrix to use in the beam constraint.  We use the identical
procedure that is used to beam constraint the real data. 
The results are insensitive to the cuts on $D_0$ and to variations of
the correlation.
\end{itemize}

We also try setting the energy scale with the $E/p$ distribution 
before the beam constraint.  We compare the Monte Carlo distribution 
to the data distribution. 
We get a result for the energy scale which is consistent with the beam
constrained $E/p$ result.  

\item [Landau-Pomeranchuk-Migdal Effect.] Multiple scattering of the electron
can suppress the production of bremsstrahlung at low photon
energies~\cite{lpm}. Qualitatively, if the electron is disturbed while in the
``formation zone'' of the photon, the bremsstrahlung will be suppressed.  The
``formation zone'' is appreciable for the low energy bremsstrahlung.  
(Similarly, the
electron bending in a magnetic field can also suppress low energy photons, but
the CDF magnet is not strong enough for this to be significant.)  SLAC has
measured this effect for $25$ GeV electrons.  The suppression of
bremsstrahlung depends on the density of the material and occurs below 
$y \simeq 0.01$ for gold and $y=0.001$ for carbon, 
where $y$ is the fraction of the
electron energy taken up by the photon. The average density of material in the
CDF detector before the CTC is closer to carbon than gold, and since we have a
cutoff at $y=0.001$, we are in effect simulating $100\%$ suppression for the
carbon case.  This is a negligible effect on $E/p$.  Any effect, if
there were, will make the discrepancy bigger.

\item [Synchrotron Radiation.] 
We considered the possibility that secondary particles, such as
synchrotron photons, may interact in the drift chamber, generating
spurious hits and biasing the electron momentum measurement.  To
estimate the effect of synchrotron photons, we used a simple Monte
Carlo simulation to convolute the synchrotron radiation spectrum for
35~GeV electrons with the photoelectric absorption length in
argon/ethane.  Assuming each absorbed photon to produce one drift
chamber hit (except for the merging of nearby hits due to finite pulse
widths), electron and photo-electron hits were fed to a hit-level
drift chamber simulation and processed by the full track
reconstruction software.  The predicted bias in beam-constrained
momenta due to synchrotron photons was $\sim -0.02\%$, more than an
order of magnitude too small to explain the energy scale discrepancy.
We performed a second study, using a GEANT-based detector simulation
under development for a future run of the CDF experiment.  We used
GEANT to simulate secondary particles near a 35~GeV electron, using
the material distribution of the upgraded detector, and transplanted
the secondaries into the same hit-level simulation used in the first
study.  The bias due to secondary particles was again $\sim -0.02\%$.
We conclude that interactions of secondary particles in the drift
chamber are unlikely to be the source of the discrepancy.

\item [Significant Energy Loss in Silicon Crystals.]  
An electron moving through
the material before the CTC will pass through $\approx 400~\mu$m of aligned
silicon crystals.  If it travels through the crystal along a major axis of
symmetry, it can potentially lose significantly more energy than is lost
through bremsstrahlung~\cite{sicrystal}.  However, in the data we
do not see any significant difference between electrons that pass through the
SVX$^\prime$ and those that do not, relative to the Monte Carlo.  
This indicates that this is not a significant effect. 

\item [Track Quality Comparison.]
In a completely data-driven study, we examined a large number
of track quality variables, such as hit residuals signed in various
ways, track $\chi^2$, and correlations between hit residuals, as well
as occupancies and pulse widths.  While we had no quantitative model
in mind to set the scale for comparisons, none of the track variables
we considered showed any significant difference between the $W$ electron
and $W$ muon samples.

\end{description}

\subsection{Other Checks}

\begin{description}

\item [Invariant Mass Measurement:] Calculating the invariant
mass of {\zee} events makes use of a different set of track parameters than
calculating $E/p$, and one could hypothesize errors in the angular variables
causing errors in the invariant mass.  We would not necessarily expect the
electron and muon invariant masses to look the same since one uses $E_T$ and
the other $p_T$. One could also imagine measurement correlations between the
different tracking parameters which have the net effect of shifting the
measured mass.  The two tracks themselves could also be correlated since
for $Z$
events they are largely back-to-back. For example, if one track enters a
superlayer on the right side of a cell, the other track will be biased to do
the same.  However, we have not been able to see any effect 
on the $Z$ mass in the data.

\item [Inner Superlayers:] 
Wires of the CTC inner superlayers have larger occupancy than those
of the outer superlayers, giving a higher probability of using wrong
hits in the inner layers.
To check this the $Z$ electron tracks are refit 
with superlayers 0 and 1 removed. While the resolution becomes 
worse, no significant change is seen in the means of $E/p$ of the
electrons or the invariant mass of $Z$ electron tracks.
Refitting is also done with the same tracks but by 
removing superlayer 5 instead of 0 and 1.  
Again no significant change was observed in the means of $E/p$, 
or the invariant mass of $Z$ electron tracks.
The mean of the $E/p$ distribution of $W$ data is checked with
the number of stereo or axial hits used in the track reconstruction. 
It is found to be insensitive to the number of hits.

\item [Coding Errors.] 
Several independent $E/p$ simulation codes produce highly 
consistent results.

\item [CEM Non-Linearity.] 
When we applied the non-linearity correction of
Section~\ref{enonlin}, the CEM energy scale factor as determined from
$E/p$ moved from $0.9963$ to $0.9948$, which makes the discrepancy 
between $E/p$
and $M_Z$ worse.  The uncertainty on the energy scale was also significantly
increased by the uncertainty on the non-linearity.  If we do not consider a
non-linearity correction, then the discrepancy between the Z mass energy
scale and the $E/p$ energy scale is closer to $3.3$ standard deviations.
The data (see Figure~\ref{f_enonlin}), 
however, support a CEM non-linearity.  

\item [Amount of Material is Incorrect.]  
To increase the fitted energy scale by $0.5$~\%, 
we would have to increase the amount of
material in the Monte Carlo by $\sim 5.6$~\% of a radiation length.  
However, the tail of the $E/p$ distribution of the $W$ data is not 
consistent with such an increase. Moreover, the low tail of the invariant 
mass distribution of $J/\psi\rightarrow \ee$ decays (see
Figure~\ref{f_mass_ee}) has been examined, 
and such an increase in the amount of material would significantly 
contradict the data. 

\item [Backgrounds are Biasing the Result.] It is possible that our estimate of
the $E/p$ shape of the background is flawed, and that there is a significant
source of non-electron background in the $E/p$ peak region that is biasing our
energy scale fit.  We consider the worst case possibility that all the
background is located at one of the edges of the $E/p$ fit region.  
To increase the $S_E(E/p)$ to 1, we would need to have about $6\%$ 
background piled up at $E/p=1.1$.  This is a factor of $\sim$17 larger
than the QCD background we have measured, and since
we expect the QCD background to be largely flat in $E/p$, we do not expect that
backgrounds are significantly biasing our result.  The agreement of the
$Z$ $E/p$ fit with the $W$ fit also indicates that the backgrounds are 
not a significant
effect in the $W$ fit.  

\item[Tracking Resolutions Not Simulated Correctly.] 
For the Monte Carlo,
we smear the track parameters according to the calculated 
covariance matrix, and we then apply the beam constraint according 
to this same covariance matrix.
Thus, in the Monte Carlo, the covariance matrix used in the beam constraint
describes the correlations and resolutions of the track parameters exactly.  On
the other hand, it is not necessarily the case for the data that the
correlations and resolutions are described correctly by the covariance matrix. 

We can measure the correlation between impact parameter and curvature by
plotting the average of $qD_0$ as a function of $E/p$.  The slope of this
plot for the data is slightly different than for the Monte Carlo. Since the
Monte Carlo covariance matrix is the same matrix that is used to beam constrain
the data, we conclude that the beam constraint covariance matrix does not
perfectly describe the underlying measurement correlations of the data.

To see how much of an effect this has on $E/p$ we run the Monte Carlo 
as follows:
We smear the Monte Carlo according to an adjusted covariance matrix, where all
the off-diagonal terms are set to $0$ except for $\sigma^2(C,D_0)$, and
which we fix according to the $W$ data.  When we apply
the beam constraint, however, we use the same covariance matrices that are used
by the data to do the beam constraint. In this way, we simulate the data more
closely: smearing according to one matrix, and beam constraining according to a
slightly different matrix. We find no effect on the average $E/p$ between $0.9$
and $1.1$. 

\item [The Solenoid May Cause Non-Linearity in Photon Response.]~~~
The solenoid
coil presents $\sim 1$ radiation length for electrons 
in $W$ and $Z$ events, and
also for any associated soft photons.   Electron energy losses in the solenoid
are not expected to affect our results since they are part of the CEM scale,
which we are fitting for.  However, it is possible that the soft photons are
not making it through the solenoid and that this is distorting the $E/p$
shape.  As a simple check, we use a formula from the PDG Full
Listings~\cite{pdg} which describes the energy loss profile of a particle as a
function of its depth in radiation lengths.  We apply this formula to all the
photons created in the Monte Carlo and reduce their energy accordingly.  This
is not a rigorous check since we are applying the formula to low
energy photons, which are in an energy region where the formula is not
necessarily accurate. We rerun the $Z$ Monte Carlo with this effect put in, and
we treat this new Monte Carlo as ``data'' and fit it with the default Monte
Carlo.  Fitting $E/p$ gives a Monte Carlo energy scale of $0.99960$, 
and fitting $M_Z$
gives a scale of $0.99935$.  We are interested in $M_Z$ relative to $E/p$, and
thus $0.99960-0.99935=0.00025 \pm 0.00015$. 
This is more than an order of magnitude too small to explain
the energy scale discrepancy.

\end{description}

\subsection{Conclusion}
\label{section:finalconclusion}

We have measured the energy scale using the peak of the $E/p$
distribution of $W$ data. The $E/p$ distribution of $Z$ events 
gives consistent results for the $E/p$ distribution of $W$ events.  
However, if we set the energy scale with $E/p$, then
the invariant mass distribution of the $Z$ events comes out 
significantly low. As
a check we have refit the Run IA data with the Run IB Monte Carlo simulation, 
and the result agrees excellently with the published results.

We have discussed several possible reasons that the $Z$ mass comes out 
wrong. The problem could be a momentum scale problem or otherwise a 
tracking problem; it could be related to our simulation of $E/p$ as 
presented in this paper; or it could be something theoretically unexpected.  
None of the plausible explanations considered here appears to be 
capable of creating a discrepancy of the magnitude observed in 
the Run IB data sample, and the source for the inconsistency 
remains an open question.

For the final $W$ mass measurement reported in this paper, we have 
used the invariant mass of the {\zee} and {\zmumu} events.  
In this way, we have 
separated our energy scale measurement from almost all questions 
associated with the $E/p$ method. 



\begin{thebibliography}{99}
\bibitem{mwave} This average includes the LEP II measurements upto
	$\sqrt s$ = 183~GeV.  Thus the ALEPH's published measurement at 
	$\sqrt s$ = 189~GeV and the OPAL's preliminary measurement at
	$\sqrt s$ = 189~GeV are not included.
\end{thebibliography}
\end{document}